\documentclass{aa830} 
\usepackage{natbib}
\usepackage{txfonts}
\usepackage{textcomp}
\usepackage{upgreek}
\bibpunct{(}{)}{;}{a}{}{,}
\usepackage[figuresright]{rotating}
\usepackage{savesym}
\usepackage{amsmath}
\usepackage{amsfonts}
\usepackage{color}
\usepackage{graphicx}
\usepackage{appendix}
\usepackage{verbatim}   % useful for program listings
\usepackage{multirow}
\usepackage{subfigure}  % use for side-by-side figures
\makeatletter
\renewcommand\aa@pageof{, page \thepage{} of 20}
\makeatother

\begin{document}


\title{Dust destruction by the reverse shock 
    in the Cassiopeia A supernova remnant}

%% You can use \email to mark an email address

\author{Elisabetta R. Micelotta\inst{1,2,3}, Eli Dwek\inst{2}, and
  Jonathan D. Slavin\inst{4}}
\offprints{E. R. Micelotta}
\institute{Institut d'Astrophysique Spatiale, UMR 8617,
  Universit{\'e} Paris-Sud, 91405, Orsay, France 
\and
Observational Cosmology Lab, Code 665, NASA Goddard Space Flight
Center, Greenbelt, MD 20771, USA
\and
Department of Physics, PO Box 64, 00014 University of Helsinki,
Finland \\
Present address. \\
\email{elisabetta.micelotta@helsinki.fi} 
\and
Harvard-Smithsonian Center for Astrophysics, 60 Garden Street, Cambridge, MA 02138, USA
}

\date{Received 12 September 2015 / Accepted 6 February 2016}

\abstract
% Context heading (optional), leave it empty if necessary: 
{ Core collapse supernovae (CCSNe) are important sources of
interstellar dust, which are potentially capable of producing
1~{M$_{\odot}$} of dust in
their explosively expelled ejecta. However, unlike other dust sources,
the dust has to survive the passage of the reverse shock, generated by
the interaction of the supernova blast wave with its surrounding
medium. Knowledge of the net amount of dust produced by CCSNe is
crucial for understanding the origin and evolution of dust in the
local and high-redshift universe.}
% Aims heading (mandatory):
{ We identify the dust destruction mechanisms in the ejecta and
derive the net amount of dust that survives the passage of the reverse
shock.  }
% Methods heading (mandatory): 
{We use analytical models for the evolution of a supernova blast wave
and of the reverse shock with special application to the clumpy ejecta
of the remnant of Cassiopeia~A (Cas~A).  We assume that the dust
resides in cool oxygen-rich clumps, which are uniformly distributed
within the remnant and surrounded by a hot X-ray emitting plasma
(smooth ejecta), and that the dust consists of silicates (MgSiO$_3$)
and amorphous carbon grains. The passage of the reverse shock through
the clumps gives rise to a relative gas-grain motion and also destroys
the clumps. While residing in the ejecta clouds, dust is processed via
kinetic sputtering, which is terminated either when the grains escape
the clumps or when the clumps are destroyed by the reverse shock. In
either case, grain destruction proceeds thereafter by thermal
sputtering in the hot shocked smooth ejecta.
}
% Results heading (mandatory): 
{We find that 11.8 and 15.9 percent of silicate and carbon dust,
respectively, survive the passage of the reverse shock by the time the
shock has reached the centre of the remnant. These fractions depend on
the morphology of the ejecta and the medium into which the remnant is
expanding, as well as the composition and size distribution of the
grains that formed in the ejecta. Results will therefore differ for
different types of supernovae. }
% Conclusions heading (optional), leave it empty if necessary:
{}

%% Keywords should appear after the \abstract command. See the
%% instructions to authors for the journal to which you are submitting
%% your paper to determine what keyword punctuation is appropriate.


 \authorrunning{E. R. Micelotta et al.}

   \titlerunning{Dust destruction in the Cas~A supernova remnant}
   \maketitle


%=========================================
% Section 1
\section{Introduction}
%=========================================

Core collapse supernovae (CCSNe) are major producers of heavy
elements and drive the chemical enrichment in galaxies.  The high
densities and preponderance of refractory elements in their ejecta
provide the necessary conditions for the formation and growth of dust.
Dust of supernova (SN) origin was found in meteorites \citep[see
reviews in][]{clayton04, zinner08}, and its presence was also
inferred from their infrared (IR) and sub-millimeter emission in young
SNe, such as SN~1987A \citep[][and references therein]{moseley89b,
dwek92, wooden97, bouchet04, indebetouw14, matsuura15}, and
in young unmixed supernova remnants (SNRs) such as Cassiopeia~A
\citep[Cas~A; ][]{lagage96, rho08, dunne09, barlow10, arendt14} or
the Crab Nebula \citep{hester08, gomez12a, temim13}.  The observations
of SNRs suggest that the typical yield of dust is about 0.1~{M$_{\odot}$},
with the largest inferred dust mass of $\sim$0.5~{M$_{\odot}$} in SN~1987A
\citep{matsuura15}. Theoretical models suggest that refractory
elements precipitate very efficiently from the gas phase, giving a
dust yield of 0.1-0.5~{M$_{\odot}$} for a typical 25~{M$_{\odot}$} progenitor star
\citep{todini01, schneider04, cherchneff10, nozawa10}. A review of the
production of dust in galaxies was presented by \citet{gall11c}.

Supernovae also destroy dust during the remnant phase of their
evolution. Most recent calculations show that in the local solar
neighbourhood SNRs destroy more dust than is produced by SNe and
asymptotic giant branch (AGB) stars combined \citep[][ and references
therein]{bocchio14, slavin15}. Similar conclusions were reached for
the Magellanic Clouds \citep{temim15}. This imbalance between the
production and destruction rate also prevails in local and
high-redshift galaxies; this suggests that dust may need to
reconstitute by accretion in the dense phases of the interstellar
medium \citep[ISM; ][]{dwek80b, valiante11, dwek11b,gall11b,
gall11a}. The problem becomes more acute in the very high redshift
universe ($z \gtrsim 6$) where large amounts of dust have been
detected \citep[$\gtrsim$ 10$^{7}${M$_{\odot}$} -- ][]{bertoldi03,
watson15}, but AGB stars have not yet evolved to form dust. If they
are to be the sole source of dust in the early universe
\cite{dwek11a}, SNe must then produce $\sim$1~{M$_{\odot}$} of
dust. Supernovae are net producers of interstellar dust in galaxies
with metallicities below $\sim$0.002 only, and SNe are capable
of producing a significant amount of dust with a dust yield of 
$\sim$0.1~{M$_{\odot}$}/SN without resorting to the need of grain growth in
the ISM \citep{dwek14}.

Calculated dust yields, and dust yields that are observationally
inferred for very young SNe, do not reflect the net amount of dust that
is ultimately injected by SNe into the ISM. The pressure of the ISM gas
that is shocked by the expanding SN blast wave generates a reverse
shock that propagates through the expanding ejecta \citep{mckee74,
truelove99}, partially destroying the newly formed dust by sputtering
\citep[][]{dwek05, bianchi07, nath08, nozawa07,silvia10,silvia12, micelotta13,
biscaro14, biscaro16}.  However, because of the complexity of the
problem, the total amount of mass destroyed is still unclear.

In this paper we calculate the mass of dust destroyed in the most
extensively studied SNR, Cas~A. We illustrate the relative importance
of the many physical processes in the shocked ejecta by generalizing the
analytical model of \citet[][hereafter TM99]{truelove99} to describe
the evolution of the forward and reverse shock in the remnant. We
adopt a set of parameters that reproduce the dynamics and the density
and temperature profile of the Cas~A ejecta. The reverse shock has
partially propagated through its ejecta \citep{gotthelf01} heating the
SN-condensed dust, which gives rise to observed mid- to far-IR emission
from the remnant \citep{ennis06,rho08,arendt14}. The mass of dust
inferred from these observations is about 0.1~{M$_{\odot}$}
\citep{arendt14}. We assume that all the dust is initially in the
clumps, and that there is no dust in the interclump region. The mass,
composition, and size distribution of the dust that survives the
passage of the reverse shock and is injected into the ISM is the
subject of this publication.

The paper is organized as follows.  We first present our
generalization of the analytical model of TM99 to the case of
non-uniform ejecta expanding into a non-uniform ambient medium
(Sect.~\ref{sec:dynamics}), and in Sect.~\ref{sec:ejecta_properties}
we provide the specific results of this model for Cas~A. We assume
that the dynamics of the reverse shock is unaffected by the presence
of clumps (clouds) in the ejecta, and adopt the parameters of
\citet{sutherland95} to describe their density. The density contrast
between the clumps and the smooth ejecta determines the velocity of
the reverse shock that propagates into the clumps.

The reverse shock propagating through the clumps has low velocity
compared to that of the reverse shock propagating through the smooth
ejecta. The passage of the reverse shock causes a relative dust-gas
motion and subjects the dust to kinetic sputtering. The amount of
grain destruction depends on the initial velocity of the grains, and
the mass column density they traverse through the clump, which in turn
depends on both the size of the grains and their position in the
cloud. In Sect.~\ref{sec:kin_sput} we present the formalism that we
use to investigate the effect of kinetic sputtering inside the clumps.

Kinetic sputtering is terminated either when the grains escape the
clump, or when the clump is destroyed by crushing and evaporation. At
the end of the kinetic sputtering process, the grains are injected
into the ambient hot gas. The average thermal kinetic energy of the
gas exceeds the kinetic energy of the grains, and grain destruction
proceeds predominantly by thermal sputtering.  In
Sect.~\ref{sec:clump_stability} we study the time it takes the dust to
travel through the cloud, and compare this to the cloud crushing and
evaporation timescales.

Any dust escaping the clumps is subjected to thermal sputtering in the
hot SN cavity. Conditions in the cavity evolve with time as the
density and temperature of the X-ray emitting gas decrease by
expansion and radiative cooling. The destruction of dust in this
evolving environment is presented in Sect.~\ref{sec:thermal_sput}. We
adopt silicates in the form of MgSiO$_3$ and amorphous carbon as the
main dust components in the ejecta, and assume that their initial size
distribution is characterized by a power law in grain
radii. Section~\ref{sec:MC_simulation} presents our Monte Carlo
approach to evaluate kinetic and thermal sputtering, together with our
results in terms of the net amount and mass distribution as a function
of grain size of the surviving dust in Cas~A.  A discussion of the
astrophysical significance of our findings is presented in
Sect.~\ref{sec:discussion}, and our results are summarized in
Sect.~\ref{sec:summary}.

%=========================================
% Section 2
\section{Dynamics}\label{sec:dynamics}
%=========================================
Cas~A is a young SNR ($\sim$333 years in 2004) currently in an
intermediate stage between the two non-radiative phases of
supernova remnants 
evolution \citep{laming03}: the \emph{ejecta-dominated} phase (ED), when the
mass of the ejecta is dominant with respect to the swept-up
circumstellar material, and the \emph{Sedov-Taylor} phase (ST), which starts
as the remnant has swept-up an amount of circumstellar material
comparable with the mass of the stellar ejecta.
 
To describe the dynamical evolution of the SNR through the ED and ST
stages, we refer to the analytical treatment of TM99. This seminal
work focused on the evolution of supernova ejecta expanding into an
uniform density ambient medium, emphasising the dependence of the ED
stage on the ejecta parameters and providing analytical expressions
that smoothly merge the blast-wave and reverse shock solutions between
the ED and ST stages. We generalize their treatment to general
power-law ambient media (described by an index $s$: $\rho(r) =
\rho_sr^{-s}$), following the indications provided in their
Appendix~A.  Then, referring to the work of \citet{laming03}, we
consider the specific case of $s=2$ appropriate for Cas~A, i.e. ejecta
expanding into a pre-supernova steady stellar wind \citep[][and
references therein]{vanDenBergh71, chevalier94, laming03}.

\subsection{Characteristic scales}

As shown by TM99, in the problem considered here the initial conditions
introduce three independent dimensional parameters: the ejecta energy, $E$,
ejecta mass, $M_{\rm ej}$ and normalization parameter for the ambient 
density, $\rho_s$. These three dimensional parameters can be combined in a unique 
way to define characteristic scales of length, time, and mass as follows:
\begin{eqnarray}\label{ch_quantities}
  R_{\rm ch} & \equiv & M_{\rm ej}^{1/(3-s)}\rho_s^{-1/(3-s)}\, , \\
  t_{\rm ch} & \equiv & E^{-1/2}\,M_{\rm ej}^{(5-s)/2(3-s)}\rho_s^{-1/(3-s)}\, , \\
  M_{\rm ch} & \equiv & M_{\rm ej}\,.
\end{eqnarray}
Additional scales can be derived from the following base set:
\begin{eqnarray}\label{more_ch_quantities}
  v_{\rm ch} &\equiv & R_{\rm ch}/t_{\rm ch}  =  (E/M_{\rm ej})^{1/2} \\
  T_{\rm ch} & \equiv & \frac{3}{16}\,\frac{\mu}{k_{\rm B}}\,v_{\rm ch}^2 
,\end{eqnarray}
where $\mu$ is the mean mass per particle
and $k_{\rm B}$ is the Boltzmann constant.
   
We adopt $s=2$, which corresponds to ejecta expanding into a
pre-supernova steady stellar wind. In this case, $\rho_s = \rho_2 =
n_0 R_{\rm b0}^2$, where $R_{\rm b0}$ is the blast-wave radius at a
given time and $n_0$ is the circumstellar density at $R_{\rm b0}$. We
can write
 \begin{eqnarray}
   R_{\rm ch} &  = & 40.74\,M_{\rm ej}\,(n_0\,R_{\rm b0}^2)^{-1}\;\;\rm pc   \\
   t_{\rm ch} & = & 5633\, E^{-1/2}\,M_{\rm ej}^{3/2}\,(n_0\,R_{\rm b0}^2)^{-1}\;\;\rm yr
,\end{eqnarray}
with $M_{\rm ej}$ in solar masses, $E$ in units of $10^{51}$ ergs,
$n_0$ in H atoms per cubic centimeter, and $R_{\rm b0}$ in parsecs.

Following TM99, we perform our calculations using the dimensionless
starred variables defined as $R^* \equiv R/R_{\rm ch}$, $t^* \equiv
t/t_{\rm ch}$, $v^* \equiv v/v_{\rm ch}$, etc. The results are
presented in usual units for clarity.

\subsection{Initial conditions}\label{sec:initial_conditions}

We assume that the SN ejecta initially expands homologously with a
maximum ejecta velocity $v_{\rm ej}$, and velocity profile $v(r) \sim
r/t$ for $r< R_{\rm ej}$, where $R_{\rm ej} \equiv v_{\rm ej} t$ is
the free-expansion radius of the outer layer of the ejecta if no
ambient medium is present. The density is given by the following
expression (Eq. A1 in TM99):
\begin{equation}\label{initial_density}
  \rho(r, t) = 
\begin{cases}
  \;\rho_{\rm ej}(v, t)  \equiv \frac{M_{\rm ej}}{v_{\rm
      ej}^3}\,f(\frac{v}{v_{\rm ej}})\,t^{-3}\;\;\; r<R_{\rm ej} \\
  \;\rho_sr^{-s} \qquad \qquad \qquad  \quad \:\;\;  r>R_{\rm ej} 
\end{cases}
,\end{equation}
where $\rho_s$ is a normalization constant and the condition $s<3$ is
required to ensure a finite ambient mass. In the initial
ejecta density distribution, the term $t^{-3}$ accounts for the free
expansion of the ejecta before encountering the circumstellar medium,
while the time-independent shape of the density distribution is
described by the \emph{structure function} $f(v/v_{\rm ej})$.  We
consider a power-law structure function expressed as
\begin{equation}\label{f_w}
  f(w) = 
\begin{cases}
  \;f_0  \qquad \qquad  0 \leq w\leq w_{\rm core} \\
  \;f_nw^{-n} \qquad \;\; w_{\rm core} \leq w \leq 1 
\end{cases}
,\end{equation}
where $w \equiv v/v_{\rm ej}$, $w_{\rm core}=v_{\rm core}/v_{\rm ej}$
and $v_{\rm core}$ is the velocity of the ejecta at the boundary
between an inner uniform \emph{core} region and an external
power-law \emph{envelope} region characterized by the index
$n$. Following \citet{chevalier03} and \citet{laming03}, we adopt for
Cas~A the value $n$=9, which provides: {\it i}) the
correct ratio between the forward and reverse shock radii, {\it ii})
the correct relationship between the forward shock and free-expansion
rates, and {\it iii}) a more plausible ejecta mass of 2~M$_{\sun}$. As
discussed by TM99, a power-law ejecta envelope with index $n>5$
requires the presence of a core in order for $M_{\rm ej}$ to be
finite.

The parameter $f_0$ can be determined imposing the continuity of $f(w)$ 
in $w_{\rm core}$, i.e.
\begin{equation}\label{f_0}
  f_0 = f_nw^{-n}_{\rm core}.
\end{equation}
The value of $f_n$ results from the condition that the integral of the mass
must equal $M_{\rm ej}$, i.e.
\begin{equation}\label{f_n}
  f_n = \frac{3}{4\pi}\,\left[\frac{1-n/3}{1-(n/3)\,w^{3-n}_{\rm core}}\right].
\end{equation}
We can introduce the following dimensionless ratio:
\begin{equation}\label{energy_ratio}
  \frac{E}{(1/2)\,M_{\rm ej}v_{\rm ej}^2} = \left(\frac{3-n}{5-n}\right)
  \left(\frac{w^{-(5-n)}_{\rm core}-n/5}{w^{-(3-n)}_{\rm core}-n/3}\right)w_{\rm core}^2
  \equiv \bar{\alpha}.
\end{equation}
If $n>5$, in the limit $w_{\rm core}\rightarrow0$, from Eq.~\ref{initial_density} we obtain
\begin{equation}\label{rho_ej}
 \rho_{\rm ej}(v, t)  \equiv \frac{M_{\rm ej}}{v_{\rm core}^3}
 \left(\frac{3}{4\pi}\frac{n-3}{n}\right)\left(\frac{v}{v_{\rm core}}\right)^{-n}\,t^{-3}
\end{equation}
while Eq.~\ref{energy_ratio} becomes
\begin{equation}\label{energy_ratio_core}
  \frac{E}{(1/2)\,M_{\rm ej}v_{\rm core}^2} = \frac{3}{5}\,\left(\frac{n-3}{n-5}\right).
\end{equation}
From this, we can derive the expression for $v_{\rm core}$ as follows:
\begin{equation}\label{v_core}
  v_{\rm core} = \left(\frac{10}{3}\frac{n-5}{n-3}\right)^{1/2}
  \left(\frac{E}{M_{\rm ej}}\right)^{1/2}.
\end{equation}
Remembering that $v_{\rm ch} = (E/M_{\rm ej})^{1/2}$  and that 
$v^*\equiv v/v_{\rm ch}$ we then obtain 
\begin{equation}\label{v_core_star}
  v^{*}_{\rm core} = \left(\frac{10}{3}\,\frac{n-5}{n-3}\right)^{1/2}.
\end{equation}

  For the density profile of the ejecta core,
  Eq.~\ref{initial_density} and Eq.~\ref{f_w} yield
\begin{equation}\label{core_density_1}
  \rho_{\rm ej}(t) = \frac{M_{\rm ej}}{v_{\rm
      ej}^3}\,f_0\,t^{-3}.
\end{equation}
To derive $f_0$, we impose the continuity between
Eq.~\ref{core_density_1} and Eq.~\ref{rho_ej}, which is valid for the
ejecta envelope, at the boundary between the inner uniform core
and the external power-law envelope, where $v = v_{\rm core}$. For
$v_{\rm core}$, we use Eq.~\ref{v_core}. Following this procedure
we obtain
\begin{equation}\label{f_0_core}
 f_0 = \left(\frac{v_{\rm ej}}{v_{\rm core}}\right)^3 \,
 \left(\frac{3}{4\pi}\frac{n-3}{n}\right)
,\end{equation}
which substituted in Eq.~\ref{core_density_1} provides the following
expression for the density profile of the ejecta core:
\begin{equation}\label{core_density_2}
  \rho_{\rm ej}(t) = \frac{M_{\rm ej}}{v_{\rm
      core}^3}\,\left(\frac{3}{4\pi}\frac{n-3}{n}\right) \,t^{-3}.
\end{equation}

% FIGURE 1  *************************************************************
%
\begin{figure}
  \begin{center}
    \includegraphics[width=\hsize]{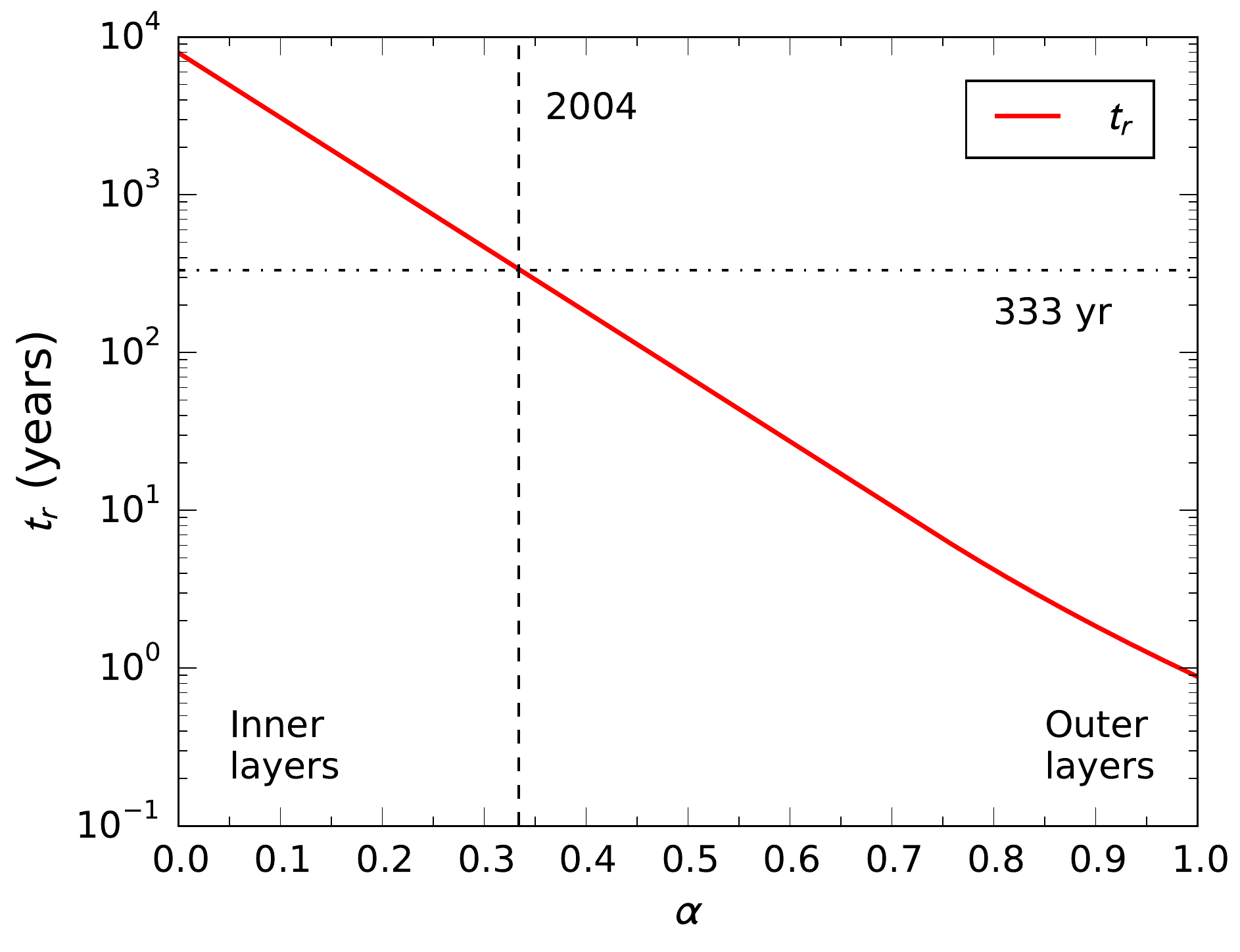}
  \end{center}
  \caption{Time, $t_{\rm r}$, after the explosion at which an ejecta
    layer characterized by the parameter $\alpha$ encounters the reverse
    shock. The outer layer has the value $\alpha$~=~1, 
    while the inner one has $\alpha$~=~0. The vertical line indicates the
    value $\alpha$~=~0.33, corresponding to the ejecta layer which
    encounters the reverse shock in 2004, and of course
    intersects the red solid curve at the time corresponding to the
    age of Cas~A in 2004, i.e. $t_{\rm r}$ = 333 yr (indicated by the
    horizontal line).}
    \label{fig:tr_vs_alpha} 
\end{figure}
% ********************************************************************

% FIGURE 2  *************************************************************
%
\begin{figure}
  \begin{center}
    \includegraphics[width=\hsize]{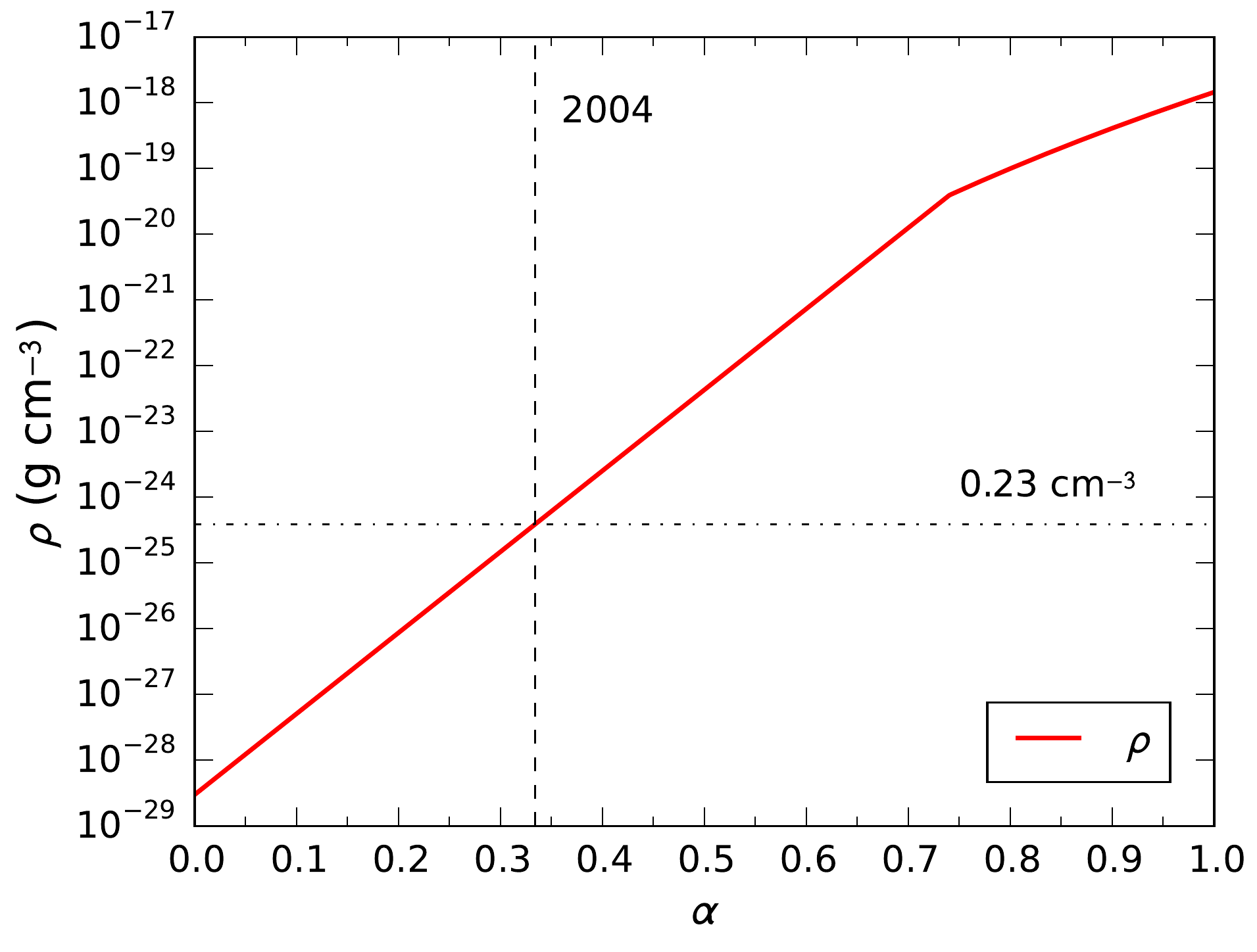}
  \end{center}
  \caption{Density structure of the ejecta calculated from
    Eq.~\ref{rho_ej} but expressed as a function of the parameter
    $\alpha$. Vertical line as in Fig.~\ref{fig:tr_vs_alpha}. The
    horizontal line indicates the mass density of the Cas~A smooth ejecta in
    2004 and the label indicates the corresponding number density.
    \label{fig:rho_vs_alpha} }
\end{figure}
% ********************************************************************

It is convenient to characterize the layers of the ejecta by a
dimensionless parameter $\alpha$, defined by $v(0) \equiv \alpha\,v_{\rm
  ej}$ ($\alpha \leq$ 1), where $v(0)$ is the initial velocity of the
layer with respect to the unperturbed ISM. The outer layer of the
ejecta expands at velocity $v_{\rm ej}$ and is therefore characterized
by $\alpha$~=~1, whereas the innermost layer has a value of
$\alpha$~=~0. A layer $\alpha$ encounters the reverse shock at time
$t_{\rm r}$ when its radius is equal to the position of the reverse
shock, i.e. when $\alpha\,v_{\rm ej}\,t_{\rm r} = R_{\rm r}(t_{\rm
  r})$. From this, we obtain
\begin{equation}\label{beta_eq}
  \alpha = \frac{R_{\rm r}}{v_{\rm ej}\,t_{\rm r}} = \frac{R_{\rm r}}{R_{\rm ej}}.
\end{equation} 
The relation between $t_{\rm r}$ and the parameter $\alpha$ is shown
in Fig.~\ref{fig:tr_vs_alpha}, while the ejecta density from
Eqs.~\ref{rho_ej} {  and \ref{core_density_2},} but expressed
as a function of $\alpha$, is reported in
Fig.~\ref{fig:rho_vs_alpha}. {The mass density of the Cas~A
  ejecta in 2004 ($\alpha$~=~0.33) is indicated by the horizontal line
  and corresponds to a number density of 0.23 cm$^{-3}$. This value is
consistent with the average density of $\sim$0.25 cm$^{-3}$ estimated
by \citet{morse04} for the smooth X-ray emitting ejecta in Cas~A. }

% FIGURE 3 *************************************************************
%
\begin{figure*}
  \begin{center}
    \includegraphics[width=0.48\hsize, height=7.cm]{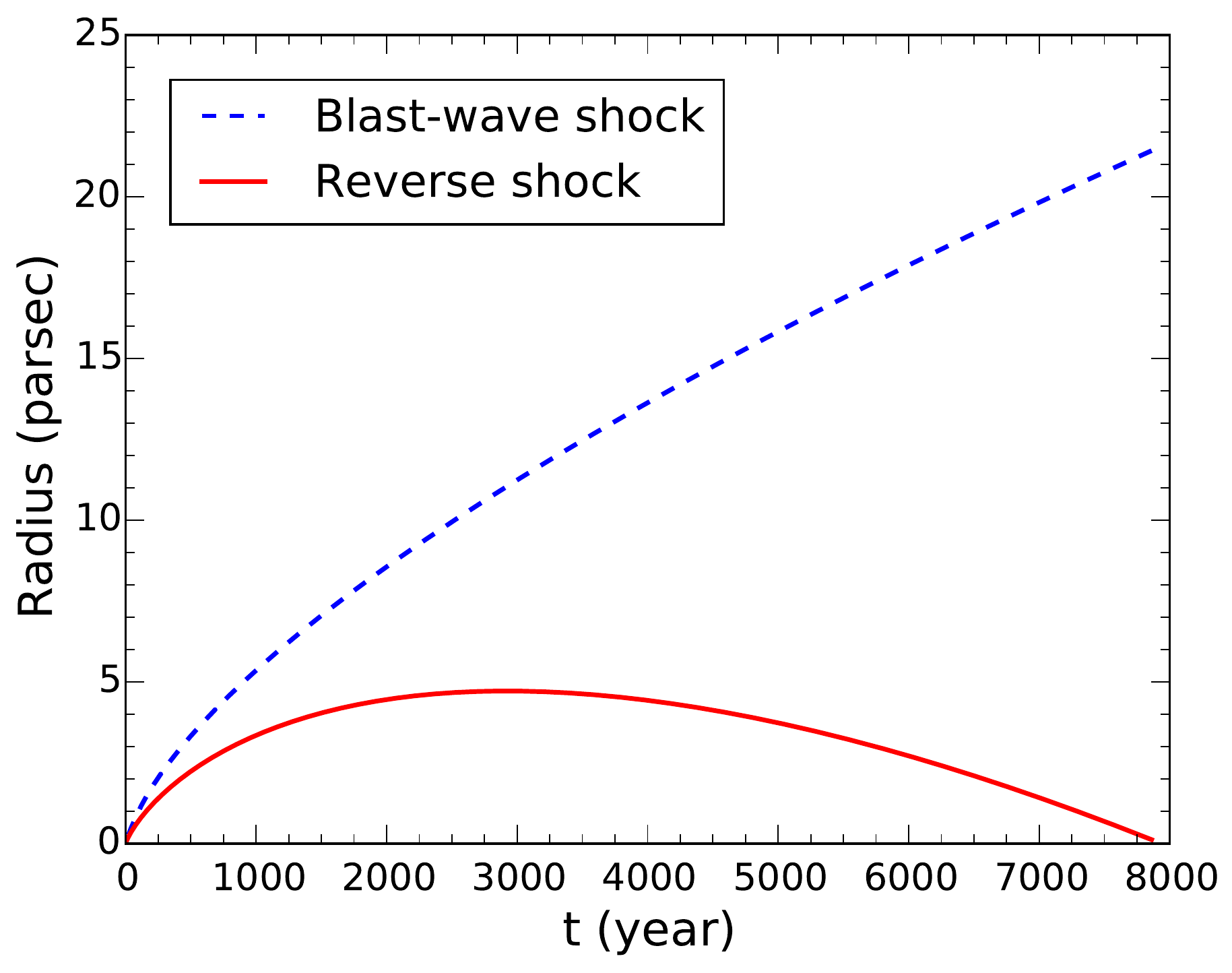}
    \includegraphics[width=0.48\hsize, height=7.cm]{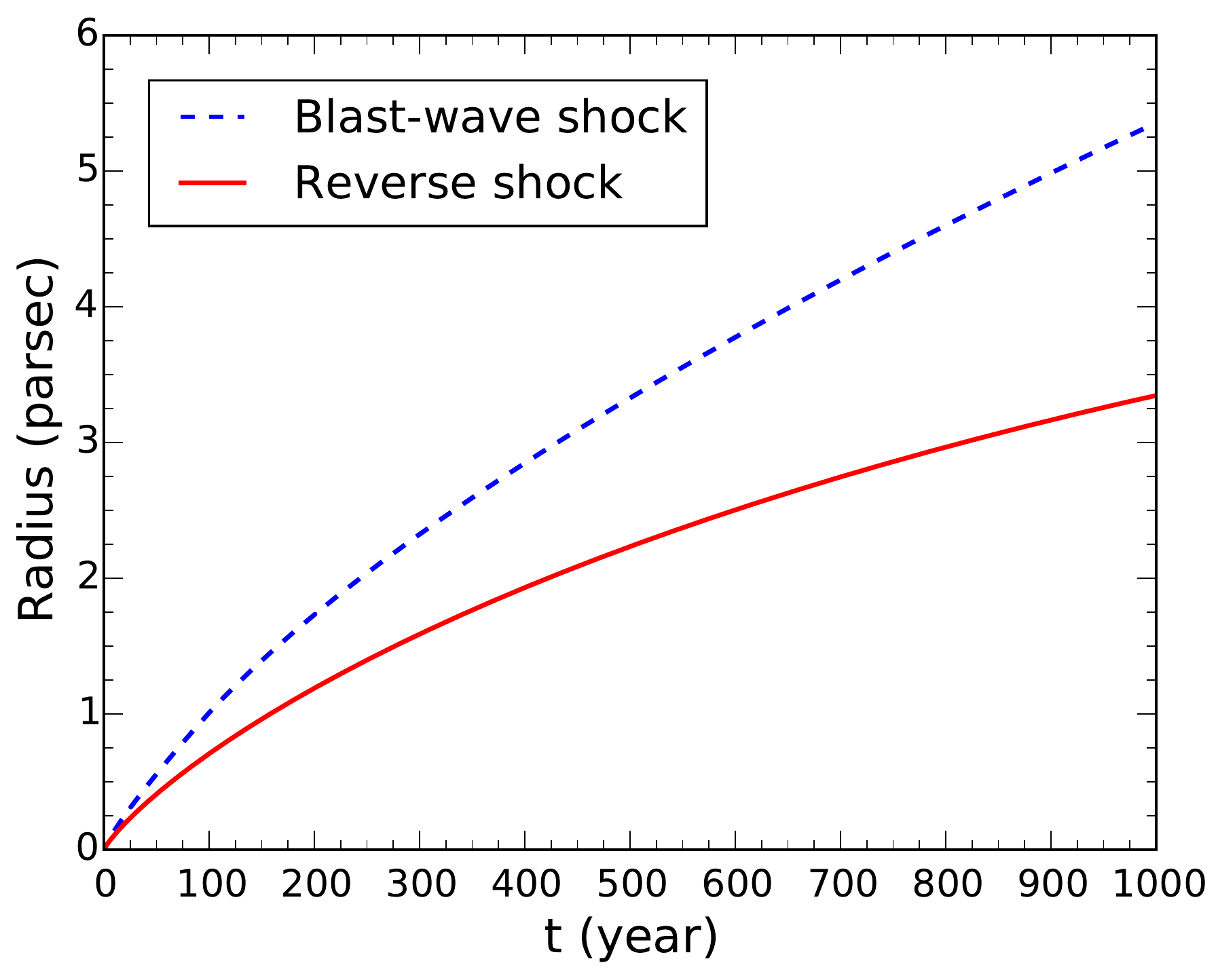}
  \end{center}
  \caption{{\it Left: }Blast-wave and reverse shock radii as a function of the
    time elapsed since explosion. {\it Right: }Zoom of left panel
    for $t \leq$~1000 yr.
\label{fig:blast_reverse_radius_plus_zoom} }
\end{figure*}
% ********************************************************************

\subsection{Final equations for the supernova shocks}\label{sec:final_shock_equations}

To derive the equations describing the evolution of the blast
wave and of the reverse shock, we use the
treatment from \citet{laming03} as a starting point. However, to develop our calculations
we fully evaluated the implications of the assumption that $w_{\rm
  core}\rightarrow0$ \citep[instead of adopting $w_{\rm
  core}\rightarrow1$ as by][]{laming03}. 

In this section, we report the final expressions that we use in
the rest of our study. For more details about the derivation of these
equations, see Appendix~\ref{app_shocks}.

We introduce the quantity $\phi$, defined as the ratio of the
pressures behind the reverse and blastwave shocks \citep{mckee74}. The
\emph{lead factor} $\ell$ is defined as the ratio of the radii of
the forward shock to the reverse shock:
\begin{equation}\label{l_definition}
  \ell(t) \equiv \frac{R_{\rm b}(t)}{R_{\rm r}(t)}
\end{equation}
Following TM99, we make the approximations 
$\phi(t) \cong \phi(0) \equiv \phi_{\rm ED}$ and 
$\ell(t) \cong \ell(0) \equiv \ell_{\rm ED}$. 
The subscript ``ED'' refers to the ejecta-dominated phase of the
evolution of the supernova remnant. With these approximations, from
Eq.~\ref{l_definition} we obtain $R_{\rm r} = R_{\rm b}/\ell_{\rm
ED}$. The quantities $\phi_{\rm ED}$ and $\ell_{\rm ED}$ are well
fitted by the following expressions (Hwang \& Laming 2011):
\begin{equation}
  \phi_{\rm ED} = \left[0.65\,-\,\exp{(-n/4)}\right]\,\sqrt{1-\frac{s}{3}},
\end{equation}
\begin{equation}
  \ell_{\rm ED} = 1\,+\,\frac{8}{n^2}\,+\,\frac{0.4}{4-s}. 
\end{equation}

During the ejecta-dominated phase, the reverse shock propagates
through the ejecta envelope before reaching the ejecta core. The
blast-wave radius during the initial envelope phase is given by the
following expression: 
\begin{equation}\label{blast_radius_envelope_eq}
  R^{*}_{\rm b} =
  \left\{v^{*^{n-3}}_{\rm core}\,
  \frac{(3-s)^2}{n(n-3)}\,\frac{3}{4\pi}\,\frac{\ell_{\rm ED}^{n-2}}
       {\phi_{\rm
           ED}}\right\}^{\frac{1}{n-s}}\;t^{*^{\frac{n-3}{n-s}}},
\end{equation}
while the velocity of the forward shock is written as
\begin{equation}\label{blast_velocity_envelope_eq}
  v^{*}_{\rm b} = \frac{n-3}{n-s}\,\frac{R^{*}_{\rm b}}{t^{*}}.
\end{equation}
This comes directly from Eq.~\ref{blast_radius_envelope_eq} using
the definition of velocity, $v^{*}_{\rm b} = {\rm d}R^{*}_{\rm b}/{\rm d}t^*$.

During the core phase, when the reverse shock propagates through the
ejecta core, the blast-wave radius is given by
%\begin{equation}
\begin{align}\label{blast_radius_core_eq}
%\begin{split}
  R^{*}_{\rm b} & = 
  \Bigg \{\left[\left(\frac{(3-s)^2}{n(n-3)}\,\frac{3}{4\pi}\,\frac{l_{\rm
        ED}^{n-2}} {\phi_{\rm ED}}\right)\left(v^{*}_{\rm
      core}\,t^{*}_{\rm conn}\right)^{n-3}\right]^{\frac{5-s}{2(n-s)}}
   \nonumber  \\  &  +\,\xi_s^{1/2}\left(t^{*} -
  t^{*}_{\rm conn}\right)\Bigg \}^{\frac{2}{5-s}}
%\end{split}
\end{align}
%\end{equation}
and the blast-wave velocity is written as
\begin{equation}\label{blast_velocity_core_eq}
  v^{*}_{\rm b} = \frac{2}{5-s}\,\xi_{s}^{1/2}\,R_{\rm
    b}^{*^{\frac{s-3}{2}}}
\end{equation}
with
\begin{equation}
  \xi_s = \frac{(5-s)(10-3s)}{8\pi}
\end{equation}
and
\begin{align}\label{t_conn_star}
  t^{*}_{\rm conn}  =
  \left(\frac{n-3}{n-s}\sqrt{2\pi\frac{5-s}{10-3s}}\right)^{\frac{2(n-s)}{(5-n)(3-s)}}
   \left(v^{*}_{\rm core}\,\ell_{\rm ED}\right)^{\frac{(n-s)(5-s)}{(5-n)(3-s)}}t^{*^{\frac{5-s}{5-n}}}_{\rm core},
\end{align}
where $v^{*}_{\rm core}$ comes from Eq.~\ref{v_core_star} and
$t^{*}_{\rm core}$ is the time at which the reverse shock hits the
ejecta core, i.e.
\begin{equation}\label{t_core_star}
   t^{*}_{\rm core} = \left[\frac{\ell_{\rm ED}^{s-2}}{\phi_{\rm ED}}
     \frac{3}{4\pi}\frac{(3-s)^2}{n(n-3)}\right]^{1/(3-s)}\frac{1}{v^{*}_{\rm core}}.
\end{equation}

Equations~\ref{blast_radius_envelope_eq} and
\ref{blast_velocity_envelope_eq} are valid for $t^{*} \le t^{*}_{\rm
  conn}$, while Eqs.~\ref{blast_radius_core_eq} and
\ref{blast_velocity_core_eq} are valid for $t^{*} > t^{*}_{\rm
  conn}$. The reason for the introduction of the time $t^{*}_{\rm
  conn}$, its definition and derivation can be found
in Appendix~\ref{app_shocks}.

The equations for the reverse shock are the following. During the
envelope phase, i.e. for $t^{*} \le t^{*}_{\rm core}$, the reverse
shock radius is written as
\begin{equation}\label{reverse_radius_envelope_eq}
    R_{\rm r}^{*} = \frac{R_{\rm b}^{*}}{\ell_{\rm ED}}
\end{equation}
and the reverse shock velocity (in the frame of the unshocked ejecta)
is written as
\begin{equation}\label{reverse_velocity_envelope_eq}
  v_{\rm r}^{*}  = \frac{3-s}{n-3}\,\frac{v_{\rm b}^{*}}{\ell_{\rm ED}}.
\end{equation}
During the core phase, i.e. for $t^{*} > t^{*}_{\rm core}$, we have for the
reverse shock radius the following expression:
\begin{align}\label{reverse_radius_core_eq}
  R_{\rm r}^{*} = \left[\frac{R_{\rm b}^{*}(t^{*} = t^{*}_{\rm
core})}{\ell_{\rm ED}\,t^{*}_{\rm core}} -
\frac{3-s}{n-3}\,\frac{v_{\rm b}^{*}(t^{*} = t^{*}_{\rm
core})}{\ell_{\rm ED}} \ln{\frac{ t^{*}}{t^{*}_{\rm
core}}}\right]\,t^{*},
\end{align}
and the reverse shock velocity is given by
\begin{equation}\label{reverse_velocity_core_eq}
   v_{\rm r}^{*}  = \frac{3-s}{n-3}\,\frac{v_{\rm b}^{*}(t^{*} = t^{*}_{\rm core})}{\ell_{\rm ED}}.
\end{equation}
We assume that inside the core the reverse shock velocity remains
constant and equal to the value for $t^{*} = t^{*}_{\rm core}$.

\begin{figure}
  \begin{center}
    \includegraphics[width=\hsize]{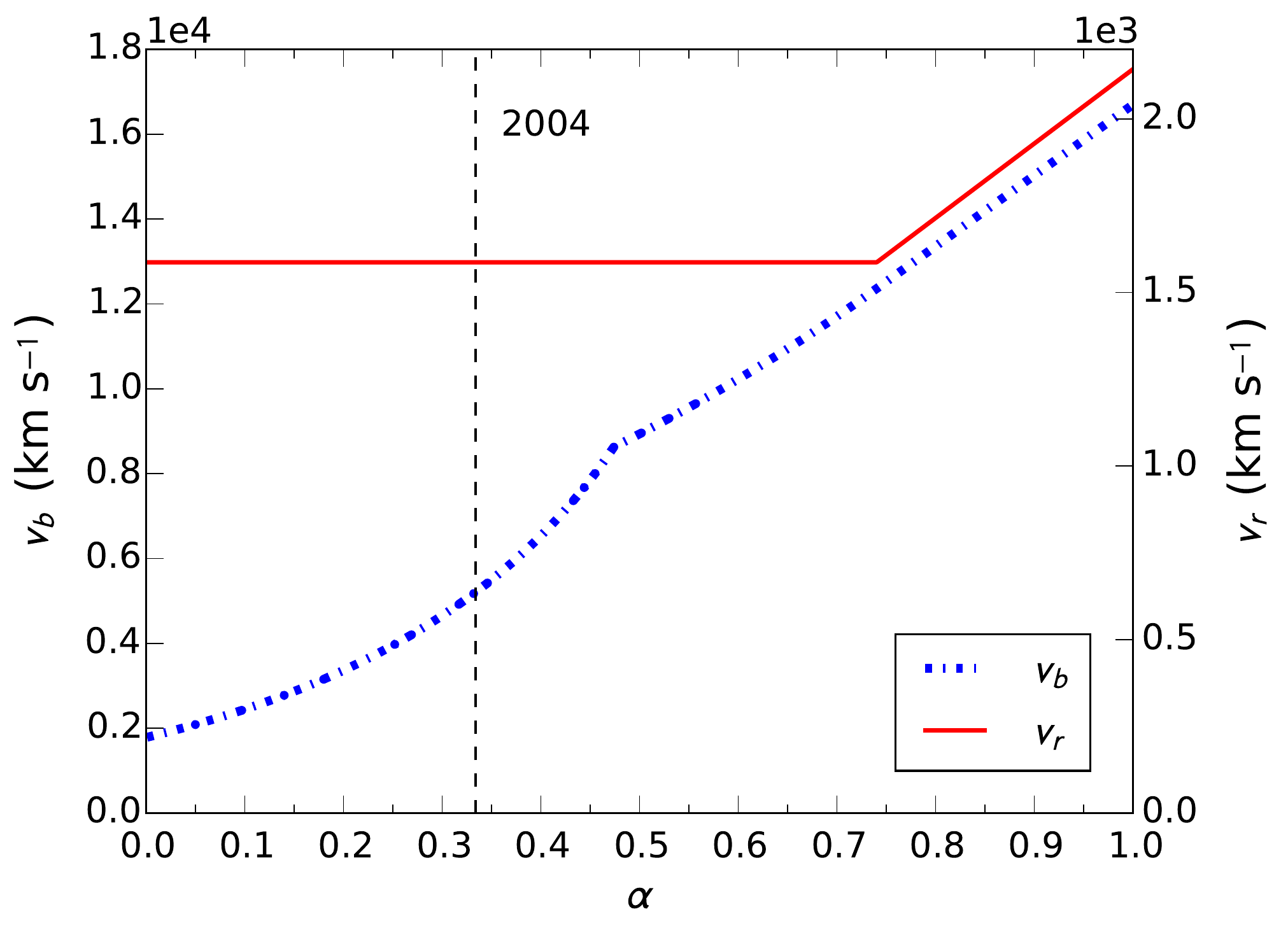}
  \end{center}
  \caption{Blast-wave and reverse shock velocities ($v_{\rm b}$
      and $v_{\rm r}$, respectively) as a function of the parameter
      $\alpha$. The reverse shock velocity (right y-axis, in units of
      10$^3$~km~s$^{-1}$) is calculated in the frame of the unshocked ejecta
      ahead of it. We assume that after the reverse shock has entered the
      ejecta core ($\alpha <$ 0.75), its velocity remains constant and equal
      to the value for $t^{*} = t^{*}_{\rm core}$
      (Eq.~\ref{reverse_velocity_core_eq}, see text). For each value of
      $\alpha$, the curve for $v_{\rm b}$ (left y-axis, in units of
      10$^4$~km~s$^{-1}$) provides the velocity of the blast-wave shock at
      the moment when the reverse shock hits a layer $\alpha$ of the
      ejecta. Vertical line as in
      Fig.~\ref{fig:tr_vs_alpha}.\label{fig:vrvb_vs_alpha} }
\end{figure}

The radius of the blast-wave shock (Eqs.~\ref{blast_radius_envelope_eq} and
\ref{blast_radius_core_eq}) and of the reverse shock
(Eqs.~\ref{reverse_radius_envelope_eq} and
\ref{reverse_radius_core_eq}) as a function of the time elapsed since
the progenitor of 
Cas~A exploded as a supernova are shown in
Fig.~\ref{fig:blast_reverse_radius_plus_zoom}.
The velocity of the blast-wave shock (Eqs.~\ref{blast_velocity_envelope_eq}
and \ref{blast_velocity_core_eq}) and of the reverse shock
(Eqs.~\ref{reverse_velocity_envelope_eq} and
\ref{reverse_velocity_core_eq}) as a function of the parameter
$\alpha$ are shown in Fig.~\ref{fig:vrvb_vs_alpha}.

% FIGURE 4 *************************************************************
%
% ********************************************************************

%=========================================
% Section 3
\section{Application to Cas~A}\label{sec:ejecta_properties}
%=========================================

% TABLE 1 ************************************************************************
\begin{table*}
\begin{center}
\caption{\label{tab:ejecta_properties} Physical properties of Cas~A}
\begin{tabular}{l l l l l}
\hline
\hline
\noalign{\vskip 1mm}
   \multicolumn{1}{c}{Parameter} & \multicolumn{1}{c}{Symbol (units)} & \multicolumn{1}{c}{Calculated/}  & \multicolumn{1}{c}{Observed/}  & \multicolumn{1}{c}{References}
   \\ 
              &     &  \multicolumn{1}{c}{adopted}   &    \multicolumn{1}{c}{estimated}  & \\
              &     &   \multicolumn{1}{c}{value}    &     \multicolumn{1}{c}{value}    &  \\
\noalign{\vskip 0.5mm}
\hline
\noalign{\vskip 1mm}
Progenitor mass       & (M$_\sun$)  &  19$^a$  & 15 -- 25 &
[1, 2]  \\
Distance            & (kpc)   & 3.4$^a$   &  3.4  & [3]  \\
Age (2004)           & (yr)  &  333$^a$   & 333 & [4] based on [5] \\
Explosion energy      & E (10$^{51}$ erg)  &  2.2$^b$   & 2.0 & [6]  \\
Ejecta mass            & $M_{\rm ej}$ (M$_\sun$)  & 2.0$^b$   & 2.2 &
[7, 8] \\
Forward shock radius           & $R_{\rm b}$ (pc)  & 2.50$^{c*}$  &
2.32 -- 2.72  & [9, 10, 11]  \\
Reverse shock radius            & $R_{\rm r}$ (pc) & 1.71$^{c*}$   & 1.52 --
2.01 &  [9, 10, 11, 12] \\
Forward shock velocity        & $v_{\rm b}$ (km s$^{-1}$) & 5226$^{c*}$   &
4000 -- 6000 & [13]  \\
Reverse shock velocity            & $v_{\rm r}$ (km s$^{-1}$) & 1586$^{c*}$
& 1000 -- 2000  & [6, 12]  \\
Circumstellar density (\emph{pre-shock})   & $n_0$ (cm$^{-3}$)  & 2.07$^b$
& 1.99 & [8]  \\
Density contrast  clumps/smooth ejecta  & $\chi$  & 100$^a$   & $\approx$100  & [14]  \\
Ejecta clumps density (\emph{pre-shock})          & $n_{\rm cloud}$
(cm$^{-3}$)  & 100$^a$   & 100 -- 300 & [14, 15]  \\
Smooth ejecta density (\emph{pre-shock})  & $n_{\rm smooth}$ (cm$^{-3}$)  & 1.0$^c$   &  0.1 -- 10 & [12]  \\
Ejecta clumps diameter  & 2$R_{\rm cloud}$ (cm) & 1.5$\times$10$^{16}$$^a$
& (1 -- 5)$\times$10$^{16}$ & [16]\\
\noalign{\vskip 0.5mm}
\hline
\end{tabular}
\end{center}
\tablefoot{
 \tablefoottext{a}{Adopted.}
 \tablefoottext{b}{Allows a good match between measured and calculated
   values.}
 \tablefoottext{c}{Calculated -- this work.} 
 \tablefoottext{*}{Value in 2004.} \\
   {  References:} [1]~\citet{young06},
   [2]~\citet{vink96}, [3]~\citet{reed95}, [4]~\citet{hwang12},
   [5]~\citet{thorstensen01},  [6]~\citet{laming03},
   [7]~\citet{willingale02}, 
    [8]~]\citet{willingale03}, [9]~\citet{hwang12}, 
    [10]~\citet{helder08}, [11]~\citet{gotthelf01}, 
    [12]~\citet{morse04}, [13]~\citet{delaney03}, 
    [14]~\citet{sutherland95}, [15]~\citet{docenko10}, 
    [16]~\citet{fesen11}. }
\end{table*}
% ******************************************************************************

\subsection{Ejecta geometry and physical
  properties}\label{ejecta_geo_sec}

The Cas~A supernova remnant is the result of the explosion of a Type
IIb supernova \citep{krause08} with a progenitor mass estimated
between 15 and 25 M$_\sun$ \citep{young06, vink96}. The remnant is
located at a distance of 3.4 kpc \citep{reed95} and its age was
$\sim$333 years in 2004 (date of the Chandra observations which have
been used to determine some of the parameters of Cas~A used in this work). 
From our calculations (Sect.~\ref{sec:final_shock_equations}) we
obtain the following values for the radius and velocity of the forward
and reverse shock of Cas~A in 2004: $R_{\rm b}$ = 2.5 pc,
$v_{\rm b}$ = 5226 km s$^{-1}$, $R_{\rm r}$ = 1.71
pc and $v_{\rm r}$ = 1586 km s$^{-1}$. Our numbers are
consistent with the values derived from observations: $R_{\rm b}$ =
2.32 -- 2.72 pc, $v_{\rm b}$ = 4000 -- 6000 km s$^{-1}$, $R_{\rm r}$ = 1.52 --
2.01 pc, $v_{\rm r}$ = 1000 -- 2000 km s$^{-1}$ (see
Table~\ref{tab:ejecta_properties} for the corresponding references).

We obtained this nice match  adopting the ejecta mass $M_{\rm
ej}$~=~2~M$_\sun$ \citep[consistent with the value of 2.2 M$_\sun$
inferred by][]{willingale02, willingale03}, and the explosion energy $E$ =
2.2$\times$10$^{51}$ erg, in agreement with the amount of energy
expected from a core-collapse SN \citep{laming03}, and the circumstellar (preshock)
density $n_0$ = 2.07 H atom cm$^{-3}$, which is compatible with the
value of 1.99 cm$^{-3}$ from \citet{willingale03}.

The density distribution of the supernova ejecta is described in a
more realistic and observationally motivated way by a series of
overdense clouds embedded into a smooth and tenuous medium
\citep[e.g.][]{peimbert71, kamper76, chevalier78, chevalier79,
fesen01, rho09, rho12, wallstrom13}. For the ejecta clouds, we assume
a pre-shock density $n_{\rm cloud}$ = 100 cm$^{-3}$ with a density
contrast with respect to the smooth component $\chi \,=\,n_{\rm
cloud}/n_{\rm smooth}$ = 100. These values allow to reproduce the
optical spectra of the fast moving knots (FMKs) in Cas~A in a fairly
accurate way \citep{sutherland95}. For the smooth ejecta, therefore,
we obtain the pre-shock density $n_{\rm smooth}\, = \, n_{\rm cloud}/
\chi$ = 1.0 cm$^{-3}$, which is consistent with the ambient density of
0.1 -- 10 cm$^{-3}$ (average $\sim$0.25 cm$^{-3}$) estimated for the
X-ray emitting gas on the bases of ram pressure arguments
\citep{morse04}. This value for the smooth ejecta density should be
considered as indicative. To evaluate the effect of dust sputtering we
use the density structure shown in Fig.~\ref{fig:rho_vs_alpha}, which
has been calculated with our model for the evolution of the Cas~A
ejecta. Our model predicts for 2004 a number density of 0.23
cm$^{-3}$, which is very close to the average value estimated by
\citet{morse04}.

The most recent determinations of the amounts of dust produced in
Cas~A were carried out by \citet{barlow10} and \citet{arendt14}. The warm dust
emission at 5 -- 35 \textmu m observed by Spitzer is mostly generated by
$\sim$0.04 M$_\sun$ of mainly silicate dust. Most of the dust mass is
associated with a colder dust component that dominates the broadband
emission at wavelengths $\gtrsim$ 60 \textmu m and is associated with the
Si II] 34.82 \textmu m line emission. Because of the lack of any
distinguishing dust emission features at these wavelengths, its
composition could not be uniquely determined and its mass was
estimated to be $<$ 0.1 M$_\sun$. This cold dust component has
probably not yet encountered the reverse shock. Its distribution is
very different from the warm dust that is spatially coincident with
reverse shock features. We assume that the dust, initially present
solely in the clumps, does not make the clumps affect the dynamics of
the reverse shock.

The ejecta clouds become visible due to reverse shock heating and
they constitute the bright main shell of the remnant.  Optical maps
show that the diameter of the ejecta clumps, $D=2R_{\rm cloud}$, is in
the range 0.2-1.0$\arcsec$ \citep{fesen11}, which corresponds
to (1 -- 5)$\times$10$^{16}$ cm at a distance of 3.4 kpc\footnote{{ 
  In Docenko \& Sunyaev (2010) there is a misprint in the linear size
  of the clumps, which is erroneously reported as (3 --
  15)$\times$10$^{16}$ cm instead of (1 -- 5)$\times$10$^{16}$ cm (at
  a distance of 3.4 kpc)}.}. These clumps are
bigger than the knots located outside the main shell of the remnant,
at or ahead of the forward shock front \citep{fesen06, hammell08}, whose
size is typically lower than $\sim$10$^{15}$ cm \citep{fesen11}.
\citet{kamper76} and \citet{vanDenBergh85} found that optical
clumps brighten up and gradually fade with time, and the number of
clouds visible at any time is described well by an exponential decay
that has an $e$-folding time of around 25 years \citep{kamper76}. We
adopt as a representative cloud size the intermediate value 2$R_{\rm
cloud}$ = 1.5$\times$10$^{16}$ cm $\simeq$ 0.005 pc, which provides a
cloud lifetime (see below) that is consistent with the aforementioned determination
from \citet{kamper76}.

When the reverse shock encounters an ejecta cloud, it generates a \emph
{cloud shock} which propagates into the cloud with velocity $v_{\rm
cloud} = v_{\rm r}/\sqrt{\chi}$. The clumps heated up by the shock
become visible before fading gradually away. The timescale for
this process is correlated to the time required by the cloud shock to
reach the opposite side of the clump. This latter time is given by
twice the \emph{cloud crushing time}, $t_{\rm cc} \equiv R_{\rm cloud}/
v_{\rm cloud}$ \citep{klein94}, which in addition provides an
indication of the dynamical timescale for shock-induced cloud
fragmentation.

\begin{figure}
  \begin{center}
    \includegraphics[width=0.9\hsize]{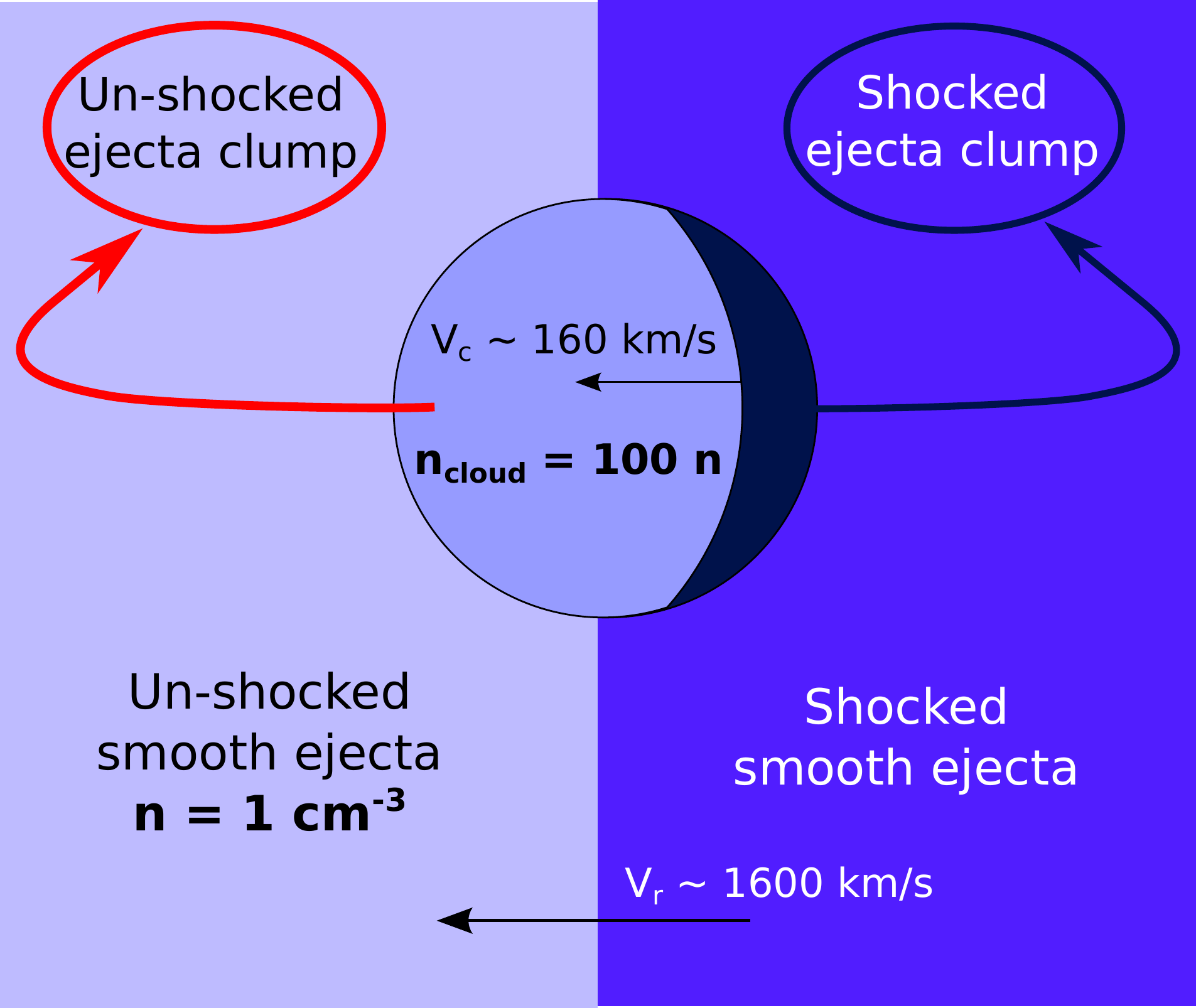}
  \end{center}
  \caption{  Schematic illustration of the propagation of the
      reverse shock (velocity $v_{\rm r} \sim 1600$ km s$^{-1}$) into the smooth
      ejecta and through an ejecta clump. The reverse shock hitting the
clump generates a cloud shock that propagates with velocity $v_{\rm
  c} \equiv v_{\rm cloud} = v_{\rm r}/\sqrt{\chi}$, where we assume
$\chi$ = $n_{\rm cloud}/n_{\rm smooth}$ = 100 and $n_{\rm cloud}$ =
100 cm$^{-3}$, which leads to $n_{\rm smooth} \equiv n$ = 1
cm$^{-3}$. Figure adapted from \citet{micelotta13a}.\label{fig:cloud} }
\end{figure}

From our calculated value $v_{\rm r}\sim 1600$ km s$^{-1}$, we obtain
$v_{\rm cloud}$ = 160 km s$^{-1}$, which gives 2$t_{\rm cc}$ =
2$R_{\rm cloud}$/$v_{\rm cloud}$ = 30 yr, in agreement with the
lifetime of the Cas~A FMKs as deduced from optical observations
\citep[][see above]{kamper76}.
The physical properties of Cas~A are summarized in
Table~\ref{tab:ejecta_properties}. Figure~\ref{fig:cloud} shows a schematic
illustration of the propagation of the reverse shock into the smooth
ejecta and through an ejecta clump.

% FIGURE 5  *************************************************************
%
% ********************************************************************

The ejecta clouds contain 80\% -- 90\% of oxygen, with a small
contribution from Ne, Si, S, Ar, and negligible abundances of hydrogen
and helium \citep{docenko10}. Optical observations have revealed some
knots made of almost pure oxygen, dubbed [\ion{O}{iii}] filaments
\citep{chevalier79}. This justifies our simplifying assumption of
ejecta clouds with a composition of 100\% oxygen.

We assume that all the dust resides in the oxygen rich clumps
and that it is composed of silicates (MgSiO$_3$) and amorphous carbon.
Because the initial dust grain size distribution of the
newly formed dust is not  determined well, as a test case we adopt the
classical MRN power-law expression \citep{MRN77}: $g(a) \sim a^{-3.5}$ and
$a_{\rm min} < a < a_{\rm max}$, with $a_{\rm min} = $50 \AA{} and
$a_{\rm max} = $ 2500 \AA. The quantity $g(a)$ represents the number
of grains with radius $a$ in the interval ${\rm d}a$.

For the smooth (X-ray emitting) ejecta, we assume a composition of O,
Ne, Mg, Si, S, Ar, and Fe as determined by \citet{hwang12} and
reported in their Table 2. This work provides the elemental masses
with respect to oxygen, M$_i$/M$_{\rm O}$. To convert them to
M$_i$/M$_{\rm H}$, elemental masses with respect to hydrogen, we use
the equation M$_i$/M$_{\rm H}$ = (M$_i$/M$_{\rm O}$)(M$_{\rm
O}$/M$_{\rm H}$), where the protosolar M$_{\rm O}$/M$_{\rm H}$ =
8.59$\times$ 10$^{-3}$ is given by Table~1.4 in \citet{draine11} and
based on \citet{asplund09}. The values of all masses are reported in
Table~\ref{tab:ejecta_masses}.

% TABLE 2 ************************************************************************
\begin{table}
\begin{center}
\caption{\label{tab:ejecta_masses} Elemental masses in the smooth
  ejecta of Cas~A}
\begin{tabular}{l l c}
\hline
\hline
\noalign{\vskip 1mm}
   Element & M$_i$/M$_{\rm O}$\tablefootmark{a} & M$_i$/M$_{\rm H}$\tablefootmark{b} \\ 
\noalign{\vskip 0.5mm}
\hline
\noalign{\vskip 1mm}
O             & 1.00  &    8.59$\times$10$^{-3}$ \\
Ne            & 0.015   &  1.28$\times$10$^{-4}$ \\
Mg            & 0.0039  &  3.35$\times$10$^{-5}$ \\
Si            & 0.021   &  1.80$\times$10$^{-4}$ \\
S             & 0.011  &   9.45$\times$10$^{-5}$\\
Ar            & 0.0056   & 4.81$\times$10$^{-5}$ \\
Fe            & 0.057 \tablefootmark{c}  &  4.90$\times$10$^{-4}$ \\
\noalign{\vskip 0.5mm}
\hline
\end{tabular}
\end{center}
\tablefoot{
\tablefoottext{a}{Derived from X-rays observations \citep[][Table 2]{hwang12}.} \\
\tablefoottext{b}{Calculated from M$_i$/M$_{\rm O}$ -- see text for details.} \\
\tablefoottext{c}{Sum of two iron components: 0.041$+$0.016 \citep{hwang12}.} \\
}
\end{table}
% ******************************************************************************

\subsection{ Temperature evolution of the ejecta}\label{ejecta_temp_sec}

When a shock propagates through a gas, it heats the gas located right
beyond the shock front up to a
temperature $T$ given by the following expression:
\begin{equation}\label{temperature_eq}
  T = \frac{3}{16}\, \frac{\mu}{k_{\rm B}}\, v_{\rm s}^2
,\end{equation}  
where $\mu$ is the mean molecular mass per particle and $v_{\rm s}$
is the velocity of the shock. For a mixture of neutral elements, $\mu$
is written as
\begin{equation}\label{mu_neutral_eq}
  \mu_{\rm neutral} = \sum_i\, \frac{{\rm M}_i}{{\rm M_H}} \times \rm amu 
,\end{equation}
where amu is the atomic mass unit. For a mixture of ionized elements
in different ionization states $j$, we have
\begin{equation}\label{mu_ionized_eq}
  \mu_{\rm ionized} = \frac{\mu_{\rm neutral}}{ion}. 
\end{equation}
The quantity $ion$ represents the total number of particles (nuclei
plus electrons) per hydrogen atom and can be calculated from the
following relation:
\begin{equation}\label{ion_eq}
  ion = \sum_i\, \frac{{\rm N}_i}{{\rm N_H}}\, \left[\sum_j\, f_{ij}
    \left(1+N_{\rm e}^{ij} \right) \right]
.\end{equation}
The index $i$ runs over the elemental species and the index $j$ runs over
the ionization states. For each species $i$, the quantity ${\rm
N}_i/{\rm N_H}$ = (${\rm M}_i/{\rm M_H}$)/($\langle m_i \rangle$/{\rm
amu}) represents the number of atoms per H atom, where $\langle m_i
\rangle$/{\rm amu} is the mean mass in atomic units. The number of
electrons ejected per atom $i$ in the ionization state $j$ is given by $N_{\rm
e}^{ij}$ and the fraction of such atoms by $f_{ij}$.

In the ejecta clouds, the rise of temperature is due to the cloud
shock, which has velocity $v_{\rm cloud}$ = 160~km~s$^{-1}$. As
mentioned above we assume a composition of pure oxygen, and an
ionization state of +2. This implies that
$ion$ = 1+2 = 3 and $\mu$ = 5.34$\times$amu = 8.9$\times$10$^{-24}$
g. Using the above equations, we obtain the cloud temperature $T_{\rm
  cloud}\sim$~3$\times$10$^6$~K. This is consistent
with the results from \citet{sutherland95} and \citet{borkowski90},
who obtained temperatures of 2.75$\times$10$^6$~K and
3.46$\times$10$^6$~K for cloud shock velocities of 150~km~s$^{-1}$
and 170~km~s$^{-1}$, respectively.

Equation~\ref{temperature_eq} assumes equilibration between the
ion and electron temperatures. The work of \citet{itoh81} on the
emission from oxygen-rich supernova ejecta (the case of Cas~A) shows
that in the immediate post-shock region the electron temperature,
$T_{\rm e}$, is much lower than the ion temperature, $T_{\rm
ion}$. Figure 2 however indicates that $T_{\rm e}$ and $T_{\rm ion}$
equilibrate over a scale length of $\sim$4$\times$10$^{13}$ cm, which
corresponds to only $\sim$0.27\% of the diameter of our clumps.  This
is confirmed by recent shock calculations performed for cases
comparable to those studied here, which show that both the
equilibration and the cooling length scales are very short,
$\sim$2$\times$10$^{13}$ cm at most even for the case of $T_{\rm
e}$/$T_{\rm ion}$= 0.01 at the shock front (J. Raymond, private
communication). Such results imply that temperature equilibration
occurs quickly; these results are in agreement with \citet{vanAdelsberg08} and
\citet{ghavamian07}, who both find that full equilibration occurs in
non-radiative shocks (not necessarily strongly cooling) for shock
velocities below $\sim$400 km~s$^{-1}$. The assumption of temperature
equilibration appears therefore justified in the present case.  
 
The shock propagating into the smooth ejecta is the reverse shock
with velocity $v_{\rm r}$ = 1600~km~s$^{-1}$. To calculate the
resulting temperature of the shocked smooth ejecta, we would need the
ionization state of the pre-shock gas, which is pre-ionized by the
incoming reverse shock. Unfortunately, there are no determinations of
such an ionization state. To make an estimate, we use the results from
the photo-ionization and shock modelling code Mappings III
\citep{allen08}.  For the elemental composition given by
\citet{hwang12}, we adopt for each element the ionization state
calculated by Mappings III for a gas of solar abundances, with number
density $n$ = 1 cm$^{-3}$,  pre-ionized by a shock with velocity of
1000 km s$^{-1}$ (Model M).
We adopt a transverse magnetic field B = 1 \textmu G, to be
consistent with the model of \citet{sutherland95} who assumed this
value for the ejecta clumps; we make the {  assumption} that the
pre-shock magnetic field is the same in both the clumps and smooth
ejecta. Such a magnetic field does not affect the temperature or
dynamics, but it would imply betatron acceleration of the charged
grains, which would affect the cloud shock. In a following
study, we will investigate the modifications of the velocity profiles
of the grains induced by betatron acceleration and their impact on the
dust sputtering process.

The set of parameters described above is the closest match to Cas~A
among those available from the code. The resulting ionization states
are the following: O$^{6+}$, Ne$^{5+}$, Mg$^{7+}$, Si$^{8+}$,
S$^{8+}$, Ar$^{6+}$, Fe$^{7+}$. Because the maximum shock velocity
considered by Mappings III is lower than the velocity of the reverse
shock in Cas~A ($v_{\rm r}$ = 1600~km~s$^{-1}$), we expect that the
resulting ionization states are underestimated. With these
parameters, the temperature of the shocked smooth ejecta from
Eq.~\ref{temperature_eq} is $T_{\rm smooth}$ = 1.38$\times$10$^8$~K.
Even if we consider the extreme (and unrealistic) case of complete
pre-ionization of the pre-shock gas, the temperature is only
reduced to 1.02$\times$10$^8$~K. 

It should be noted that in the smooth ejecta the hypothesis of
equal ion and electron temperatures may not be fully
justified. \citet{yamaguchi14} show that in the reverse shock of Tycho
there is some electron heating via ion-electron Coulomb collisions,
but that it is weak. The results from \citet{vanAdelsberg08} would
indicate that the electron-to-ion temperature ratio, $\beta \equiv
T_{\rm e}/T_{\rm ion}$, has a value of $\sim$3 -- 8\% for a shock speed of 1600 km
s$^{-1}$. In Cas~A the electron temperature derived from X-ray
observations is $\sim$1.7$\times$10$^7$~K \citep[ $k_{\rm B}\,
T$~=~1 -- 2 eV; ][]{hwang09}. This number combined with the values of
$\beta$ reported above gives $T_{\rm ion}$ = (2 -- 5.7)$\times$10$^8$
K, which is higher than the temperature derived from
Eq.~\ref{temperature_eq}. The implications of this situation will be
evaluated in a follow-up work, while for the present study we maintain
the hypothesis of ion-electron temperature equilibration in the smooth ejecta.

The temperature of the gas in which the dust is embedded determines
the type and level of processing of the grains. It is therefore
important to evaluate the cooling of both the cloud and smooth
ejecta. To determine the cooling time as a function of temperature,
$\tau_{\rm cool}(T)$, we adopt the formalism from
\citet{sutherland93}. For a gas composed of only one elemental species
in a single ionization state, this gives
\begin{eqnarray}\label{tau_cool_eq}
  \tau_{\rm cool}(T) &  = & \frac{\mathcal{U}}{n_{\rm e}\, n_{i}\, \Lambda
    (T)}  =  \frac{(3/2)\, \mathcal{N} \, k_{\rm B}\, T}{n_{\rm e}\, n_{i}\,
    \Lambda (T)} \\ 
  & = & \frac{3}{2}\frac{(n_{\rm e}+n_{i})\, k_{\rm B}\, T}{n_{\rm e}\, n_{i}\, \Lambda (T)}.
\end{eqnarray}
In this set of equations, $\mathcal{U}$ is the internal energy,
$\Lambda(T)$ the cooling function and $\mathcal{N} = n_{\rm e} + n_{i}$,
where $n_{\rm e}$ and $n_{i}$ are the number density of electrons and ions,
respectively. 
Because $n_{\rm e} = An_{i}$, where $A$ is the number of electrons ejected
from each atom (i.e. its ionization state), Eq.~40 can be rewritten
in a more general form as
\begin{equation}\label{tau_cool_general_eq} \tau_{\rm cool}(T) =
\frac{3}{2}\frac{A+1}{A}\frac{k_{\rm B}\, T}{n_{i}\, \Lambda (T)}.
\end{equation}
We have assumed a composition of pure oxygen for the ejecta clumps,
which implies that $n_i$ = $n_{\rm O}$ = $n_{\rm cloud}$.
For the cooling function of the shocked ejecta clouds, we use the
values computed by \citet[][see their Fig.~2]{sutherland95} for
$v_{\rm cloud}$ = 150 km s$^{-1}$, which is very close to our own
value of 160 km s$^{-1}$. To our knowledge, this study provides the
only available estimate of the cooling function of an oxygen-rich
shocked gas.
To determine $n_{\rm e}$, we would need the oxygen ionization structure for
a 160 km s$^{-1}$ cloud shock, which is not available. Instead, we use
the ionization structure of the 200 km s$^{-1}$ shock model from
\citet{sutherland95} shown in their Fig.~3 (lower panel), which gives
O$^{5+}$ as the dominant ionization state. This results in $A$=5,
$n_{\rm e} = 5n_{i}$ and $\mathcal{N} = 5n_{i} + n_{i} =
6n_{i}$. \citet{docenko10} performed their own calculation of the
post-shock oxygen ion distribution for a 200 km s$^{-1}$ cloud shock
obtaining results similar to \citet{sutherland95}, but only if the
spectroscopic symbols in the original figure are increased by
unity. \citet{docenko10} assumed a misprint. In this case, the
dominant ionization state would be O$^{6+}$, which leads to $A$=6,
$n_{\rm e} = 6n_{i}$, and $\mathcal{N} = 7n_{i}$.

\begin{figure}
  \begin{center}
    \includegraphics[width=\hsize]{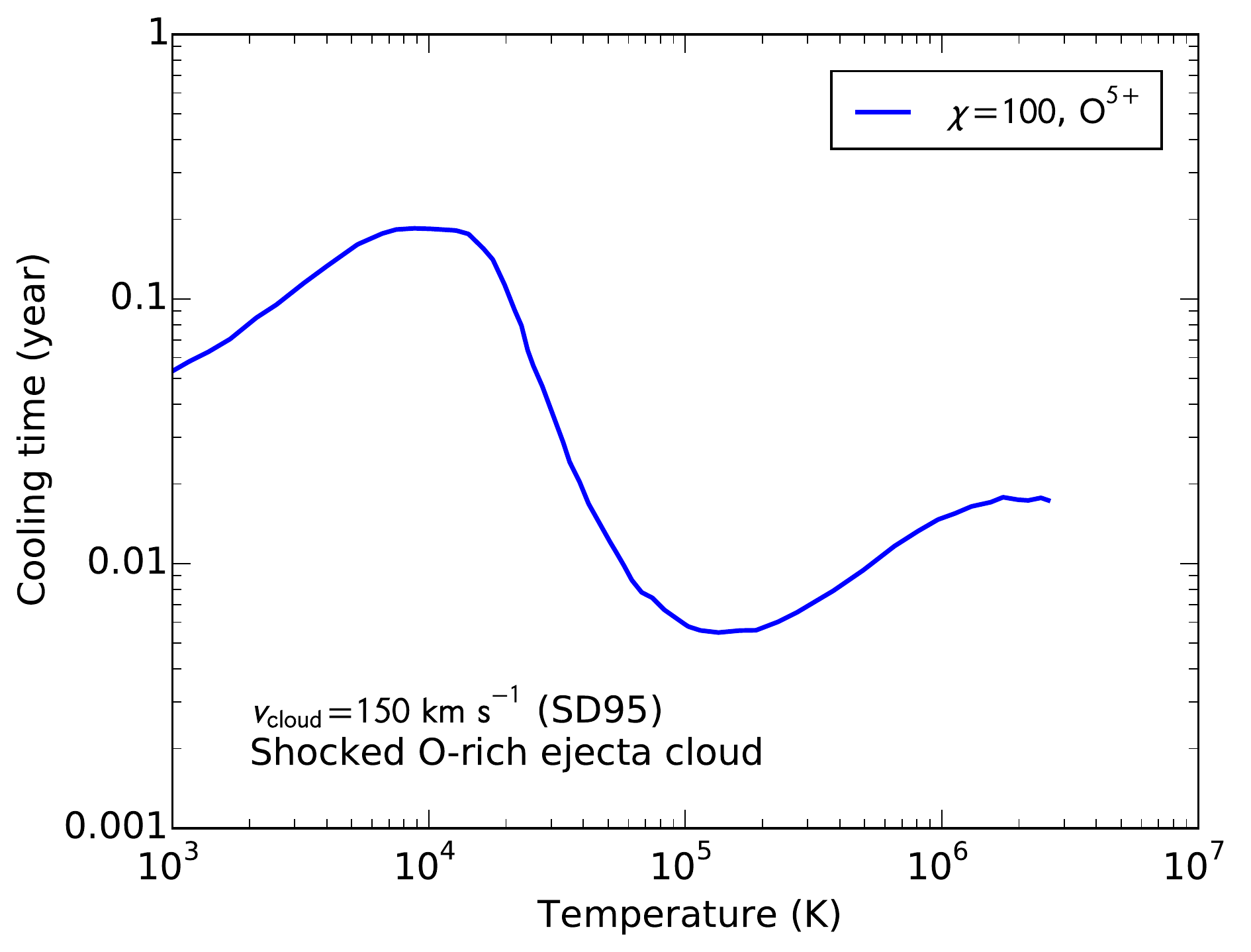}
  \end{center}
  \caption{Cooling time as a function of temperature for a shocked
    O-rich ejecta cloud in Cas~A, calculated from
    Eq.~\ref{tau_cool_shocked_clump_eq}. \label{fig:cool_time_clumps} }
\end{figure}

\begin{figure}
  \begin{center}
    \includegraphics[width=\hsize]{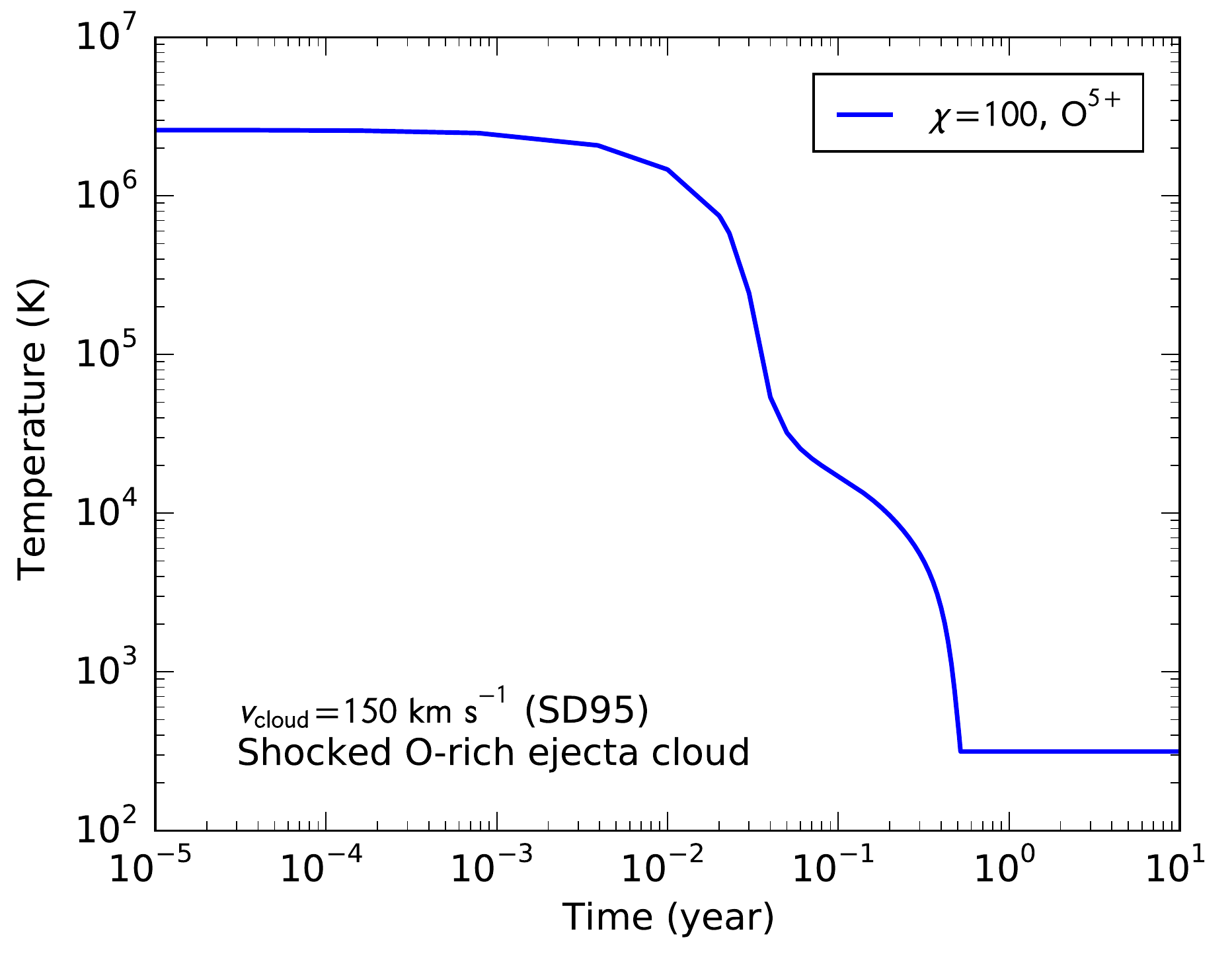}
  \end{center}
  \caption{Temperature evolution as a function of time for a shocked
    O-rich ejecta cloud in Cas~A, calculated from
    Eq.~\ref{temperature_evol_eq} using the cooling time from
    Eq.~\ref{tau_cool_shocked_clump_eq}. 
\label{fig:temp_vs_time_clumps} }
\end{figure}

For our shocked ejecta cloud, the number density to include in
Eq.~\ref{tau_cool_general_eq} is that in the immediate post-shock
gas, which is increased by a factor of 4 with respect to the unshocked
gas. As a result, for the number density of the shocked cloud we have
$n_{i}^{\rm {sc}} =$ 4$n_{i}$.  The cooling time of a shocked
ejecta cloud is then given by Eq.~\ref{tau_cool_general_eq} modified
as follows
 \begin{equation}\label{tau_cool_shocked_clump_eq}
   \tau_{\rm cool}^{\rm sc}(T) = 
  \frac{3}{2}\frac{A+1}{A}\frac{k_{\rm B}\, T}{4\,n_{i}\, \Lambda (T)}.
\end{equation}
The cooling time $\tau_{\rm cool}^{\rm sc}(T)$ from
Eq.~\ref{tau_cool_shocked_clump_eq} is shown in
Fig.~\ref{fig:cool_time_clumps} assuming O$^{5+}$ as the
dominant ionization state. If in the plot showing the oxygen ionization
structure for the 200 km s$^{-1}$ cloud shock model \citep[Fig.~3, lower
panel in ][]{sutherland95} there is indeed the misprint indicated
by \citet{docenko10},
the dominant ionization state for oxygen
would be O$^{6+}$. In this case, the cooling time for the shocked cloud
is given again by Eq.~\ref{tau_cool_shocked_clump_eq} with the same
cooling function $\Lambda (T)$, but adopting $A$=6. The curves for
O$^{5+}$ and O$^{6+}$ are indistinguishable (the ratio between the two
is 3\%), so we consider negligible the variation induced by an eventual
misprint and follow \citet{sutherland95} adopting O$^{5+}$ as the
dominant ionization state produced by a 200 km s$^{-1}$ shock
propagating into an oxygen-rich cloud. To determine the temporal
evolution of the gas temperature in the shocked cloud, we have to
solve the following differential equation:
\begin{equation}\label{temperature_evol_eq}
 \frac{{\rm d}T}{{\rm d}t} = - \frac{T}{\tau_{\rm cool}^{\rm sc}(T)},
\end{equation}
with $\tau_{\rm cool}^{\rm sc}(T)$ from
Eq.~\ref{tau_cool_shocked_clump_eq}. The solution of the equation is
shown in Fig.~\ref{fig:temp_vs_time_clumps}.  Because of the extreme
metallicity, it takes only six months for the temperature of the
ejecta in the clouds to go from $\sim$3$\times$10$^6$~K to 300
K. While the cloud shock progresses through the clump, the layer
beyond the shock front with a temperature close to $\sim$10$^5$~K,
i. e. high enough for thermal sputtering to be effective, has a
thickness of $\sim$2$\times$10$^{14}$ cm; this layer is only 1.4\% of
the diameter of the clump. The small amount of dust in such a thin
layer allows us to neglect thermal sputtering and to assume that,
while residing in the ejecta clumps, the dust is affected only by 
inertial (kinetic) sputtering (see Sect.~\ref{sec:kin_sput}).

\begin{figure}
  \begin{center}
    \includegraphics[width=\hsize]{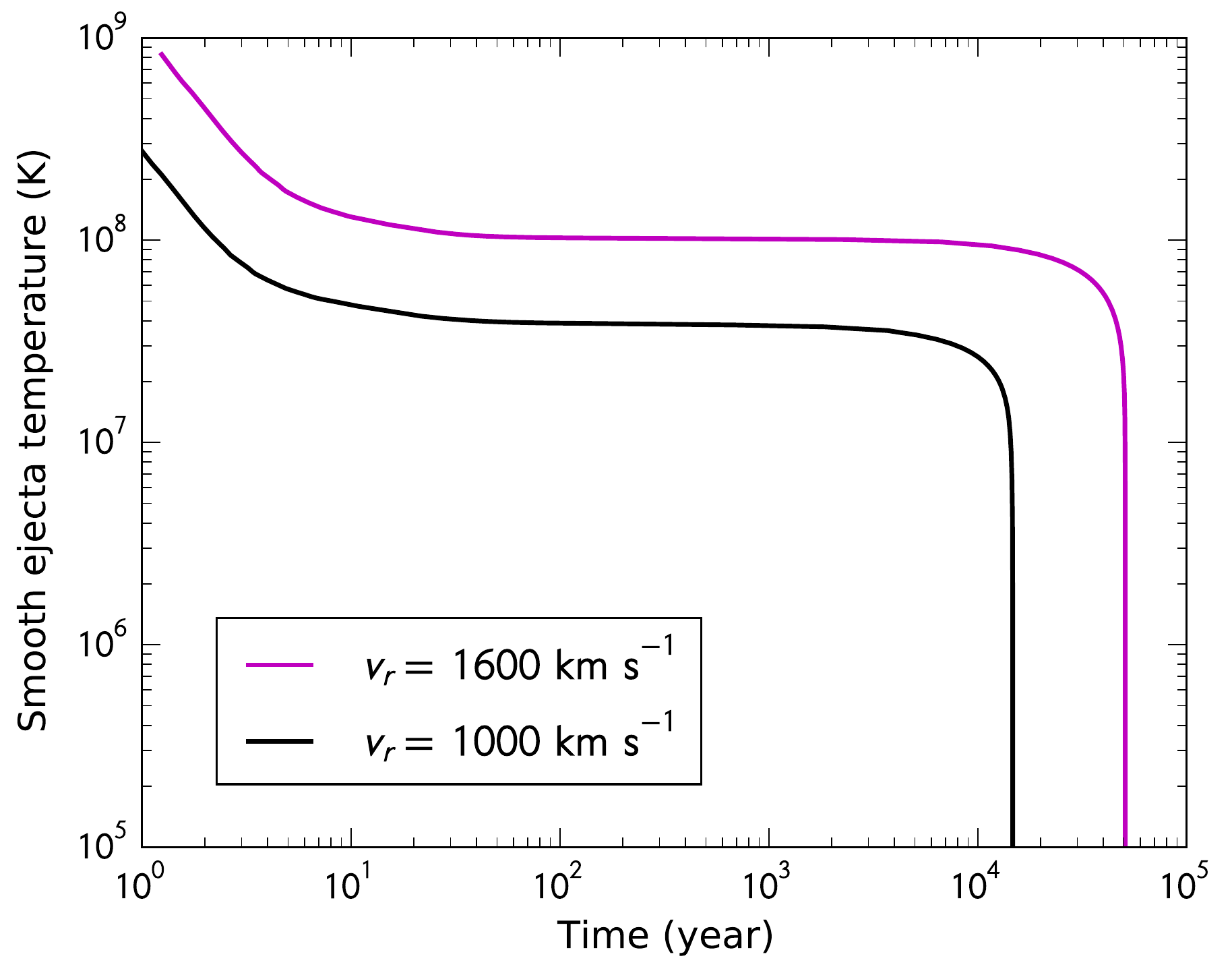}
  \end{center}
  \caption{Temperature evolution as a function of time in the shocked
    smooth ejecta, calculated for two shock velocities under
    non-equilibrium ionization (NEI) conditions, using the Mappings III
    code. 
    \label{fig:temp_vs_time_smooth}}
\end{figure}

To estimate the temperature evolution of the shocked smooth ejecta, we
used Mappings III with appropriate modifications of the input
parameters, to obtain the closest match with the observed properties of
Cas~A. The calculations are performed under non-equilibrium ionization
(NEI) conditions. We adopted the elemental composition determined by
\citet{hwang12} for the smooth ejecta in Cas~A and run the code for a
shock with the velocity of the reverse shock, $v_{\rm r}$ = 1600 km
s$^{-1}$. This value is above the nominal limit of the code (1000 km
s$^{-1}$) and is expected to increase the pre-ionization state of the
pre-shock gas. However, we have shown early on in this section that
even a complete pre-ionization does not significantly modify the
resulting post-shock temperature. We consider therefore that the
calculation could provide a reasonable estimate of the temperature
evolution for the faster shock as well. The result is shown in
Fig.~\ref{fig:temp_vs_time_smooth} and is consistent with our
previous calculation. With the same ejecta composition but a shock
velocity of 1000~km~s$^{-1}$, we obtain a post-shock temperature of the
order of 10$^7$~K, which is consistent with the value of the electron
temperature derived from X-ray observations \citep{hwang09}. The same
calculations performed using the updated code Mappings
IV\footnote{\url{http://miocene.anu.edu.au/miv/}}, give very similar
results. As can be seen in Fig.~\ref{fig:temp_vs_time_smooth}, in
both cases, after a phase of rapid cooling ($\sim$10 years) the
temperature remains fairly constant at $\sim$10$^8$~K and
$\sim$10$^7$~K for more than 10$^4$ years before dropping abruptly.

% FIGURE 6 *************************************************************
%
% **********************************************************************

% FIGURE 7 *************************************************************
%
% **********************************************************************

% FIGURE 8 *************************************************************
%
% **********************************************************************

%===================================================================
% Section 4
\section{Grain destruction by the reverse shock: Kinetic sputtering}\label{sec:kin_sput}
%===================================================================

The rate at which a dust grain is sputtered as a result of its relative motion
through a gas, a process also known as \emph{kinetic} or \emph{inertial
sputtering}, is given by the sum over colliding gas species $i$
\citep[e.g.][]{dwek92a}, i.e.
\begin{equation}\label{eq:dm_dt}
  \frac{{\rm d} m_{\rm gr}}{{\rm d} t} = 2\,\pi\, a_{\rm gr}^2\, m_{\rm sp}\, \sum_i n_i\,
v_{\rm gr}\, Y_i\,(E_i=m_i\,v_{\rm gr}^2/2)
,\end{equation} 
where $a_{\rm gr}$ is the grain radius; $m_{\rm sp}$ is the average mass of
the atom/molecule sputtered from the grain, $v_{\rm gr}$ the grain
velocity relative to the ambient gas, which is equivalent to the
  velocity of the incident projectile seen by a target grain
  considered stationary; $n_i$ is the number density of
the different gas species; $m_i$ is their atomic mass; and $Y_i$, the
sputtering yield, is the number of atoms/molecules ejected from the
grain per incident projectile of composition $i$
\citep{tielens94,nozawa06}. The factor of 2 in Eq.~\ref{eq:dm_dt} corrects
the yield, which is measured for normally incident projectiles on a
target material for a distribution of incident angles. 

% TABLE 3 ************************************************************************
\begin{table}[t]
\begin{center}
\caption{\label{sput_param_tab} Dust grain properties for sputtering calculations.}
\begin{tabular}{c c c c c c c}
\hline
\hline
\noalign{\vskip 1mm}
         Material           &  $\rho_{\rm gr}$  &  $U_0$  &  $M$ & $Z$
         &  $K$ & $m_{\rm sp}$  \\  
                    &  (g cm$^{-3}$)   &  (eV)  & (amu) &  &
   &  (amu)  \\ 
\noalign{\vskip 0.5mm}
\hline
\noalign{\vskip 1mm}
MgSiO$_3^a$                     & 2.65  &  6.0  &  20 & 10 & 0.1  & 23  \\
C$^b$            & 2.20   &  4.0 &  12 & 6  & 0.61 & 12  \\
\noalign{\vskip 0.5mm}
\hline
\end{tabular}
\end{center}
\tablefoot{
\tablefoottext{a}{\citet{tielens94},}
\tablefoottext{b}{\citet{nozawa06} and references therein -- Table
  2.} \\
{{\bf Parameters} -- $\rho_{\rm gr}$: mass density of the grains;
  $U_0$: surface binding energy; $M$ and $Z$: atomic mass and atomic
  number  of the target material, respectively; $K$: sputtering constant; and $m_{\rm
    sp}$: average mass of sputtered atoms.} \\}
\end{table}
% ******************************************************************************

For the sputtering yield $Y_i$ we adopt the expression given by
Eq.~11 in \citet{nozawa06}. This is the same provided by
\citet{tielens94} except for the different formula used for the function
$\alpha_i$, which appears in the yield and allows for a better
agreement with sputtering data \citep[for details see][and
references therein]{nozawa06, tielens94}.

For simplicity, we assume that all the grains in the SN ejecta are
made of silicates (in the form of MgSiO$_3$) and amorphous carbon (see
Sect.~\ref{ejecta_geo_sec}).  For these two kinds of grains, we adopt the
sputtering parameters summarized in Table~\ref{sput_param_tab}. The
quantities $M$ and $Z$ are the atomic mass and atomic number of the
target material, respectively, $U_0$ is the surface binding energy,
defined as the minimum energy which is necessary to transfer into the
target to remove an atom from the top surface layer.  The
constant $K$ enters in one of the terms of the expression for the
sputtering yield $Y_i(E)$ and has been determined via a comparison
with laboratory experiments. Following \citet{tielens94} and
\citet{nozawa06}, for MgSiO$_3$ we adopt the sputtering parameters of
SiO$_2$, which can be considered a good representative of silicates
and for which experimental sputtering data are available. The mass of
the ejected species is given by the average value $m_{\rm sp}$ = 23
amu. For carbonaceous grains, we assume a composition of pure carbon,
thus there is no hydrogenation following collisions because
oxygen is the only projectile present in the clouds. As a consequence,
the only sputtered atoms are carbon atoms, therefore, $m_{\rm sp}$
= 12 amu.

\begin{figure*} \centering
  \includegraphics[width=0.48\hsize]{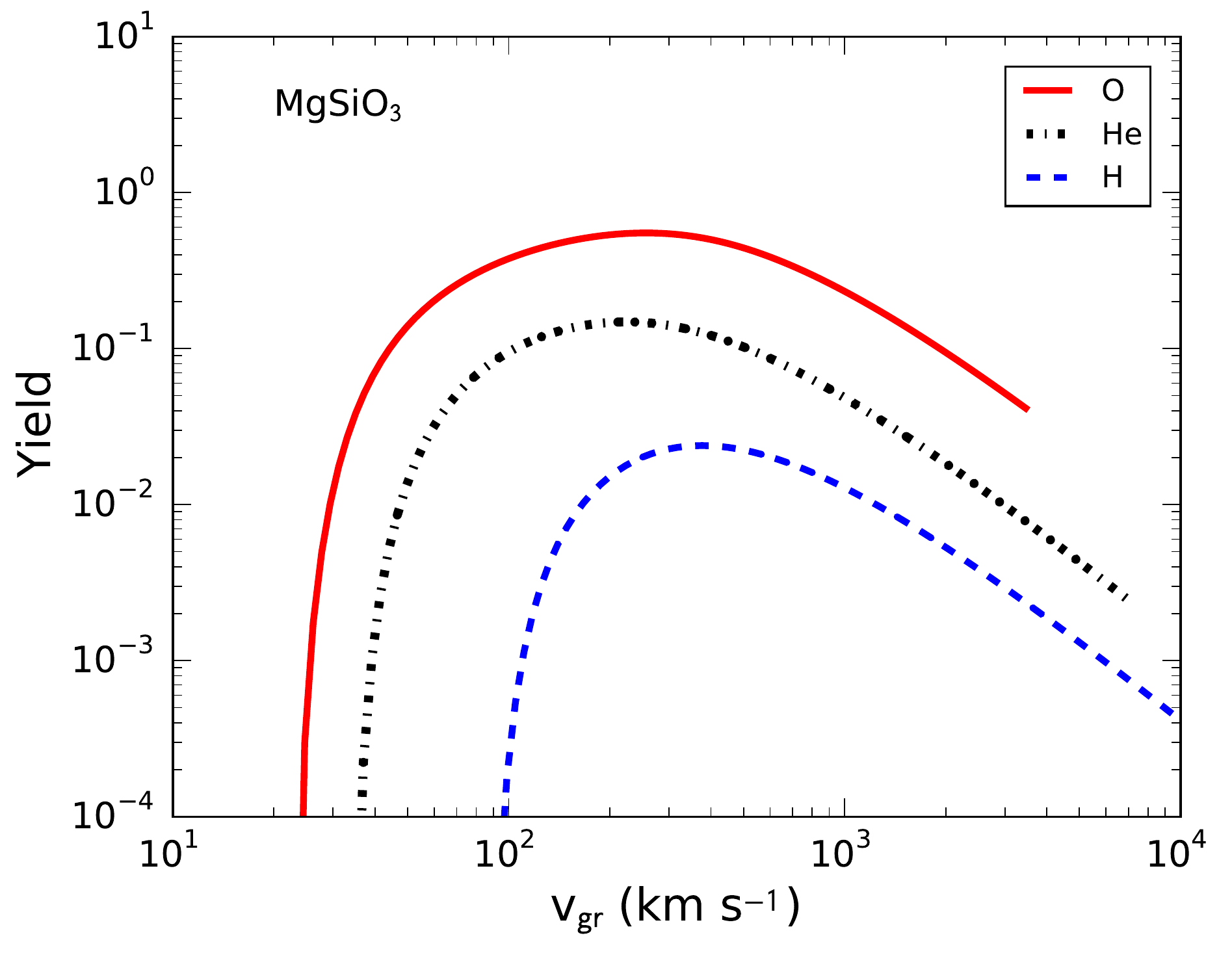}
  \includegraphics[width=0.48\hsize]{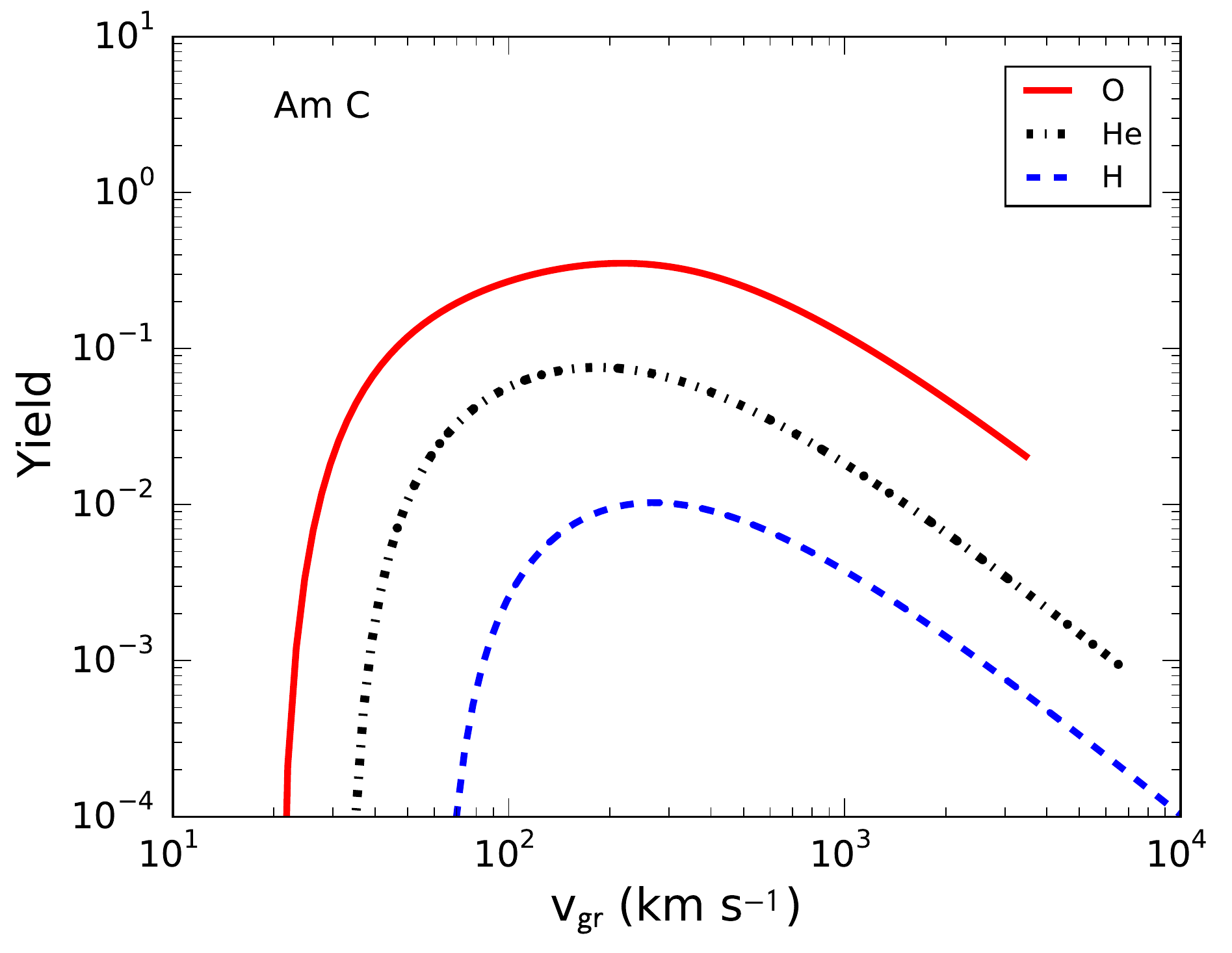}\\
  \includegraphics[width=0.48\hsize]{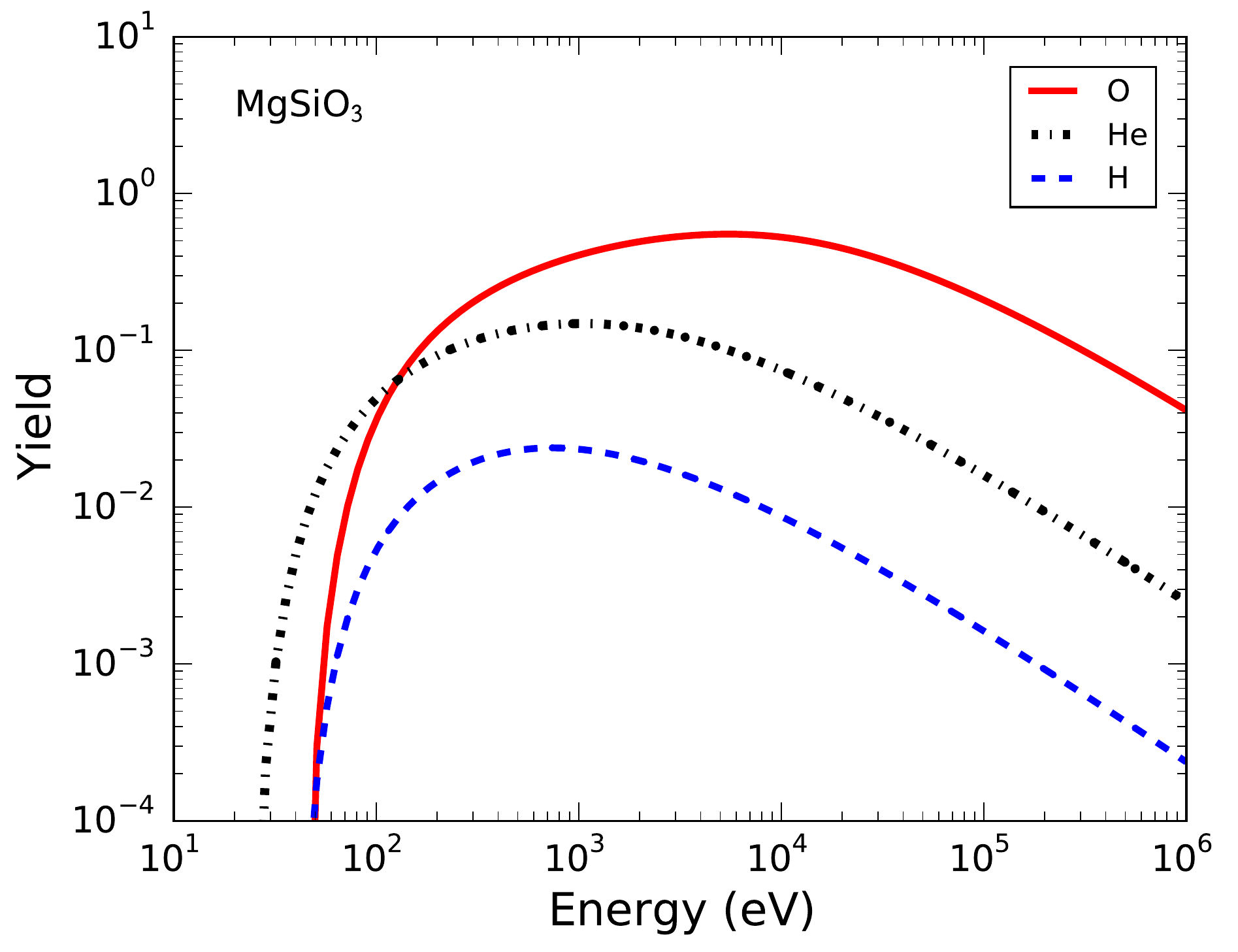}
  \includegraphics[width=0.48\hsize]{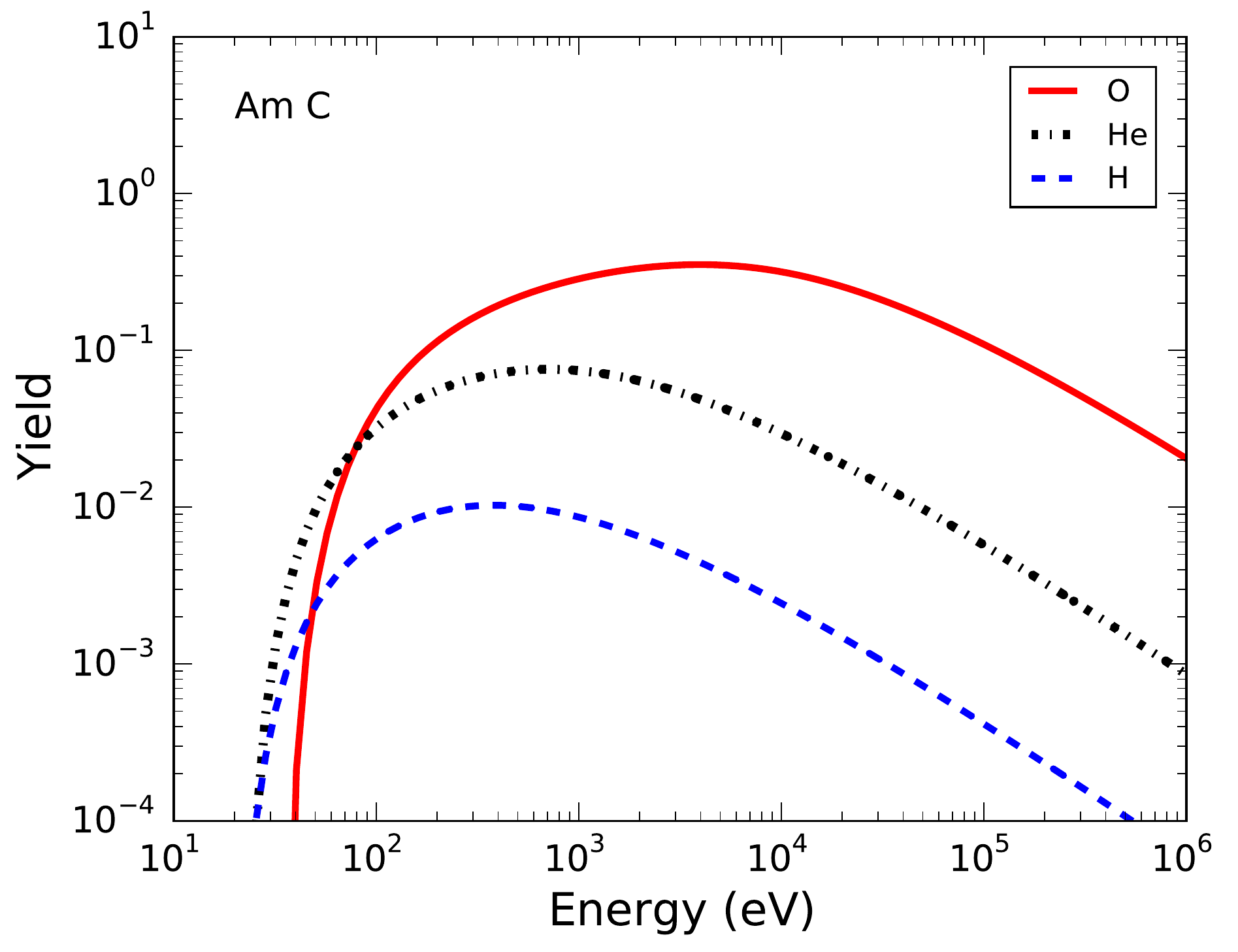}
   \caption{Sputtering yields of silicate (left column) and
     amorphous carbon grains as a function of the relative grain-gas
     velocity (top row) and corresponding kinetic energy for the incident
     projectiles: oxygen, present in the ejecta clouds, helium, and
     hydrogen. {  The sputtering yields were calculated using Eq.~11 in
     \citet{nozawa06}.}}
\label{fig:kin_sput}
\end{figure*}

Figure~\ref{fig:kin_sput}
presents the sputtering yield of amorphous carbon (Am C) and silicate (MgSiO$_3$)
grains as a function of the velocity and kinetic energy of the
incident projectiles, calculated using Eq.~11 in \citet{nozawa06}.

% FIGURE 9 *************************************************************
%
% **********************************************************************

Equation~\ref{eq:dm_dt} assumes that the thermal velocities of the
colliding atoms are negligible compared to the grain's velocity.  The
rate at which a grain slows down as a result of collisional drag
forces is given by $({\rm d}v_{\rm gr}/{\rm d}t) = F_{\rm drag}/m_{\rm
gr}$. For grains moving through a cold gas (2$kT << m_{\rm gr}v_{\rm
gr}^2$), the plasma correction to the drag force
\citep[][Sect.~6]{baines65} can be neglected and $F_{\rm drag}$ is
expressed as
\begin{equation}\label{eq:drag} 
  F_{\rm drag} = \pi a_{\rm gr}^2\, v_{\rm gr}^2 \sum_i\, n_i\, m_i
\end{equation} 
so that
\begin{equation}\label{eq:dv_dt} 
 \frac{{\rm d} v_{\rm gr}}{{\rm d} t} = \frac{3}{4}\, \frac{v_{\rm gr}^2}
{\rho_{\rm gr}\, a_{\rm gr}}\, \gamma\, \sum_i n_i\, m_i
,\end{equation} 
where $\rho_{\rm gr}$ is the mass density of the grain.\\

Combining Eqs.~\ref{eq:dm_dt} and \ref{eq:dv_dt} we can write the
erosion rate of the dust mass as it slows down through the gas as
 \begin{equation}\label{eq:dm_dv} 
   \frac{{\rm d} m_{\rm gr}}{{\rm d} v_{\rm gr}} = 
   \frac{m_{\rm gr}}{v_{\rm gr}}\, \frac{m_{\rm sp}}{m_{\rm H}}\,
       \frac{2}{\gamma}\, \frac{\sum_i\, n_i\, Y_i(E)}{\sum_i\,
         (m_i/m_{\rm H})\, n_i}
 ,\end{equation} 
which can readily be solved for the grain mass
$m_{\rm gr}(v_{\rm gr})$, i.e.
\begin{equation}\label{eq:m_m} 
  \frac{m_{\rm gr}(v_{\rm gr})}{m_{\rm gr}(0)} = 
  \exp\left[\frac{2}{\gamma}\, \frac{m_{\rm sp}}{m_{\rm H}}\,
    \int_{v_{\rm gr}(0)}^{v_{\rm gr}(t)}\, \frac{\sum_i n_i\, Y(E)}{\sum_i
      \, (m_i/m_{\rm H})\, n_i}\, \frac{{\rm d}v_{\rm gr}}{v_{\rm gr}}\right],
\end{equation}
where $m_{\rm gr}(0)$ and $v_{\rm gr}(0)$ are the initial grain mass and velocity, respectively, and $m_{\rm gr}(v_{\rm gr})$ is the
grain's mass at velocity $v_{\rm gr}(t)$.

The column density of gas traversed by a grain is
\begin{equation}\label{eq:n} 
  N_{\rm gas}(t) = n_{\rm gas} \int_{v_{\rm gr}(0)}^{v_{\rm gr}(t)}\,
  v_{\rm gr}(t')\, {\rm d}t'.
\end{equation} 
Using Eq.~\ref{eq:dv_dt}, we obtain that
\begin{eqnarray}\label{eq:dng_dv1}
  \frac{{\rm d}N_{\rm gas}(v_{\rm gr})}{{\rm d}v_{\rm gr}} & = &  \left[\frac{{\rm d}N_{\rm gas}(v_{\rm gr})}{{\rm d}t}\right]\, \left[\frac{{\rm d}(v_{\rm gr})}{{\rm d}t}\right]^{-1} ,\\ \nonumber
  & = & (n_{\rm gas}\, v_{\rm gr})\, \frac{4}{3 \gamma}\,  \frac{a_{\rm gr}(t)}{v_{\rm gr}^2}\, \frac{\rho_{\rm gr}}{\sum_i \rho_i}
\end{eqnarray}
where $\rho_i \equiv n_i\, m_i$.\\
Defining a characteristic slowing down time, $\tau_{\rm sd}$ as
\begin{equation}\label{eq:taud}
  \tau_{\rm sd} \equiv \frac{4}{3}\, \left[\frac{a_{\rm gr}(0)}{v_{\rm
gr}(0)}\right]\, \left(\frac{\rho_{\rm gr}}{\sum_i \rho_i}\right),
\end{equation}
we obtain that
\begin{equation}\label{eq:dng_dv2}
  \frac{{\rm d}N_{\rm gas}(v_{\rm gr})}{{\rm d}v_{\rm gr}} =
  \frac{1}{\gamma}\left[\frac{a_{\rm gr}(t)}{a_{\rm gr}(0)}\right]\,
  \left[\frac{v_{\rm gr}(0)}{v_{\rm gr}(t)}\right]\, \tau_{\rm sd}\, \sum_i
  n_i \;,
\end{equation}
where $n_{\rm gas} \equiv \sum_i n_i$. Since $[a_{\rm gr}(t)/a_{\rm
gr}(0)] = [m_{\rm gr}(t)/m_{\rm gr}(0)]^{1/3}$, Eq.~\ref{eq:dng_dv2}
can be integrated to give
\begin{equation}\label{ng} 
  N_{\rm gas}(v_{\rm gr})=\frac{1}{\gamma}\,N_0\, \int_{v_{\rm
      gr}(0)}^{v_{\rm gr}}\, F(v'_{\rm gr})\, \frac{{\rm d}v'_{\rm gr}}{v'_{\rm gr}},
\end{equation} 
where 
\begin{equation}
\label{fv} 
F(v_{\rm gr}') = \exp\left[\frac{2}{3\gamma}\,
\frac{m_{\rm sp}}{m_{\rm H}}\, \int_{v_{\rm gr}(0)}^{v_{\rm gr}'}\,\frac{\sum_j n_j\,
Y(E)}{\sum_i (m_i/m_{\rm H})\, n_i}\, \frac{{\rm d}v}{v}\right]
\end{equation}
and where $N_0$ is the column density, defined as
\begin{equation}
\label{n0}
N_0 \equiv v_{\rm gr}(0)\, \tau_{\rm sd}\, \sum_i n_i = \frac{m_{\rm gr}(0)}{\pi a_{\rm gr}^2(0)}\,\frac{\sum_i n_i}{\sum_i \rho_i}\;,
\end{equation}
so that the mass of gas contained within the volume swept up by the
grain is equal to the initial mass of the grain if it was not eroded by the gas. 

\begin{figure} \centering
  \includegraphics[width=\hsize]{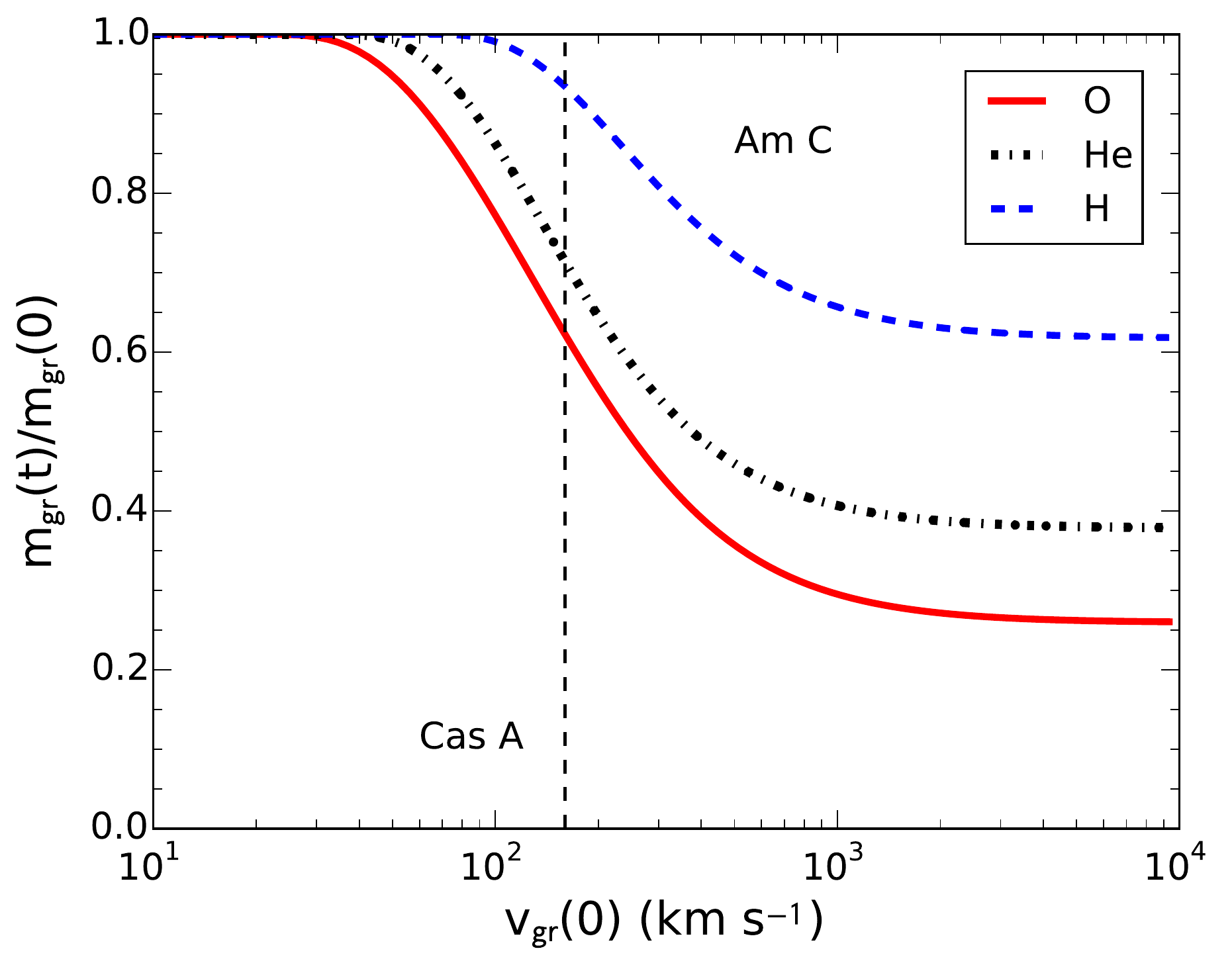}
  \includegraphics[width=\hsize]{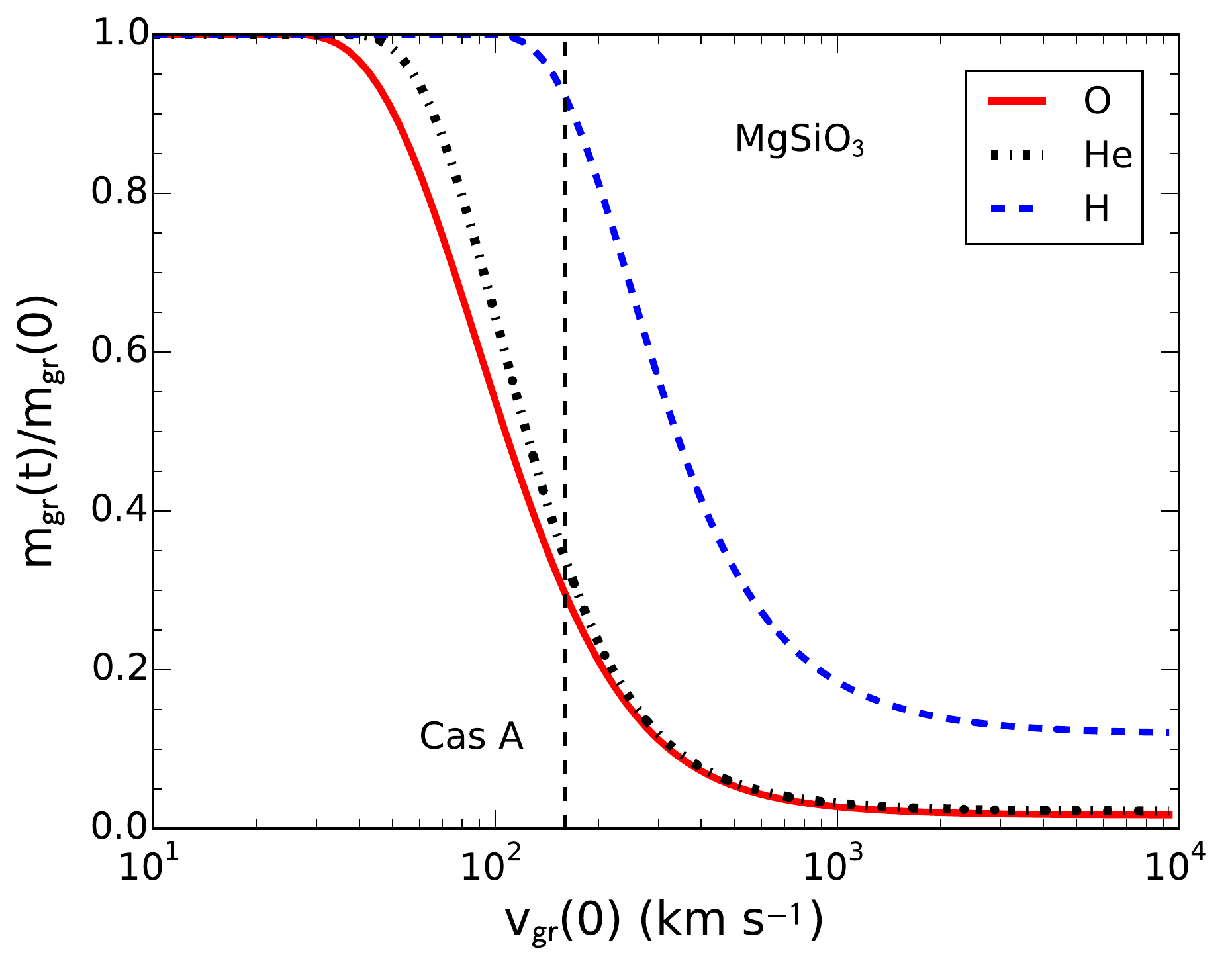}
  \caption{Mass fraction of a carbon (top panel)
and MgSiO$_3$ dust grain surviving the erosion by kinetic sputtering as a
function of its initial velocity $v_{\rm gr}(0)$, as it traverses a gas of
pure H, He, and O composition with density $n_{\rm gas}$ = $n_{\rm
cloud}$ = 100 cm$^{-3}$. The vertical line indicates the initial grain
velocity for Cas~A, $v_{\rm gr}(0)$ = $v_{\rm cloud}$ = 160 km s$^{-1}$.}
\label{fig:mass_ratio}
\end{figure}

Figure~\ref{fig:mass_ratio} shows which fraction of the initial
  mass of a dust grain survives the effect of kinetic sputtering. The
  mass fraction, represented as a function of the initial velocity of
  the grain $v_{\rm gr}(0)$, has been calculated after the grain has
  slowed down to zero, assuming $n_{\rm gas}$ = $n_{\rm
cloud}$ = 100 cm$^{-3}$. The mass fraction does not depend on the
initial grain size, since the same atoms that slow the grain down are
also eroding it by sputtering.

% FIGURE 10 *************************************************************
%
% **********************************************************************

% FIGURE 11 *************************************************************
%
 \begin{figure} \centering
  \includegraphics[width=\hsize]{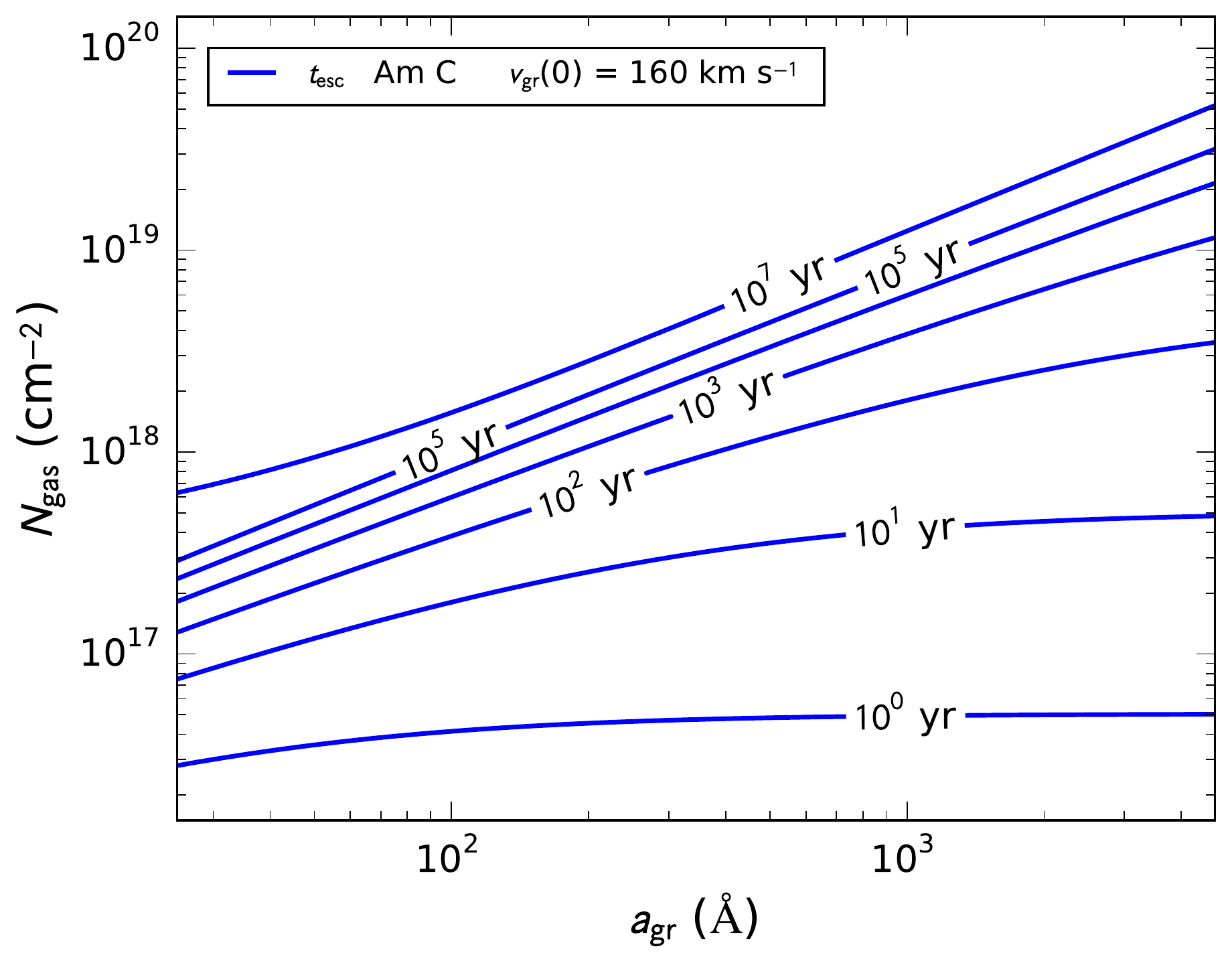}
   \includegraphics[width=\hsize]{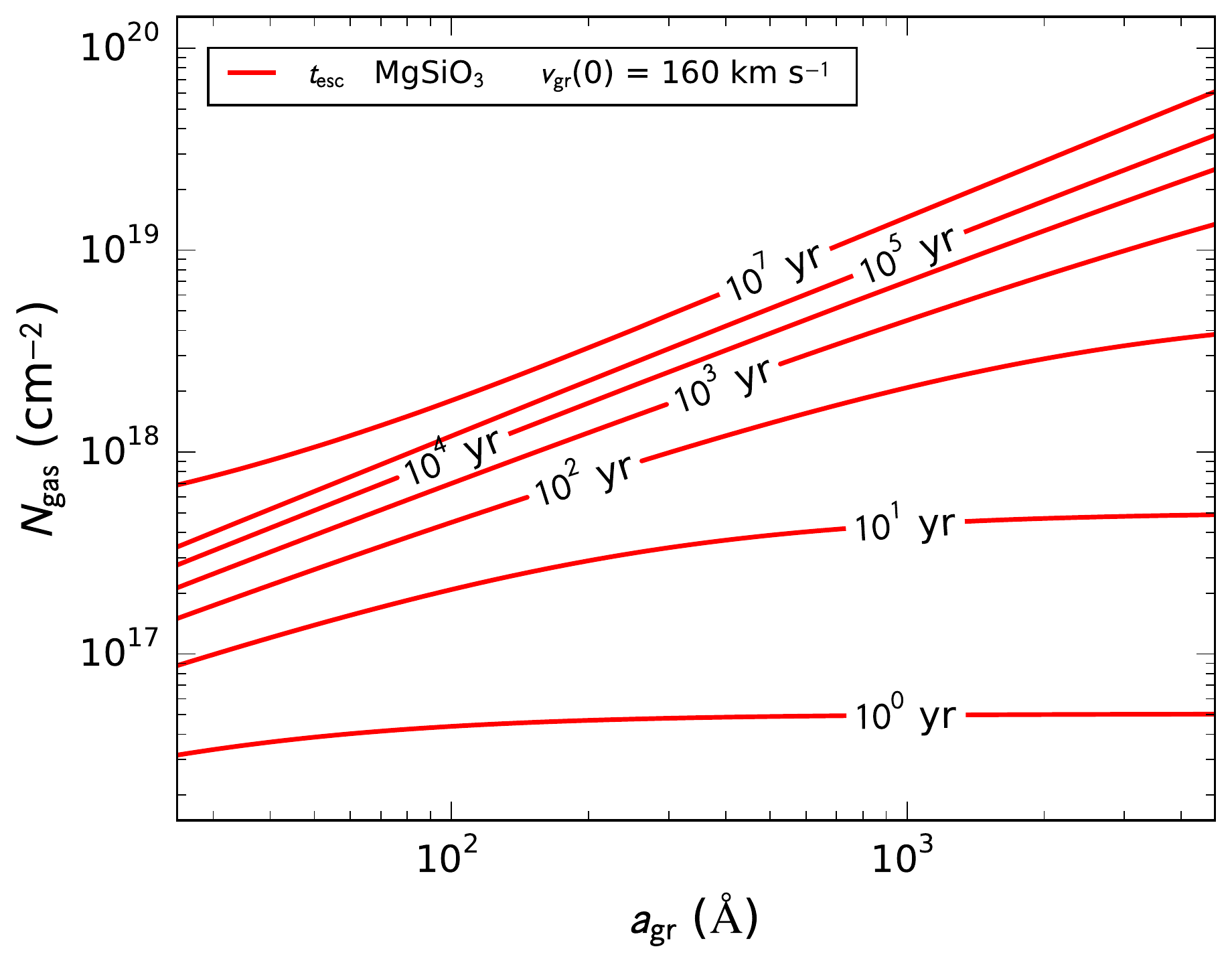}
   \caption{Contour plot showing the time $t_{\rm esc}$ required by a
     dust grain with initial velocity $v_{\rm gr}(0)$ = 160 km s$^{-1}$ to
     traverse a column density of gas $N_{\rm gas}$, as a function of the
     grain radius $a_{\rm gr}$.}
   \label{contour}
\end{figure}
% **********************************************************************

To calculate the time $t_{\rm esc}$ required by a dust grain to traverse a
  column density of gas $N_{\rm gas}$, we use the expression ${\rm
    d}N_{\rm gas} = n_{\rm gas}\, v_{\rm gr}(N)\,{\rm d}t$, from which we
  derive
\begin{equation}\label{Eq:t_esc}
  t_{\rm esc} = \int_0^{N_{\rm gas}} \frac{{\rm d}N'}{n_{\rm gas}\,
    v_{\rm gr}(N')}         
,\end{equation} 
where we obtained $v_{\rm gr}(N')$ numerically from Eq.~\ref{ng}. 
Figure~\ref{contour} is a contour diagram depicting $t_{\rm esc}$ as a
function of the dust grains radius and column density of gas
traversed. We calculated the curves for the initial velocity
$v_{\rm gr}(0)$ = $v_{\rm cloud}$ = 160 km s$^{-1}$. This is appropriate
for Cas~A, where the cloud shock is highly radiative, therefore
the post-shock gas is compressed by a factor $>> 4$ and the relative
gas-grain velocity is close to $v_{\rm cloud}$ (in an adiabatic shock
the relative velocity would be (3/4) $v_{\rm cloud}$).

Given an initial dust grain velocity,
$v_{\rm gr}(0)$, the amount of destruction taking place in a clump 
depends on the column density traversed by the grains after the
passage of the shock, which in turns depends on the size of the grains
and their location in the cloud. The grains, because of their large inertia,
move ballistically through the shock, slipping through the gas at a
relative velocity close to $v_{\rm cloud}$. We assume that the shock is
planar, so that all the shocked dust grains move perpendicular to the
shock front. Big grains sitting close to the edge
of the clumps are able to escape without experiencing substantial
sputtering, while smaller grains located deeper inside the cloud are
eroded and/or stopped before reaching the surface. 

In Appendix~\ref{app_column_density} we report an estimate of the
amount of kinetic sputtering occurring in the clumps based on a
statistical approach. However, this approach is only valid for the
largest grain sizes.  A Monte Carlo approach is required to properly
evaluate the effect of inertial sputtering in the ejecta clouds, which
takes all grains sizes and positions inside the clumps into
account. Methods and results of our simulation are discussed in
Sect.~\ref{sec:MC_simulation}.

%===================================================================
\section{Clump survival: Instabilities and evaporation}\label{sec:clump_stability}
%===================================================================

Inertial sputtering occurs in the ejecta clouds and the
corresponding amount of destruction depends on how long the grains
stay inside the clouds. In the present case, we assume that the
residence time depends on three phenomena. The first is the capability
of the grains to escape from the cloud owing to their ballistic
velocities;  we discussed this in Sect.~\ref{sec:kin_sput}. The
two other phenomena that we consider are the dynamical fragmentation
of the clouds and the thermal evaporation of the clouds, which we discuss
below.

The cloud crushing time, $t_{\rm cc} \equiv R_{\rm cloud}/
v_{\rm cloud}$ \citep{klein94}, provides an indication of the dynamical
time required for the cloud shock to fragment the cloud. { For $R_{\rm
cloud}$ = 7.5$\times$10$^{15}$ cm and $v_{\rm cloud}$ = 160 km
s$^{-1}$}, we obtain $t_{\rm cc}$ = 15 yr. For the destruction time,
$t_{\rm dest}$, we adopt the value of 3.5$t_{\rm cc}$ found by
\citet{klein94} for density contrasts between 10 and 100 (we have
100). This destruction time is defined as the time at which the mass
of the core of the cloud has been reduced to a fraction 1/$e$ of the
initial cloud mass; it is assumed that the cloud develops a core-plume
structure following the passage of the shock. We assume that after
$t_{\rm dest}$ = 3.5$t_{\rm cc}$ the ejecta cloud is destroyed and
dispersed. At that moment, the fresh dust still residing in the cloud
will have been released into the smooth ejecta.

A competing process for the disruption of the clumps is thermal
evaporation. Thermal evaporation occurs because electron thermal
conduction transfers heat from the hot smooth ejecta to the outer
layers of the colder clouds.  The heated material has an excess of
pressure, which then drives an outflow. \citet{cowie77} examined the
steady evaporation of spherical clouds including the effects of
saturation of the conductive heat flux. Saturation of the heat flux
occurs when the heat flux predicted based on the Spitzer conductivity
and the temperature gradient exceeds that which can be carried by the
electrons given their density and thermal speed. \citet{dalton93}
found an analytical solution for cloud evaporation that includes cases
of highly saturated conduction which applies for the Cas~A
knots. \citet{dalton93} ignore radiative cooling, which may be
important at the high densities and enormous cooling coefficients in
the Cas~A clouds. However, the contribution of radiative processes
decreases progressively with increasing saturation. For the present
work, we decided to focus on the effects of saturation and we will
include radiative evaporation in a follow-up paper. Using the results
from \citet{dalton93} for the relation between the saturation
parameter, $\sigma_0$, and mass loss rate, $\dot M$ and taking the
highly saturated limit we find
\begin{equation}
\dot M(\mathrm{M}_\odot/\mathrm{Myr})
= 1.978\times10^{13}\, \mu^{7/12}\, (P/10^4 \mathrm{k_B})^{5/6}\, 
R_\mathrm{cloud}^{11/6}\,(\mathrm{pc}),
\end{equation}
where $P$ is the pressure in the hot medium, $R_\mathrm{cloud}$ is the
cloud radius (in pc), and $\mu$ is the mean mass per particle in
g. There is no explicit dependence on temperature, although there is a
dependence on pressure, which depends on the temperature if density is
held constant as it is in Fig.~\ref{contour}.  For a cloud of pure
O$^{5+}$, $\mu$ = 4.45$\times$10$^{-24}$ g.  We take as the
evaporation timescale the quantity $t_\mathrm{sat} =
M_\mathrm{cloud}/\dot M$, which is then written as
\begin{equation}
t_\mathrm{sat} = 4.420\, n_\mathrm{cloud}\, (P/10^4 \mathrm{k_B})^{-5/6}\,
R_\mathrm{cloud}^{7/6} \,(\mathrm{pc})\;\; \mathrm{Myr},
\end{equation}
where $n_\mathrm{cloud}$ is the cloud space density (in cm$^{-3}$).  The high
pressure in the remnant reduces the evaporation timescale well below typical
ISM values, but it is still generally larger than the cloud crushing time in
this context.  Thus the cloud is torn apart before it can evaporate,
although evaporation could play a role in destroying some cloud fragments.

\begin{figure*} \centering
  \includegraphics[width=0.47\hsize]{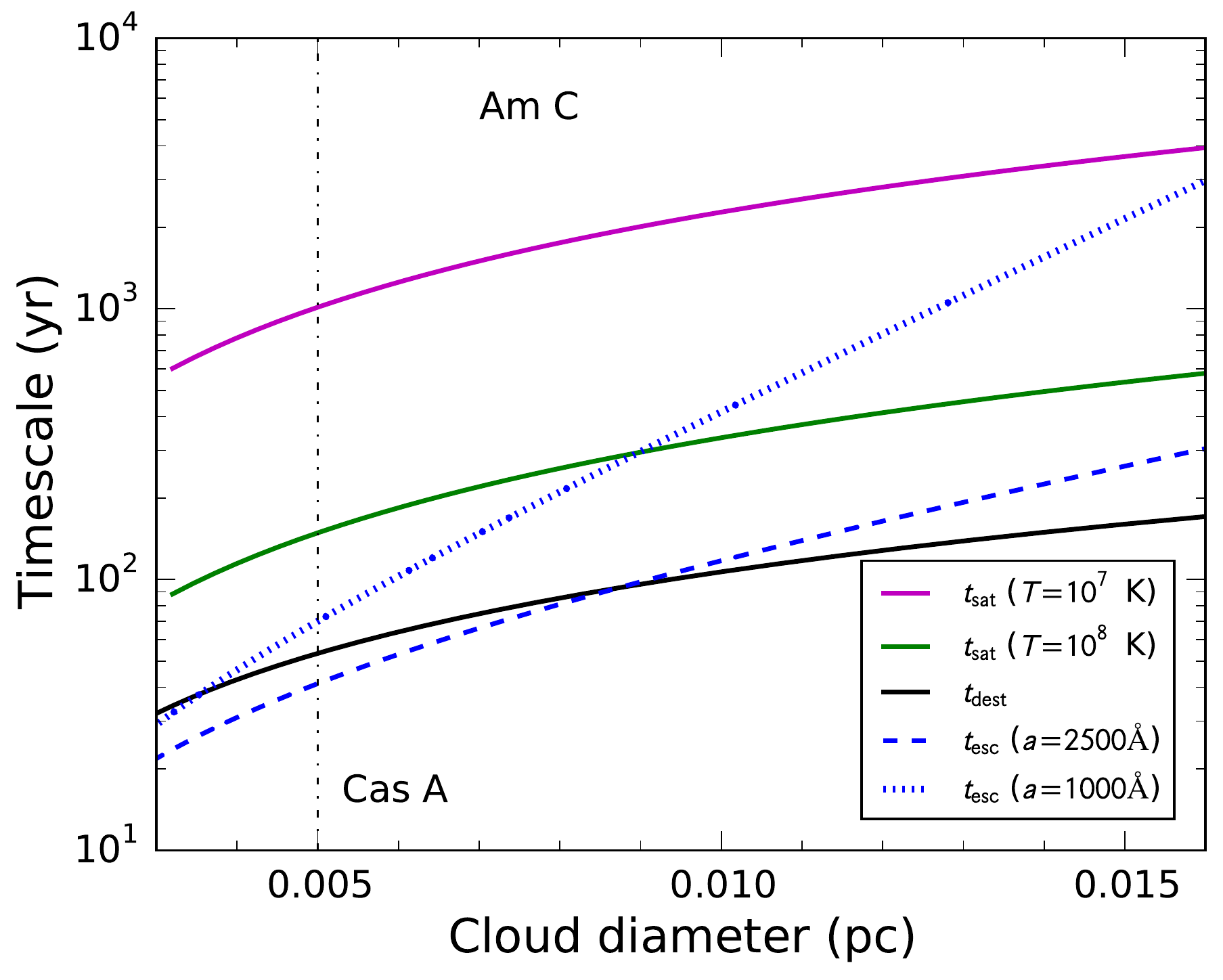}
  \includegraphics[width=0.47\hsize]{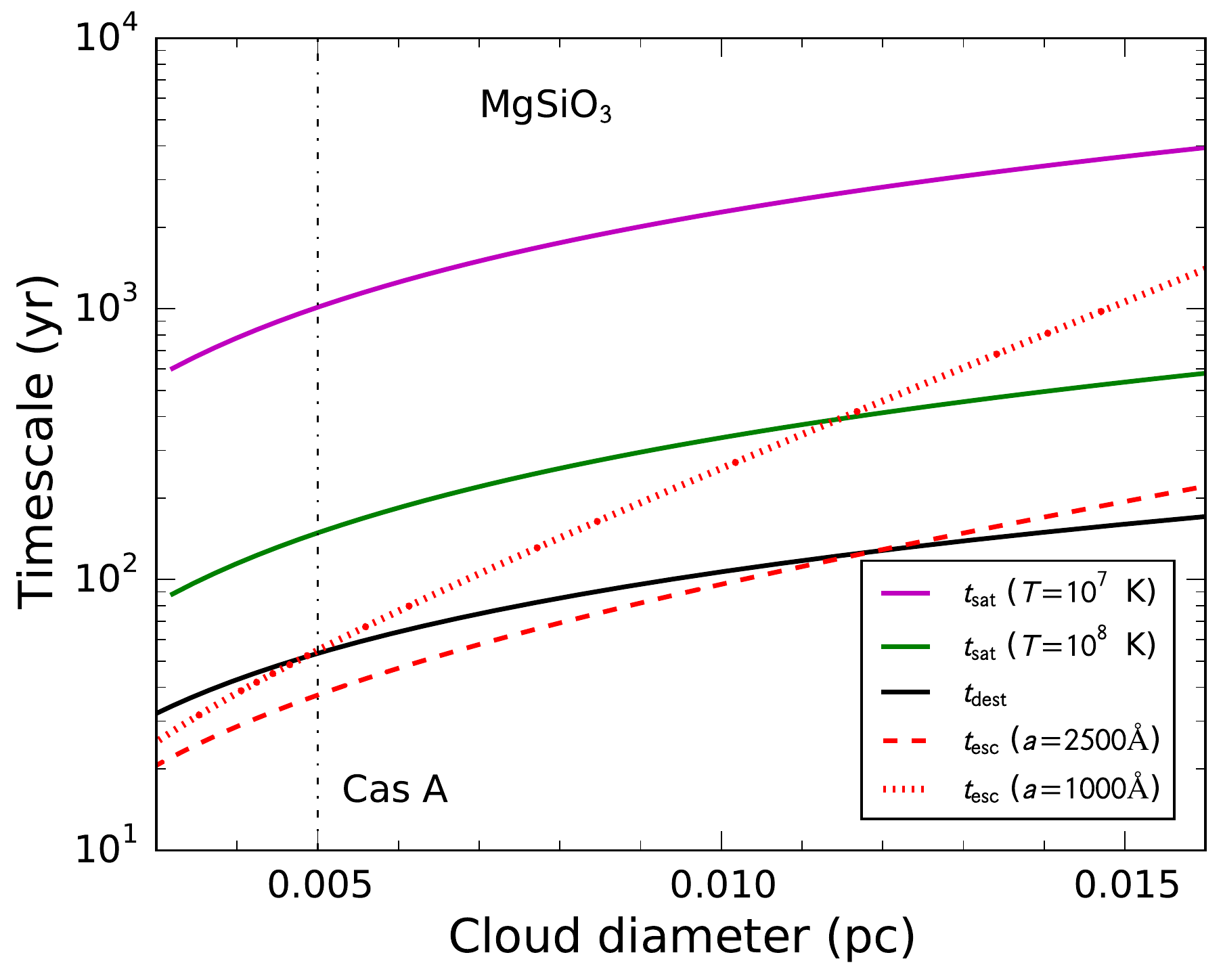}
  \caption{Comparison between timescales relevant for the processing
    of carbon (left panel) and silicate dust grains in an ejecta cloud:
    cloud destruction time $t_{\rm dest}$, cloud evaporation time
    $t_{\rm sat}$, calculated for two temperatures of the gas surrounding
    the clouds, and  dust escape time $t_{\rm esc}$, calculated for two
    grain sizes. Both $t_{\rm esc}$ and $t_{\rm dest}$ has been
    calculated assuming $v_{\rm gr}(0)$ = $v_{\rm cloud}$ = 160 km
    s$^{-1}$. All timescales are represented as a function of the
    diameter of the clump.  The vertical line indicates the diameter
    adopted for the ejecta clouds in Cas~A. See text for details.}
\label{fig:timescale_comparison}
\end{figure*}

The timescales discussed above are shown in
Fig.~\ref{fig:timescale_comparison} as a function of the diameter of
the clouds, for carbon and silicate grains. We calculated the
evaporation time $t_{\rm sat}$ for the two temperatures $T$=10$^7$~K
and $T$=10$^8$~K. For the escape time $t_{\rm esc}$, the two curves
correspond to two grain sizes: 2500 \AA\ (the upper limit of our grain
size distribution, see Sect.~\ref{sec:MC_simulation}) and 1000 \AA,
an intermediate value. For each cloud diameter, the value of $t_{\rm
esc}$ in the figure represents the time required to cross that
diameter, which is the maximum path that a grain has to traverse to
escape the cloud. Therefore, $t_{\rm esc}$ is the maximum escape
time. We calculated both $t_{\rm esc}$ and $t_{\rm dest}$ assuming
$v_{\rm gr}(0)$ = $v_{\rm cloud}$ = 160 km s$^{-1}$, where $v_{\rm
gr}(0)$ is the initial velocity of the grains. The value 160 km
s$^{-1}$ is the cloud shock velocity corresponding to the constant
reverse shock velocity in the ejecta core, which is also the lowest
velocity reached by the reverse shock in Cas~A (see
Fig.~\ref{fig:vrvb_vs_alpha}). From the definition of $t_{\rm dest}$ =
3.5$t_{\rm cc}$ = $R_{\rm cloud}/v_{\rm cloud}$ and $t_{\rm esc}$
(Eq.~\ref{Eq:t_esc}) it follows, therefore, that for each cloud size,
the values of $t_{\rm dest}$ and $t_{\rm esc}$ shown in
Fig.~\ref{fig:timescale_comparison} are upper limits.

From Fig.~\ref{fig:timescale_comparison} and from the considerations
above it follows that, for Cas~A, the trade is between the escape time
and the destruction time. For both carbon and silicate dust, grains
that are sufficiently big are able to escape before the cloud is
disrupted, even if they are located far from the surface. These
fugitive grains experience thermal sputtering in the hot phase of the
ejecta before their counterparts ejected at time $t_{\rm dest}$. The
injection of dust in the hot medium is a continuous process depending
on the grain size and position inside the clouds. To properly
investigate such a process, an analytical approach is not applicable,
therefore we decided to perform a Monte Carlo simulation; this is
described in Sect.~\ref{sec:MC_simulation}.

% FIGURE 12 *************************************************************
%
% **********************************************************************

%=========================================
\section{Grain destruction by the reverse shock: Thermal sputtering}\label{sec:thermal_sput}
%=========================================

After a dust grain is ejected from a cloud, because of its ballistic
velocity or because of the destruction of the cloud, it will find
itself in the smooth hot ejecta. Here the temperature is much higher
than in the clouds, therefore the dominant erosion mechanism is
\emph{thermal sputtering}, where the velocity of the ions is
determined by the temperature of the shocked gas (thermal motion). The
effect of thermal sputtering has been calculated for two temperatures
of the smooth ejecta: $T \sim$ 10$^{8}$ K (from
Eq.~\ref{temperature_eq}) and $T \sim$ 10$^{7}$ K (for comparison).

The variation of the grain radius, $a_{\rm gr}$, due to thermal
sputtering is given by Eq.~4.20 in
\citet{tielens94},
\begin{equation}\label{rate_av_eq}
  \frac{1}{n_{\rm H}} \frac{{\rm d}a_{\rm gr}}{{\rm d}t} = \frac{m_{\rm
      sp}}{\rho_{\rm gr}} \sum A_i \left \langle Y_i\,v \right \rangle
,\end{equation}
where $\langle Y_i\,v \rangle$ is the sputtering yield of ion $i$
averaged over the Maxwellian distribution ($Y_i$ as in Sect.~\ref{sec:kin_sput}),
$v$ is the velocity of the ions, $n_{\rm H}$ is the
hydrogen number density of the smooth ejecta, and $A_i$ is the abundance
of ion $i$ with respect to hydrogen. As discussed in
Sect.~\ref{ejecta_temp_sec}, we assume for the smooth ejecta a
mixture of O, Ne, Mg, Si, S, Ar, and Fe as determined by 
\citet{hwang12}. Because of the compression due to the
shock, the density of the smooth ejecta must be increased by a factor of
four with respect of its pre-shock value.
The Monte Carlo approach used to evaluate thermal sputtering across
the remnant is described in Sect.~\ref{sec:MC_simulation}.

\section{A Monte Carlo approach to evaluate kinetic and thermal
  sputtering: Methods and results}\label{sec:MC_simulation}

For each ejecta cloud, we consider a population of 3$\times$10$^{6}$
particles of each dust grain type (carbon and silicate) with a given
size distribution and homogeneously distributed over the entire volume
of the cloud. We follow the journey of each particle to
determine its final size (and consequently mass) as a result of
kinetic sputtering (in the ejecta clumps) and thermal sputtering (in
the smooth ejecta).  As described in Sect.~\ref{sec:ejecta_properties},
we adopt the classical MRN
expression for the dust grain size distribution \citep{MRN77}.

Following the encounter with the reverse shock, the dust grains
inside the cloud acquire a relative velocity with respect to gas. This
velocity decreases asymptotically to zero because of gas-grain collisions,
which at the same time are responsible for the progressive erosion of
the grains (kinetic sputtering). Depending on their initial size and
position inside the cloud, some of the grains are able to escape
into the smooth ejecta before the cloud gets disrupted. These fugitive
grains therefore experience thermal sputtering for a longer time
than their counterparts trapped inside the cloud.

The basic equation governing our Monte Carlo simulation, which allows
us to calculate the final radius $a_{\rm final}$ of a dust grain that
experienced the processing previously described, is the following:
\begin{equation}\label{eq:a_final}
  a_{\rm final}(t) = K_{\rm s}(t)\,a_{\rm initial} - T_{\rm s} \times
\left[\frac{1}{(t_{\rm r} + t_{\rm inj})^2} - \frac{1}{t^2}\right].
\end{equation}
The simulation starts when the reverse shock touches the outer layer
of ejecta ($\alpha$ = 1, $\sim$0.9 years after explosion, see
Fig.~\ref{fig:tr_vs_alpha}), and stops when the reverse shock reaches
the centre of the remnant ($\alpha$ = 0, $\sim$8000 years after
explosion).  In Eq.~\ref{eq:a_final}, $a_{\rm initial}$ is the size of
the grain before processing; $t_{\rm r}$ is the time when the reverse
shock hits an ejecta cloud located in a layer $\alpha$ (see
Sect.~\ref{sec:initial_conditions}); and $t_{\rm inj}$ is the time
interval (starting from $t_{\rm r}$) after which the grains are
expelled from the clouds and injected into the smooth ejecta and is
given by the relation $t_{\rm inj}$ = Min($t_{\rm esc}$, $t_{\rm
dest}$). For each layer, we computed $\alpha$, $t_{\rm esc}$ and
$t_{\rm dest}$ using the value for $v_{\rm gr}(0)$ = $v_{\rm cloud}$
corresponding to that layer and derived from $v_{\rm r}$ (see
Fig.~\ref{fig:vrvb_vs_alpha}), and we appropriately calculated
$t_{\rm esc}$ from Eq.~\ref{Eq:t_esc} for each grain size and position
inside the cloud (Sect.~\ref{sec:kin_sput}).

The term $K_{\rm s}(t)$ is the reduction factor of the grain radius
due to kinetic sputtering. We derived this term from Eq.~\ref{eq:m_m}
(to the power 1/3 to convert from mass to radius), where we derived the final
velocity of the grain at time $t$, $v_{\rm gr}(t)$,
inverting Eq.~\ref{ng}, and the column density $N_{\rm gr}$ comes from
inversion of Eq.~\ref{Eq:t_esc}. It is useful to remember here that
the effect of kinetic sputtering is of reducing the mass of each dust
grain by the same factor, regardless of {  its initial} size
(Eq.~\ref{eq:m_m}). Because of the relationship between grain
mass and grain radius, the same statement holds true for the grain
radius as well, therefore $a_{\rm final}(t)/a_{\rm initial}$ is
independent from $a_{\rm initial}$. Kinetic sputtering is switched
OFF for $t>t_{\rm inj}$, when the grains leave the clouds.

\begin{figure}
  \begin{center}
    \includegraphics[width=\hsize]{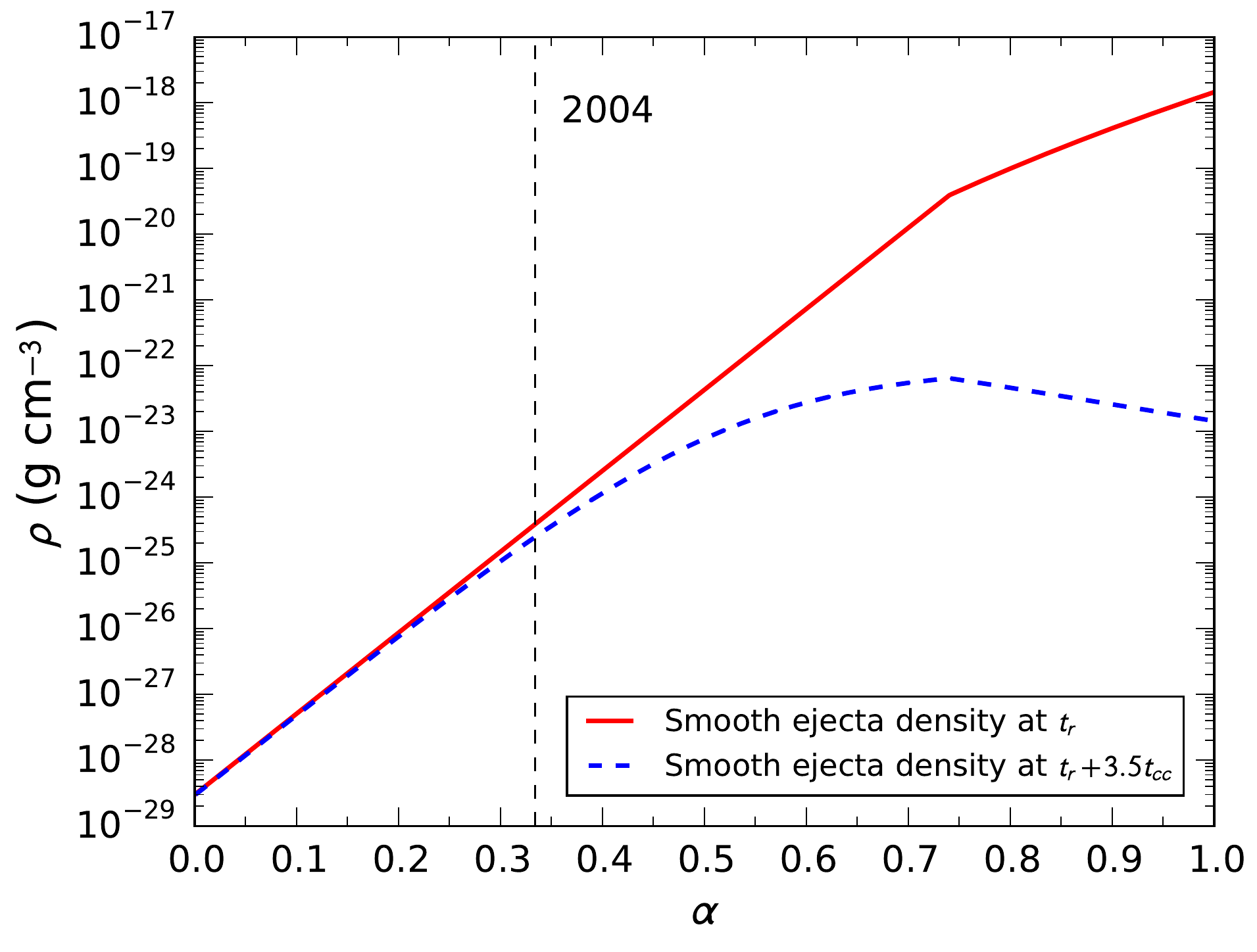}
  \end{center}
  \caption{Density structure of the smooth ejecta as a function of
    the parameter $\alpha$. Solid red line: density at the time
    $t_{\rm r}$ when the layer $\alpha$ encounters the reverse shock
    (same as Fig.~\ref{fig:rho_vs_alpha}); dashed blue line: density
    at the time $t_{\rm r}$ + 3.5$t_{\rm cc}$, when the ejecta cloud is
    dispersed and the dust still in there is injected into the smooth
    phase. {  Vertical line as in Fig.~\ref{fig:tr_vs_alpha}.} 
    \label{fig:rho_vs_alpha_2} }
\end{figure}

The term $T_{\rm s}\times [...]$ accounts for the effect of thermal
sputtering; this process removes a layer of equal thickness from each
grain, regardless of its size. This implies of course that the smaller
grains are more affected than the bigger grains. We obtain
\begin{equation}\label{eq:T}
  T_{\rm s} = \frac{n(t)\,t^{3}}{2}\times \left(\frac{m_{\rm
      sp}}{\rho_{\rm gr}} \sum A_i \left \langle Y_i\,v \right \rangle  \right),
\end{equation}
where the term in parenthesis comes from Eq.~\ref{rate_av_eq} and the
density of the expanding smooth ejecta, $n(t)$ has been calculated
from Eq.~\ref{rho_ej} and \ref{core_density_2}. Because of the
expansion of the supernova remnant, the density of the smooth ejecta
changes while inertial sputtering is occurring inside the
clouds. Therefore, the dust is progressively injected into a medium
whose density changes continuously. Figure~\ref{fig:rho_vs_alpha_2}
illustrates this phenomenon. The solid red line represents the density
of the smooth ejecta at the time $t_{\rm r}$ when each layer $\alpha$
encounters the reverse shock (same curve as in
Fig.~\ref{fig:rho_vs_alpha}). The dashed blue line shows the density
of the same layer $\alpha$ but at the time $t_{\rm r}$ + 3.5$t_{\rm
cc}$, when the clumps in that layer are destroyed and the dust
still residing in them is released into the smooth phase of the
ejecta. In the outer layers ($\alpha$ close to one) the difference is
very pronounced, and decreases progressively towards the inner layers
($\alpha$ close to zero). This has to be taken into account to
properly evaluate the effect of thermal sputtering, which strongly
depends on the density of the medium.

Using Eq.~\ref{eq:a_final} we have evaluated the fraction of dust
which is expected to survive processing while the reverse shock
progresses towards the centre of the remnant, expressed in terms of
the parameter $\alpha$. We assume a homogeneous distribution of ejecta
clouds in each layer. The fraction of surviving dust is calculated for
the entire remnant, i.e. this calculation includes not only the amount of dust
remaining after processing down to (but excluding) layer $\alpha$, but
also the unprocessed dust located inside and inwards of layer
$\alpha$. Because we stopped our simulation at $\alpha$ = 0, we did
not consider further processing of the dust due to the secondary
blast-wave shock which generates when the reverse shock bounces at the
centre of the remnant \citep{truelove99}.

% FIGURE 13 *************************************************************
%
% ********************************************************************

% FIGURE 14 *************************************************************
%
\begin{figure*} \centering
  \includegraphics[width=0.49\hsize]{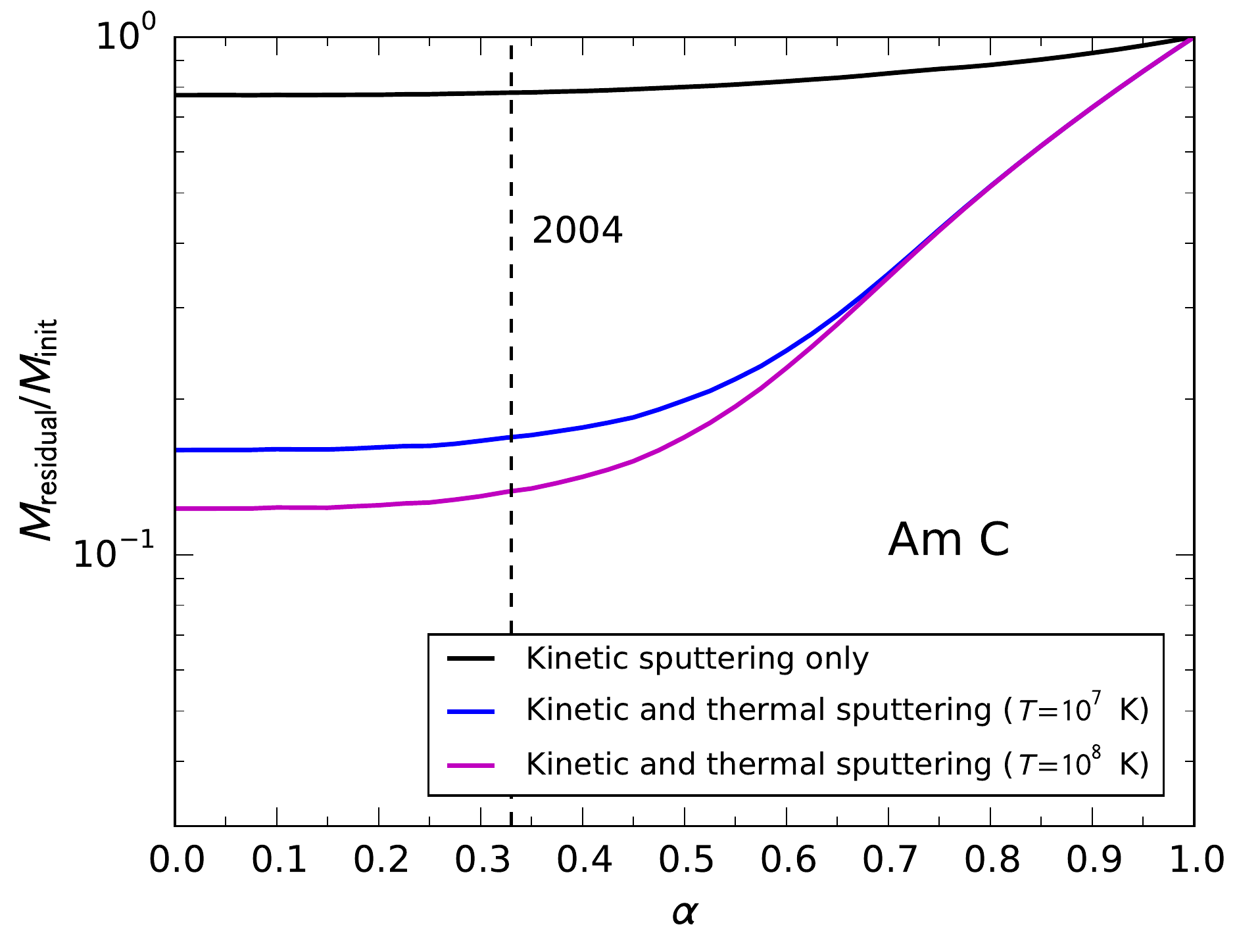}
  \includegraphics[width=0.49\hsize]{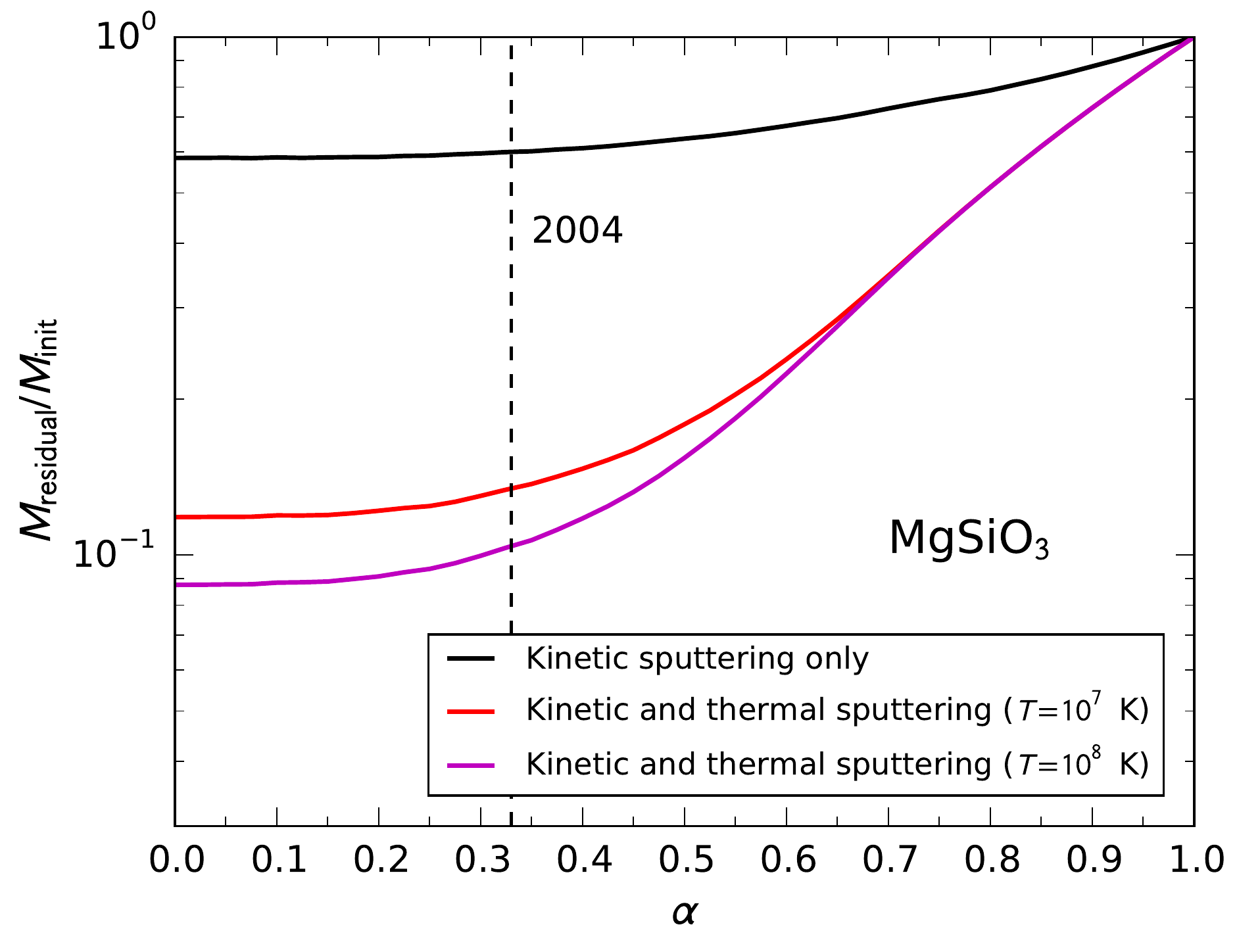}
  \caption{Fraction of surviving carbon (left panel) and silicate
      dust over the entire Cas~A remnant as a function of
      $\alpha$. For each value of $\alpha$ the curve provides the fraction
      of remaining dust, which is given by the grains able to survive
      processing located in the outer layers down to $\alpha$ plus the
      unprocessed dust residing in the inner layers. The three curves show
      the effect of kinetic sputtering only and of kinetic+thermal, the
      latter evaluated for two temperatures of the smooth ejecta: $T
      =$10$^{7}$ K and $T =$10$^{8}$ K.  }
    \label{fig:surviving_mass}
\end{figure*}
% ********************************************************************

% FIGURE 15 *************************************************************
%
 \begin{figure*} \centering
  \includegraphics[width=0.49\hsize]{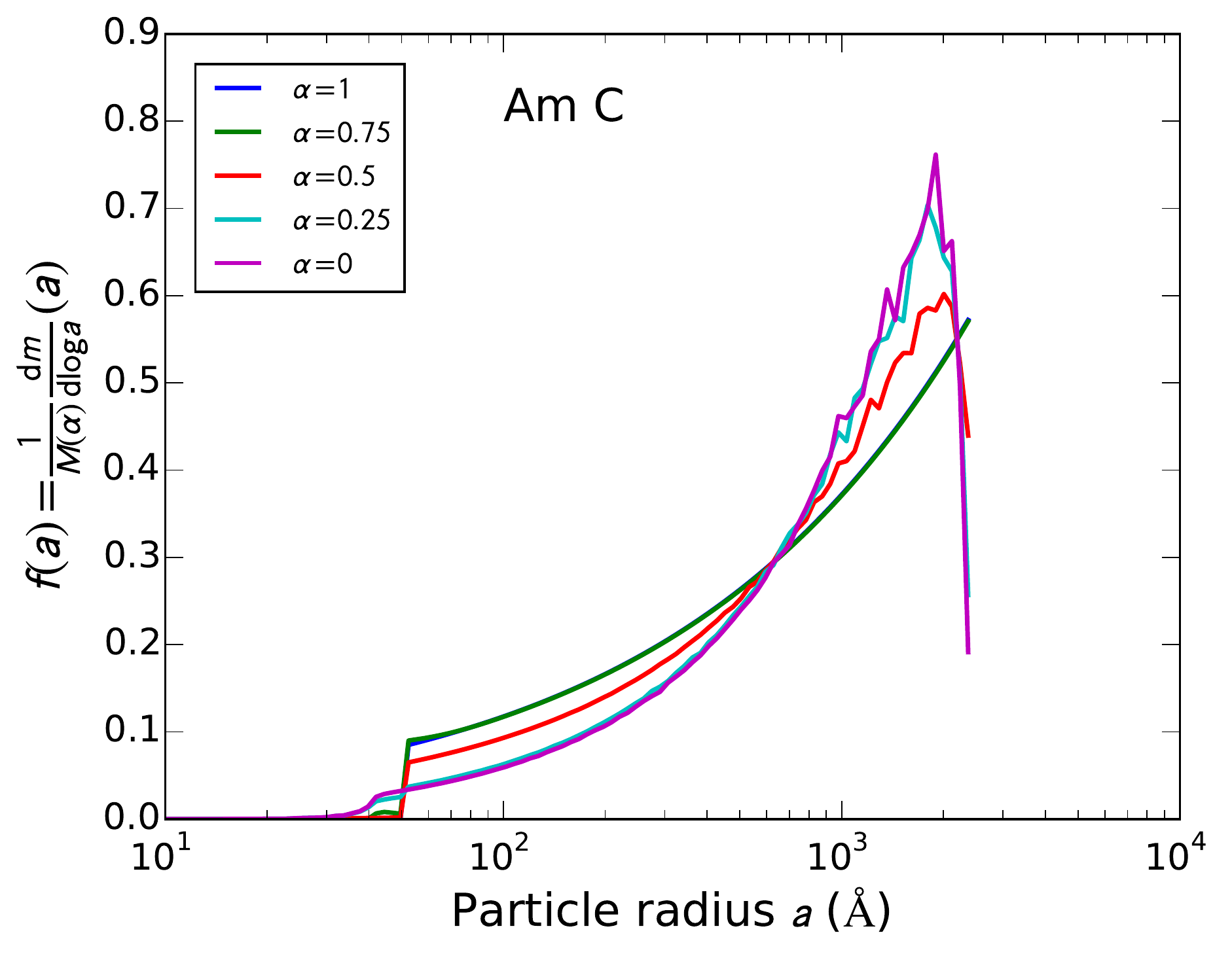}
  \includegraphics[width=0.49\hsize]{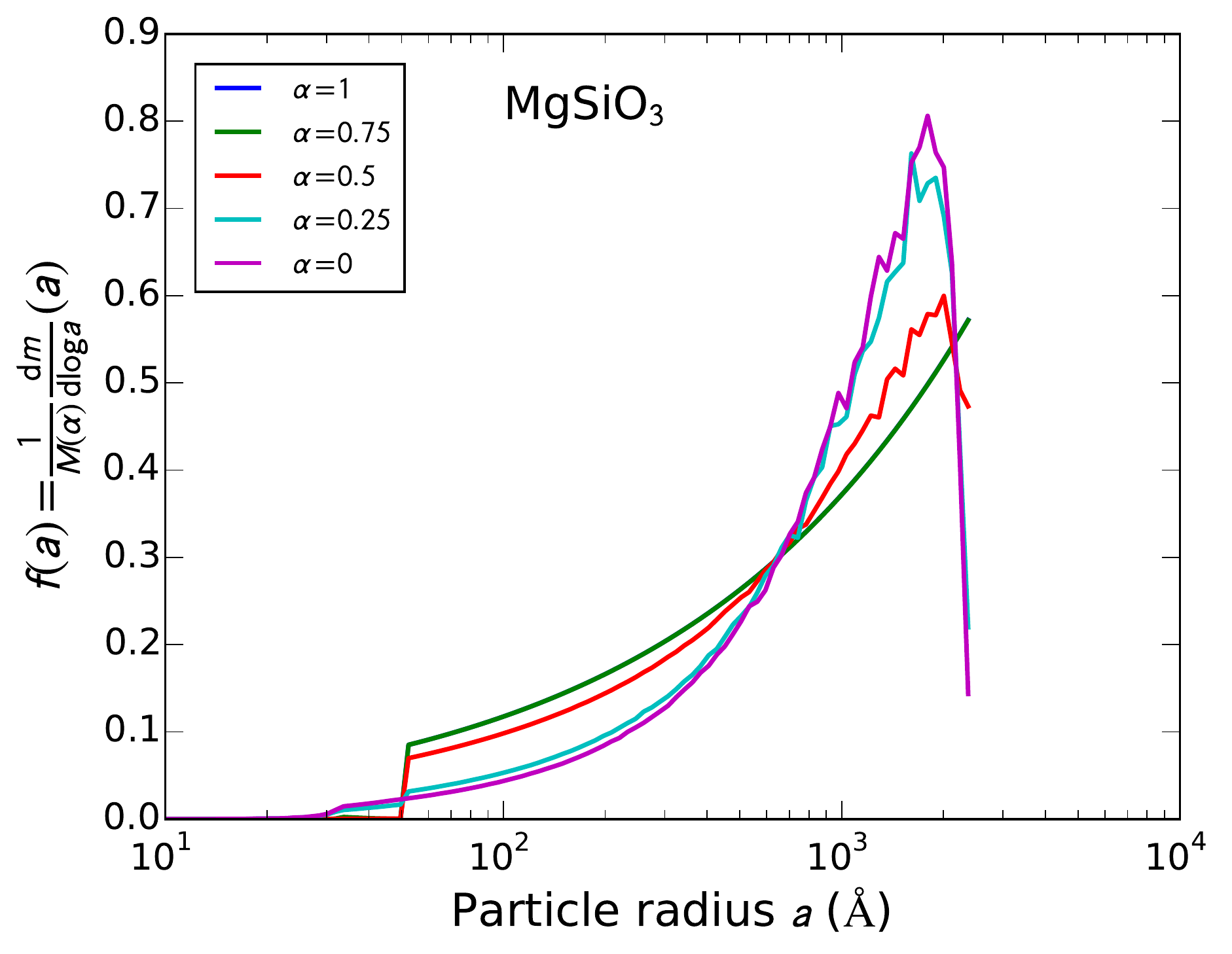}
  \caption{Surviving mass distribution as a function of the
      grain radius from Eq.~\ref{eq:mass_distr_norm}, for carbon (left
      panel) and silicate dust, {evaluated for temperatures of the
      smooth ejecta between 10$^{7}$ K and 10$^{8}$ K.} The
      distributions were normalized to the total surviving dust
      mass (with $M(\alpha) \equiv M_{\rm residual}$) so that the
      area under each curve equals 1.}
\label{fig:mass_distribution}
\end{figure*}
% ********************************************************************

Our results are shown in Fig.~\ref{fig:surviving_mass}, depicting the
evolution of the fraction of surviving dust, $M_{\rm residual}/M_{\rm init}$ as a
function of $\alpha$ for Am C and MgSiO$_3$ grains. In each panel, the
top curve illustrates the effect of kinetic sputtering only; this sputtering 
occurs inside the clumps. The two other curves refer to the combined
effect of kinetic and thermal sputtering, and this latter is evaluated
for two temperatures of the smooth ejecta: 10$^{7}$ K and 10$^{8}$
K. We notice that the effect of sputtering is important over the
entire remnant. At the end of the processing ($\alpha$ = 0), kinetic
sputtering has contributed significantly to particle erosion,
removing 20\% of the mass of carbon grains and 40\% of the mass of
silicate grains (note that the plots are in semi-logarithmic scale). Thermal
sputtering reduces the final amount of surviving dust even more dramatically.

Silicate grains appear more fragile than carbonaceous grains. This
behaviour results from the combination of different factors: their
slightly higher sputtering yield; the average mass of the sputtered
particles, almost twice that for Am C (23 amu versus 12 amu); and
the tendency of a larger fraction of silicate grains to escape the
ejecta cloud before disruption, which exposes them to thermal
sputtering for a longer time.

To provide a feeling for the preferred grain sizes of the surviving
dust, in Fig.~\ref{fig:mass_distribution} we plot the dust mass
distribution as a function of the grain radius, which is normalized to the
total surviving mass for different values of $\alpha$, i.e.   
\begin{equation}\label{eq:mass_distr_norm}
  f(a) = \frac{1}{M(\alpha)}\,\frac{{\rm d}m}{{\rm d}\log
    a}(a)\;\;\;\; {\rm with} \;\; M(\alpha) = \int {\rm d}\log a\frac{{\rm d}m}{{\rm d}\log a}
,\end{equation}
where $M(\alpha) \equiv M_{\rm residual}$
(Fig.~\ref{fig:surviving_mass}). The initial mass distribution of
carbon and silicate dust ($\alpha$ = 1, no processed layers) is
modified into the final mass distribution ($\alpha$ = 0, all layers
processed). We evaluated the mass distribution for two
temperatures of the smooth ejecta, $T$ = 10$^{7}$ K and $T$ = 10$^{8}$
K, and the resulting curves are almost indistinguishable. We decided
to show the evolution corresponding to $T$ = 10$^{8}$ K, but the
curves are indeed representative of the temperature range 10$^{7}$ --
10$^{8}$ K.  Both dust types show a progressive modification towards
their final distributions, which are very similar and cover the
original size range (50 -- 2500 \AA) with a small leakage down to 30
\AA. The differences reflect the tendency of MgSiO$_3$ grains to
dissociate more easily than carbonaceous grains.

%=======================================
% Section 8
\section{Discussion and conclusions}\label{sec:discussion}
%=======================================

In Table~\ref{final_mass_tab} we report the percentage of surviving
dust mass for Am C and MgSiO$_3$ grains evaluated at two specific
times during the evolution of the remnant: in 2004 ($\alpha$ = 0.33),
when some of the astronomical observations used in this work were
carried out, and after the reverse shock reached the centre of the
remnant ($\alpha$ = 0).

In 2004, the amounts of surviving dust are of the same order of
magnitude, but we can notice the difference between the two type of
grains and the increased destructive effect of the higher
temperature (see Fig.~\ref{fig:surviving_mass}). \citet{arendt14}
estimated the amount of dust that should be currently present in Cas
A: $\lesssim$ 0.1 M$_\sun$ (cold unshocked ejecta) + $\sim$0.04
M$_\sun$ (shocked circumstellar/interstellar and ejecta regions). If
we assume that this amount of dust represents the percentage of
surviving mass that we calculate, we can derive the corresponding
amount of dust that Cas~A should have produced. Taking a
representative value of 13.5\%, to average over dust type and
gas temperature, we obtain an initial mass of dust of about 1
M$_\sun$, which reduces to $\sim$0.8 M$_\sun$ if only silicates are
considered. These values are larger than the current estimate of 0.1 -- 0.5
M$_\sun$ \citep{todini01, schneider04, cherchneff10, nozawa10}.
The situation is similar after the reverse shock has swept
the entire cavity of the SNR, reaching the centre ($\alpha$ = 0). This
is not surprising because, after 2004, the fraction of surviving dust
mass tends to reach a plateau, as shown in Fig.~\ref{fig:surviving_mass}. 

We compared our findings with the results from \citet{bianchi07},
calculated for homogeneous (smooth) ejecta with a post-shock
temperature of the order of 10$^{7}$ -- 10$^{8}$~K expanding in a
homogeneous medium, for a progenitor star of 15 -- 25 M$_\sun$ and a
density of the surrounding ISM of 10$^{-24}$ g cm$^{-3}$. These values
are the closest to the parameters that we adopt for Cas~A: same
mass range for the progenitor star and a circumstellar medium with
density $n_0$ = 2.071 H atom cm$^{-3}$, corresponding to $\rho_0$ =
3.6$\times$10$^{-24}$ g cm$^{-3}$. We evaluated  the
effect of the two temperature values separately. We assume that the amount of
surviving dust calculated by \citet{bianchi07}, shown in their
Fig.~5, refers to a mixture of amorphous carbon and Fe$_3$O$_4$,
whose pre- and post-shock size distributions are reported in their
Fig.~4.

From Fig.~5 in \citet{bianchi07} we derive that the amount of dust
surviving the passage of the reverse shock is around 7.5\%. This is
around a factor of (1.2 -- 2.1) lower than our findings.
Some considerations have to be made when comparing these
results. \citet{bianchi07} adopted for Cas~A homogenous smooth
ejecta only (no clumps), expanding into a uniform medium. This
choice has two implications. The first is a different velocity
profile for the reverse shock than the more realistic case of
inhomogeneous smooth ejecta (uniform core + power-law envelope)
expanding into a power-law ambient medium, which we considered in this work. 
The second is that the absence of oxygen-rich, fast-cooling ejecta
clumps means that the dust spends all its life in the highly
destructive hot smooth ejecta. In our case the grains reside for
some time in the protective environment represented by the cold
clumps, where they are processed by kinetic sputtering, which is less
efficient in destroying the dust.

A comparison of our results with previous work is complicated by
a number of factors: the different approaches (numerical or
analytical) and descriptions of the supernova ejecta and ambient
environment adopted in the various studies, and the different choices
for the grain destruction processes, grain composition and
size distribution. Following the
quantitative description of \citet{dwek05}, \citet{nozawa07}
calculated the dust survival in the SN ejecta, exploring various
explosion energies, ambient media densities, and grain sizes and
composition.  Using numerical models, they find that the fraction of
the dust mass destroyed ranges from 0.2 to 1.0, depending on model
parameters. Using the analytical model of \citet{truelove99} and the
SN dust composition and size distribution of \citet{todini01},
\citet{bianchi07} found that only between 2 and 20\% of the dust mass
survives the passage of the reverse shock depending on the parameters
adopted. Exploring a wide range of dust parameters and ambient gas
density profiles, \citet{nath08} found that the fractional mass of
grains destroyed is $\lesssim 20$\%, and they provide a detailed
discussion of why their results are significantly lower than those
found by \citet{nozawa07} or \citet{bianchi07}. Recently,
\citet{biscaro16} derived a dust mass survival fraction between 6 and
11\%, using the analytical model of \citet{truelove99} and the SN dust
composition and size distribution of \citet{biscaro14}.

All models discussed above \citep[except ][]{biscaro16}
assumed that the dust grains are embedded in homogeneous
ejecta. \citet{silvia10, silvia12} performed 3D numerical simulations
of the interaction of planar shocks with dense spherical clumps for a
range of shock velocities, clump-ejecta density contrasts, and clump
metallicities. Taking the effects of cloud crushing into account, they
calculated the amount of grain destruction as a function of their
parameter grid using a SN dust compositions with the size distribution
calculated by \citet{nozawa03}. Silvia et al. find a dust mass survival
fraction between 7 and 99\% depending on the parameters of the
simulation and the dust composition. These results however are not
included in an evolutionary model of the reverse shock.
Our surviving fractions are significantly below those of
\citet{silvia10}, although the differences in approaches prevent direct
comparison. 

Our estimate of the surviving dust mass fraction has to be considered
a lower limit because we focused on the phenomena able to destroy the
dust. We did not consider conditions which could help the dust to
survive, for instance we set to zero the differential velocities that
grains of different size retain once they have leaved the ejecta
clouds. These velocities could allow the dust to escape the shell of
shocked ejecta before suffering excessive thermal sputtering.  Our
present work does not include betatron acceleration of the charged
grains induced by the magnetic field nor radiative evaporation of the
ejecta clumps. The impact of all these phenomena on the survival of
the dust will be evaluated in a follow-up paper.

An interesting fact to point out is that if most of the C and Si
in the FMKs is in dust, and if the dust destruction in such clumps is
mainly due to kinetic sputtering at low temperatures induced by the
cloud shock, then there would be little emission in lines such as the
UV lines of \ion{C}{iii}], \ion{C}{iv}, \ion{Si}{iii}] and
\ion{Si}{iv} or the [\ion{Fe}{ii}] lines even
if the abundances of those elements are substantial. The UV is, of
course, not observable in Cas~A because of the heavy galactic
reddening, but spectra of the SNRs N132D (in the LMC) and
1E0102.2-7219 (in the SMC) are available \citep{blair00}. These
spectra show some carbon, a little silicon, and some magnesium, providing
indicative lower limits to the carbon grain destruction or gas phase
abundance.

% TABLE 4 ************************************************************************
\begin{table}[t]
\begin{center}
\caption{\label{final_mass_tab} Percentage of dust mass, $M_{\rm
  residual}/M_{\rm init}$, surviving the passage of the reverse shock in
  Cas~A, evaluated in two different stages of supernova evolution
  identified by the corresponding values of $\alpha$. The result from  \citet{bianchi07} is reported for comparison.}
\begin{tabular}{c|c|c|c|c|c}
\hline
\hline
\noalign{\vskip 1mm}
 $T^a$  &  \multicolumn{2}{c|}{$\alpha$ = 0.33$^b$}  &  \multicolumn{2}{c|}{$\alpha$ = 0$^c$}  &  BS07$^d$
 \\
\noalign{\vskip 1mm}
 ($K$)  &  Am C   &  MgSiO$_3$  &  Am C  & MgSiO$_3$  & C+Fe$_3$O$_4$ \\
\noalign{\vskip 0.5mm}
\hline
\noalign{\vskip 0.5mm}
         $\sim$10$^{7}$   & {  16.9}\%  & {  13.4}\%  & { 
           15.9}\% & {  11.8}\%
         & \multirow{2}{*}{$\sim$7.5\%}  \\  
         $\sim$10$^{8}$   & {  13.3}\%  & {  10.4}\%  & { 
           12.3}\% & {  8.7}\% &   \\ 
\noalign{\vskip 0.5mm}
\hline
\noalign{\vskip 0.5mm}
\hline
\end{tabular}
\end{center}
\tablefoot{
\tablefoottext{a}{Temperature of the smooth ejecta.} \\
\tablefoottext{b}{Corresponding to 2004 (333 years after explosion).} \\
\tablefoottext{c}{After the reverse shock has reached the centre of
  the remnant ($\sim$8000 years after explosion).} \\
\tablefoottext{d}{Derived from \citet{bianchi07}: gas temperature 
between 10$^{7}$ and 10$^{8}$ K; surviving mass calculated after the reverse
shock has hit the centre of the remnant. } \\
}
\end{table}
% ******************************************************************************

%======================================
% Section 9
\section{Summary}\label{sec:summary}
%======================================

In this work, we approach the still-open question of the survival of
freshly formed dust against processing by the reverse shock inside the
cavity of supernova remnants, where the dust is actually formed. 
We have developed analytical or semi-analytical expressions for the
relevant quantities, which we implemented into a Monte Carlo
simulation. We focus on the specific case of the well-studied
supernova remnant Cas~A. Our main findings are summarized below.
\begin{itemize}
  \item[$\bullet$] Following the seminal work of \citet{truelove99}
    further developed by \citet{laming03}, we derived
    analytical expressions to describe the evolution of the forward
    and reverse shock in Cas~A, considering inhomogeneous ejecta
    (uniform core+power-law envelop) expanding into a non-uniform (power-law)
    ambient medium. This configuration is a closer match to the physical
    situation of the remnant.
  \item[$\bullet$] We adopted a set of parameters that allows us
    to reproduce the dynamics, and the density and temperature profile
    of the Cas~A ejecta, as deduced by recent observations. In particular,
    we assume that the dust forms in dense clumps immersed in smooth and
    tenuous ejecta.
  \item[$\bullet$] The ejecta in the clumps are heated up by the cloud
    shock generated by the impact between the clumps and the reverse
    shock. Because of their high oxygen content, the ejecta cool down
    quickly to temperatures below 1000 K. Under these conditions,
    kinetic sputtering is at work.
  \item[$\bullet$] For the smooth ejecta, we consider two
    temperatures: $\sim$10$^{8}$ K (from our calculations) and
    $\sim$10$^{7}$ K (average value from observations). Once the dust
    grains are injected into the smooth hot ejecta (because of their
    ballistic velocities or due to cloud destruction), they are processed
    by thermal sputtering. For the destruction of the clouds, we
      compared the effects of dynamical instabilities and thermal
    evaporation.
  \item[$\bullet$] We adopted a Monte Carlo approach to derive
    the mass and size distribution of the processed dust during the
    progression of the reverse shock inside the remnant. The sputtering
      of dust is calculated across the remnant and during its evolution,
      making use of the appropriate profiles that we have derived for relevant
      quantities, such as the velocity of the reverse shock and the
      ejecta density.
  \item[$\bullet$] Our simulation starts when the reverse shock
    touches the outer layer of ejecta ($\alpha$ = 1, $\sim$0.9 years after
    explosion), and stops when the reverse shock reaches the centre of the
    remnant ($\alpha$ = 0, $\sim$8000 years after explosion). 
  \item[$\bullet$] The fraction of surviving dust mass decreases while
    the reverse shocks proceeds towards the centre of the remnant and
    more ejecta are processed. Silicate grains (MgSiO$_3$) are more
    prone to erosion than amorphous carbon (Am C) grains. 
  \item[$\bullet$] The effect of rising the temperature of the smooth
    ejecta from 10$^{7}$ K to 10$^{8}$ K becomes noticeable in the
    inner layers of the ejecta (decreasing values of $\alpha$):
    $\alpha \lesssim$ 0.55 for both carbon and silicate dust.
  \item[$\bullet$] For 2004 ($\alpha$ = 0.33), we calculate a survival
    fraction of 16.9\% and 13.3\% for Am C, and of 13.4\% and 10.4\% for
    MgSiO$_3$, when the grains are immersed in a gas with a temperature of
    10$^{7}$ K and 10$^{8}$ K respectively. About 0.1 M$_\sun$ of dust is
    currently observed in Cas~A. If all this dust represents the surviving
    fraction, then about 0.8 -- 1 M$_\sun$ of dust must have formed in the
    ejecta. This suggests that almost all refractory elements condensed
    onto dust, implying a condensation efficiency that is somewhat higher
    than in models that predict the formation of 0.1 -- 0.5 M$_\sun$ of
    dust in CCSNe ejecta.
  \item[$\bullet$] After the reverse shock has reached the centre of
    the remnant ($\alpha$ = 0), we obtain the following survival
    fraction: 15.9\% for Am C and 11.8\% for MgSiO$_3$ for a gas
    temperature of 10$^{7}$ K, 12.3\% for Am C and 8.7\% for MgSiO$_3$
    when the gas temperature is 10$^{8}$ K. Kinetic sputtering destroys by
    itself 20\% of the carbon and 40\% of the silicate grains.
\end{itemize}

The surviving mass fractions depend on the morphology of the ejecta
and medium the remnant is expanding into, as well as the composition
and size distribution of the grains that formed in the
ejecta. Results, therefore, differ for different types of supernova. The
framework that we have developed to study dust processing and survival
in Cas~A can be adapted to other supernova remnants, both young (SN
1985A, the Crab Nebula, Kepler and Tycho) and more evolved (Cygnus Loop).

\begin{acknowledgements} 
We are grateful to our anonymous referee for the careful reading and
valuable suggestions and we would like to thank Martin Laming, Richard
Arendt, Mordecai-Mark Mac Low, and Alex Hill for useful and stimulating
discussions. E. R. M. wishes to acknowledge the support from a Marie
Curie Intra-European Fellowship within the 7th European Community
Framework Programme under project number PIEF-GA-2012-328902 NANOCOSMOS.
E. D. acknowledges the support of NASA’s 13-ADAP13-0094.
\end{acknowledgements}

\bibliographystyle{aa}
%\bibliography{Bib_Cas_A}

\clearpage

\appendix

\section{Derivation of the equations for the forward and reverse
  shocks}\label{app_shocks}

The starting point to derive the equation for the blast-wave radius during the initial envelope phase
(Eq.~\ref{blast_radius_envelope_eq}) is the first line of equation A7
in TM99, which  we report here for clarity, i.e.
\begin{eqnarray}\label{blast_radius_appendix}
  R^{*(3-s)/2}_{\rm b} &  = & (3-s)\,\left(\frac{\ell_{\rm ED}}{\phi_{\rm
        ED}}\right)^{1/2} \,\times \\ \nonumber
  &  & \times \; \frac{f_n^{1/2}}{3-n}\,\left[
    1-\left(\frac{w_{\rm b}}{\ell_{\rm ED}}\right)^{(3-n)/2}\right]
  \;\;\;\;\;\; w_{\rm core} \le \left(\frac{w_{\rm b}}{\ell_{\rm ED}}\right)
  \le 1,  
\end{eqnarray}
where 
$w_{\rm b} \equiv [R_{\rm b}(t)/t]/v_{\rm ej}$ and the term in squared
brackets represents the velocity of the free expanding ejecta at a
radius equal to $R_{\rm b}$. Using Eqs.~\ref{more_ch_quantities}
 and \ref{energy_ratio} we can rewrite $w_{\rm b}$ in
terms of starred variables as follows:
\begin{equation}\label{w_b_star}
  w_{\rm b} = \frac{R^*_{\rm b}/t^*}{v^*_{\rm ej}} = \left(\frac{\bar
      \alpha}{2}\right)^{1/2} \, \frac{R^*_{\rm b}}{t^*}.
\end{equation}
Substituting Eq.~\ref{w_b_star} in Eq.~\ref{blast_radius_appendix}
 we can derive equation A8 in TM99,
which expresses $t^*$ as a function of $R^*_{\rm b}$ (below),
\begin{eqnarray}\label{t_star}
  t^*(R^*_{\rm b}) &  = & \left(\frac{\bar \alpha}{2}\right)^{1/2}
  \frac{R^*_{\rm b}}{\ell_{\rm ED}}\, \times \\ \nonumber
   &  &  \left[1-\left(\frac{3-n}{3-s}\right) \left(\frac{\phi_{\rm
          ED}}{\ell_{\rm ED}\,f_n}\right)^{1/2} R^{*(3-s)/2}_{\rm b}\right]^{-2/(3-n)}.
\end{eqnarray}
To derive Eq.~\ref{blast_radius_envelope_eq}, we consider the
case when $w_{\rm core}\rightarrow0$. This implies $f_n \rightarrow0$
(see Eq.~\ref{f_n}) and allows us to neglect the term ``1''
within square brackets in equation \ref{t_star}. For $w_{\rm
  core}\rightarrow0$, Eqs.~\ref{f_n} and \ref{energy_ratio}
lead to
\begin{equation}
  f_n \sim \frac{3}{4 \pi}\, \frac{n-3}{n}\, w_{\rm core}^{n-3}
  \;\;\;\;\;\;\;\;\; \bar \alpha \sim \frac{3}{5}\, \frac{n-3}{n-5} \,w_{\rm core}^2 .
\end{equation}
Substituting the above expressions and Eq.~\ref{v_core_star} in the
modified Eq.~\ref{t_star}, we derive $R^*_{\rm b}(t^*)$ during the
envelope phase (Eq.~\ref{blast_radius_envelope_eq}).

To derive the time $t^{*}_{\rm core}$ at which the reverse shock hits the
ejecta core, we used the fact that $v^{*}_{\rm core} = R^*_{\rm
  r}/t^{*}_{\rm core}$ and $R^*_{\rm r} = R^*_{\rm b}/\ell_{\rm
ED}$. This leads to
\begin{equation}
  t^{*}_{\rm core} = \frac{R^*_{\rm b}}{\ell_{\rm ED}\,v^{*}_{\rm core}}.
\end{equation}
Substituting $R^*_{\rm b}$ with the envelope solution at $t^* =
t^{*}_{\rm core}$ we obtained Eq.~\ref{t_core_star} for $t^{*}_{\rm core}$.

The radius of the blast wave during the core phase is given, in
implicit form, by the second line of equation A7
in TM99, which is written as
\begin{eqnarray}\label{blast_radius_appendix_core}
  R^{*(3-s)/2}_{\rm b} & = & (3-s)\,\left(\frac{\ell_{\rm ED}}{\phi_{\rm
        ED}}\right)^{1/2} \, \frac{f_0^{1/2}}{3}\,\left[
   w^{3/2}_{\rm core} -\left(\frac{w_{\rm b}}{\ell_{\rm
         ED}}\right)^{3/2}\right]  + \, \\ \nonumber
 &  & + \; \frac{f_n^{1/2}}{3-n}\left(1- w^{(3-n)/2}_{\rm core}\right)
  \;\;\;\;\;\;\;\;\;\;\;\;\, 0 \le \left(\frac{w_{\rm b}}{\ell_{\rm ED}}\right)
  \le w_{\rm core}.
\end{eqnarray}
For $s \geq 2$, the derivation of $R^*_{\rm b}$ from
Eq.~\ref{blast_radius_appendix_core} is extremely complicated, 
% Only numerically? -> no analytical solution?
thus we decided to adopt the same approach as \citet{laming03}. This
procedure extends the blast-wave solution for the envelope phase into
the core phase, matching it to the solution that is appropriate in
the Sedov--Taylor (ST) limit. During the ST phase, the mass of the
swept-up material exceeds $M_{\rm ej}$ and the remnant expands
following a self-similar motion. The appropriate solution in the ST
limit is the offset power-law given in Eq.~A12 from TM99, i.e.
\begin{eqnarray}\label{radius_ST_offset_eq}
  R^*_{\rm b}(t^*) & = & \left[R^{*(5-s)/2}_{\rm ST} +
    \xi^{1/2}_s \left(t^*-t^*_{\rm ST}\right)\right]^{2/(5-s)}\;\; {\rm
  with} \\ \nonumber
  \xi_s & = & \frac{(5-s)(10-3s)}{8\pi},
\end{eqnarray}  
which is obtained integrating the classical ST solution from a
characteristic radius $R^*_{\rm ST}$ and time $t^*_{\rm ST}$ to the
current radius and time, $R^*_{\rm b}$ and $t^*$, respectively. The
blast-wave velocity (Eq.~\ref{blast_velocity_core_eq}) comes directly
from Eq.~\ref{radius_ST_offset_eq}. To derive the time, $t^{*}_{\rm
conn}$ at which the envelope solution and the ST offset power-law
solution connect, we equated Eqs.~\ref{blast_velocity_envelope_eq} and
\ref{blast_velocity_core_eq} to derive an expression for $t^*$ as a
function of $R^*_{\rm b}$. Then, we substituted in this expression the
envelope solution for $R^*_{\rm b}$
(Eq.~\ref{blast_radius_envelope_eq}) obtaining Eq.~\ref{t_conn_star}
for $t^{*}_{\rm conn}$. The final expression for the blast-wave radius
valid for $t^* > t^{*}_{\rm conn}$ (Eq.~\ref{blast_radius_core_eq}),
is derived from ST offset power-law solution
(Eq.~\ref{radius_ST_offset_eq}) imposing the conditions $R^*_{\rm ST}
= R^*_{\rm b}(t^* =t^{*}_{\rm conn})$ (where $R^*_{\rm b}$ is the
envelope solution) and $t^*_{\rm ST} = t^{*}_{\rm conn}$, and using
for $v^{*}_{\rm core}$ the expression from Eq.~\ref{v_core_star}.

To calculate the velocity of the reverse shock in the frame of the
unshocked ejecta during the envelope phase ($t^* \leq t^{*}_{\rm
core}$), we start from the following equation:
\begin{equation}\label{v_r_eq}
  v^{*}_{\rm r} = v^*(R^*_{\rm r}, t^*) - \frac{{\rm d} R^*_{\rm r}}{{\rm d} t^*}.
\end{equation} 
The first term on the right side of the equation represents the
velocity of the freely expanding ejecta when they encounter the
reverse shock and equals $R^*_{\rm r}/t^*$, while the second term is
the velocity of the reverse shock in the rest frame. Using 
Eqs.~\ref{blast_velocity_envelope_eq} and
\ref{reverse_radius_envelope_eq}, with few manipulations we obtain
Eq.~\ref{reverse_velocity_envelope_eq} for the reverse shock
velocity.

Following \citet{laming03}, we assume that during the core phase ($t^*
> t^{*}_{\rm core}$) the reverse shock velocity remains constant
at the value given by Eq.~\ref{reverse_velocity_envelope_eq}
evaluated for $t^* = t^{*}_{\rm core}$. This leads immediately to
Eq.~\ref{reverse_velocity_core_eq}.

To derive the radius of the reverse shock during the core phase, we
rewrite Eq.~\ref{v_r_eq} as follows:
\begin{equation}
  v^{*}_{\rm r} = -t^*\,  \frac{{\rm d}}{{\rm d} t^*} \left(\frac{R^*_{\rm r}}{t^*}\right).
\end{equation}
The solution of the above equation evaluated between $t^{*}_{\rm
  core}$ and $t^*$ provides the expression for the reverse shock
radius $R^*_{\rm r}(t^*)$ (Eq.~\ref{reverse_radius_core_eq}), where
Eq.~\ref{blast_radius_core_eq} has to be used for $R^*_{\rm b}$.

\section{A statistical approach to calculate the fraction of dust
  surviving kinetic sputtering}\label{app_column_density}

% FIGURE B1 *************************************************************
%
 \begin{figure} \centering
  \includegraphics[width=3.0in]{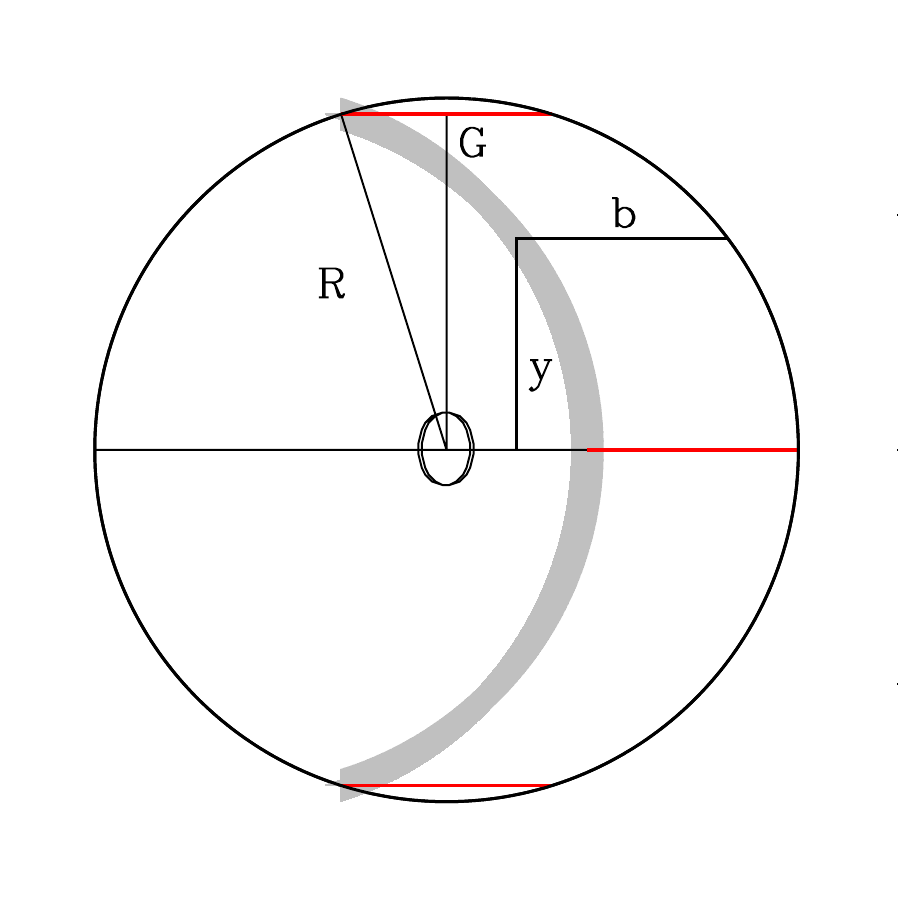}
   \caption{ Cross section of a spherical ejecta cloud of radius $R$
     in Cas~A. The grey area is the locus of all points in the cloud that
     are at the same distance $b$ (represented by the red lines) from the
     surface of the cloud.  }
\label{fig:nh}
\end{figure}
% ********************************************************************

We report here an analytical calculation
to estimate, for a given grain size, the fraction of dust able to
survive kinetic sputtering in the ejecta clouds. This statistical approach
considers a normalized distribution of purely 
geometrical column densities, while the velocity dependency, which
is related to the particle size, is given by Eq.~\ref{eq:m_m}. 

Figure~\ref{fig:nh} is a cross section of a
sphere of radius $R$ representing an ejecta cloud of Cas~A. The shock impinges
on the cloud from the right.
The grey region in the figure is the locus of all the dust
grains that are at a distance $b$ from the cloud's surface. The volume
occupied by all the grains at a vertical distance $y$ from the x-axis
and a distance between $b-{\rm d}b/2$ and $b+{\rm d}b/2$ from the surface is
given by the projected volume of $2\,\pi y\, {\rm d}y\, {\rm d}b$, where $y_{\rm m}$, the
maximum value of $y$ for a given $b$ is equal to $\overline{OG} =
[R^2-(b/2)^2]^{1/2}$. The differential volume d$V(b)$ is then
\begin{equation}
\label{vol} {\rm d}V(b) ={\rm d}b\, \int_0^{y_{\rm m}}\, 2\pi y\, {\rm
  d}y = \pi\,
[R^2-(b/2)^2]
.\end{equation} 
The distribution of distances, normalized to the
volume of the cloud is then given by
\begin{equation}
\label{ } \frac{1}{V}\, \frac{{\rm d}V(b)}{{\rm d}b}= \frac{3}{4}\,\frac{R^2-b^2/4}{R^3}
.\end{equation} 
The average column density $\left<b\right>$ is
\begin{equation}
\label{ } \left<b\right> = \frac{3}{4}\, \int_0^{2R}\,
\left(\frac{R^2-b^2/4}{R^3}\right)\, b\, {\rm d}b = \frac{3}{4}\, R
.\end{equation} 
The column density, $N$, traversed by a dust grain until it reaches
the clouds's surface is given by $N=n_{\rm gas} b$, where $n_{\rm
gas}$ is the gas density.  So the normalized distribution of column
densities is given by
\begin{equation}
\label{ } \frac{{\rm d}f(N)}{{\rm d}N} = \frac{3}{2} \, \frac{N_{\rm
    k}^2 - N^2}{N_{\rm k}^3}
,\end{equation} 
where $N_{\rm k} = 2 n_{\rm gas} R$ is the column density through the
centre of the cloud.

We let $f_{\rm M}(v_{\rm gr}(0),N) \equiv m_{\rm gr}(t)/m_{\rm gr}(0)$ be the fraction of
surviving dust, with an initial velocity $v_{\rm gr}(0)$ after it
traversed a columnn density $N$. The mass fraction of the dust in a
cloud that survives the passage of the reverse shock,
$\xi[v_{\rm cloud}(\alpha)]$ is given by the integral
\begin{equation}
\label{fracM} \xi[v_{\rm cloud}(\alpha)] = \int_0^{N_{\rm k}}\ f_{\rm M}(v_{\rm cloud}(\alpha),N)\,
\frac{{\rm d}f(N)}{{\rm d}N},
\end{equation} 
where $v_{\rm cloud}(\alpha)$ is the velocity of the cloud
shock through a clump located in the ejecta layer $\alpha$, 
and where we assumed that because of the
strong compression, the reverse shock gives all grains an initial
velocity $v_{\rm cloud}(\alpha)$, relative to the gas.

The statistical approach described above is only applicable for the
largest grain sizes, for which the maximum column density through the
clump, $2n_{\rm gas}\, R$ is smaller than that required to slow them down
to the kinetic sputtering threshold. For smaller grain sizes, a
Monte Carlo approach, as described in Sect.~\ref{sec:MC_simulation},
is needed to determine the actual amount of kinetic sputtering before
the grain exits the clump, or before the clump is destroyed by the
reverse shock.

\end{document}